%% file: Paper_QHT_final_corrected_II.tex
\documentclass[letterpaper,11pt]{article}

\usepackage{scrextend}
\usepackage{amsmath,amssymb,epsfig,bbm}
\usepackage[active]{srcltx}
\usepackage[all]{xypic}
\usepackage{MnSymbol}
\usepackage{slashed}

\usepackage{color}

\usepackage{dsfont}

\definecolor{co}{cmyk}{0,0.7,0.3,0}
\definecolor{darkgreen}{cmyk}{1,0,1,.2}
\definecolor{m}{rgb}{1,0.1,1}

\newcommand{\be}{\begin{equation}}
\newcommand{\ba}{\begin{eqnarray}}
\newcommand{\ea}{\end{eqnarray}}
\newcommand{\nn}{\nonumber}

\def\a{\alpha}

\def\d{\delta}
\def\e{\epsilon}

\def\m{\mu}
\def\n{\nu}
\def\oo{\omega}

\def\r{\rho}

\def\x{\xi}

\def\G{\Gamma}

\def\OO{\Omega}
\def\P{\Pi}

\def\ca{{\cal A}}
\def\cb{{\cal B}}

\def\cd{{\cal D}}

\def\cf{{\cal F}}

\def\ch{{\cal H}}

\def\cl{{\cal L}}

\def\co{{\cal O}}

\def\cq{{\cal Q}}

\makeatletter
\newcommand{\eqnum}{\refstepcounter{equation}\textup{\tagform@{\theequation}}}
\makeatother

\newcommand{\pa}{\partial}
\newcommand{\tr}{\mbox{tr}}

\newcommand{\C}{{\Bbb C}}

\newtheorem{thm}{Theorem}[subsection]

\newtheorem{definition}[thm]{Definition}

\newcommand{\bbC}{{\Bbb C}}
\newcommand{\bbR}{{\Bbb R}}
\newcommand{\cF}{{\cal F}}

\fontfamily{yfrak}

\begin{document}

\vskip 25mm

\begin{center}

{\large\bfseries  Quantum Holonomy Theory
}


\vskip 6ex

Johannes \textsc{Aastrup}
\footnote{email: \texttt{aastrup@math.uni-hannover.de}} \&
Jesper M\o ller \textsc{Grimstrup}\,\footnote{email: \texttt{jesper.grimstrup@gmail.com}}\\ 
\vskip 3ex  


\end{center}

\vskip 3ex

\begin{abstract}

We present quantum holonomy theory, which is a non-perturbative 
theory of quantum gravity coupled to fermionic degrees of freedom. The theory is based on a $C^*$-algebra that involves holonomy-diffeo-morphisms on a 3-dimensional manifold and which encodes the canonical commutation relations of canonical quantum gravity formulated in terms of Ashtekar variables. Employing a Dirac type operator on the configuration space of Ashtekar connections we obtain a semi-classical state and a kinematical Hilbert space via its GNS construction. 
We use the Dirac type operator, which provides a metric structure over the space of Ashtekar connections, to define a scalar curvature operator, from which we obtain a candidate for a Hamilton operator. We show that the classical Hamilton constraint of general relativity emerges from this in a semi-classical limit and we then compute the operator constraint algebra. Also, we find states in the kinematical Hilbert space on which the expectation value of the Dirac type operator gives the Dirac Hamiltonian in a semi-classical limit and thus provides a connection to fermionic quantum field theory. Finally, an almost-commutative algebra emerges from the holonomy-diffeomorphism algebra in the same limit.

\end{abstract}

\newpage
\tableofcontents

\section{Introduction}
\setcounter{footnote}{0}


The search for fundamental principles in Nature is the leitmotiv of modern physics. The aim is to explain rather than describe and what constitutes an explanation is the invocation of a principle. 
The dream is to uncover an ultimate principle behind {\it it all} -- the fundamental theory -- which will end the reductionist ladder of descent and leave 
 no door to peek behind. 

The framework of quantum holonomy theory proposes such a {\it first principle}. The theory is built over an algebra that encodes how diffeomorphisms act on spinors. Thus, the fundamental building blocks are "moving stuff in space" and as such seem immune to further reduction: the question {\it "what are diffeomorphisms made of?"} makes little sense. 

What we find is a non-perturbative and background independent  
quantum mechanics of diffeomorphisms, where we on the one hand have an algebra generated by holonomy-diffeomorphisms on a three-dimensional manifold $M$ and on the other hand conjugate operators given by canonical translations on a configuration space of connections, over which the holonomy-diffeomorphisms form a non-commutative algebra of functions. The interaction between the algebra of holonomy-diffeomorphisms, which we denote by $\mathbf{HD}(M)$, and the translation operators encodes the canonical commutation relations of canonical quantum gravity formulated in terms of Ashtekar variables \cite{Ashtekar:1986yd,Ashtekar:1987gu}. This means that we establish a direct link between an algebra generated by holonomy-diffeomorphisms and canonical quantum gravity.

This construction comes with a very high degree of canonicity: once the number of spatial dimensions is chosen it only depends on a choice of gauge group, where $SU(2)$ matches our choice of three spatial dimensions via the rotation group.

Once we have defined the algebra generated by holonomy-diffeomorphisms and translation operators, denoted $\mathbf{QHD}(M)$, the question arises whether states exist on this algebra. This issue is central to this paper. To address it we first combine the conjugate translation operators into a Dirac type operator. This constitutes a canonical metric {\it over} a configuration space of Ashtekar connections, i.e. a geometry over a configuration space of spatial geometries.
The Dirac type operator entails a flow-dependent version of the $\mathbf{QHD}(M)$ algebra and it is on this algebra that we find a semi-classical state that provides us with a kinematical Hilbert space via its GNS construction. 
Thus, each classical point gives rise to a {\it different} kinematical Hilbert space. Furthermore, we find evidence that the overlap function, which measures the transition from one semi-classical approximation to another, vanishes. This suggest that there will be no quantum interference between different semi-classical approximations. However, on certain operators in the GNS construction of a semi-classical state the expectation value can look like a different classical geometry.

The introduction of the Dirac-type operator 
opens up further avenues of analysis. We obtain a tentative candidate for a Hamilton constraint operator by constructing a scalar curvature operator and show that the  classical Hamilton constraint emerges in a semi-classical limit from this operator. This shows that the framework of quantum holonomy theory produces general relativity in a semi-classical limit.

A central test for any theory of quantum gravity is to check whether the constraint algebra closes off-shell. This determines to what extend general covariance is maintained in the quantum theory and is usually understood as a necessary requirement for the theory to be internally consistent. With a candidate for a Hamilton constraint operator we are in a position where we can perform this test. We first compute the commutator between two Hamilton constraint operators -- characterized by different choices of lapse fields -- and find that it reproduces off-shell the structure of the corresponding classical Poisson bracket up to an anomalous term, which vanishes at all finite orders in perturbation theory. This result matches the observation that the action of the diffeomorphism group on the flow-dependent $\mathbf{QHD}(M)$ algebra is not strongly continuous and thus cannot have infinitesimal generators. The exception to this is exactly the perturbative regime, where we thus find that the constraint algebra does close. Thus, we find that the 'Hamilton-Hamilton' sector of the quantum constraint algebra is free of anomalies, off-shell, in exactly the regime where the constraint algebra is a meaningful object.

This computation provides us with a candidate for a diffeomorphism constraint operator and we continue to compute the commutator between two different diffeomorphism constraint operators. As these computations are very involved we perform them with a simplified version of the Hamilton constraint operator. We believe that this simplification is the explanation for the emergence of non-physical anomalous terms in this computation. 
Nevertheless, the fact that our framework permits this type of computations is, in our opinion, very encouraging and the result calls for a computation, that involves the full Hamilton constraint operator derived from the scalar curvature operator.

In another direction of analysis we show that the construction of the Dirac type operator leads to a class of states on which its expectation value gives a spatial Dirac operator in a semi-classical limit. A certain transformation, that introduces the lapse and shift fields, entails from these expectation values the principle part of the Dirac Hamiltonian. This indicates that the theory also harbors quantized matter degrees of freedom and that these are canonical.
Also, we find that the $\mathbf{HD}(M)$ algebra descends, in a semi-classical limit, to a tensor product that involves an almost-commutative algebra. Almost-commutative algebras form the cornerstone in the formulation of the standard model of particle physics coupled to gravity in terms of non-commutative geometry \cite{Connes:1996gi,Chamseddine:2007hz}, and the emergence of such an algebra -- in company with a spatial Dirac operator -- opens up the possibility for emergent gauge degrees of freedom and a connection to the mathematics of the standard model itself. All together we find strong indications that a theory of quantum gravity based on the $\mathbf{QHD}(M)$ algebra naturally involves both additional fermionic and bosonic degrees of freedom.




The construction of the semi-classical state is carried out with a formulation of the $\mathbf{HD}(M)$ algebra in terms of infinite sequences of lattice approximations.  At the present level of analysis we are unable to determine which type of diffemorphisms this formulation is in fact capable of describing and whether we are forced to restrict ourselves to analytic diffeomorphisms. Also, more analysis is needed to describe the precise mathematical nature of the continuum limit, on which the lattice formulation of the $\mathbf{HD}(M)$ algebra depends. To be more precise, we know that the limit exist but the exact details of the limit are not completely known.


\subsection{Outline of the central idea}

The cornerstone in the theory is the algebra $\mathbf{HD}(M)$, which is generated by holonomy-diffeomorphisms on a 3-dimensional manifold $M$. A holonomy-diffeomorphism maps a connection $\nabla$ into an operator acting on spinors on $M$
$$
\ca \ni \nabla \rightarrow e^X(\nabla)
$$
where $\ca$ is the space of all smooth connections in a certain bundle and where $e^X(\nabla)$ denotes a holonomy-diffeomorphisms along a vector-field $X$. Thus, the algebra $\mathbf{HD}(M)$ depends on the manifold $M$ as well as a choice of gauge group corresponding to the bundle over $M$. We choose a 3-dimensional manifold and the group $SU(2)$ since this corresponds to canonical quantum gravity formulated in terms of Ashtekar variables\footnote{To be precise, the choice of $SU(2)$ corresponds to general relativity with an Euclidian signature. The Lorentzian signature corresponds to a connection, which takes values in the self-dual section of $\mathfrak{sl}(2,\mathbb{C})$, the Lie-algebra of $SL(2,\mathbb{C})$. We comment on this in section \ref{REAL}. } \cite{Ashtekar:1986yd,Ashtekar:1987gu}. Furthermore, we choose the trivial bundle.

Once we have the algebra $\mathbf{HD}(M)$ it is natural to consider translations on the space $\ca$ of connections
$$
U_{\oo} \xi(\nabla) = \xi(\nabla - \oo)
$$
where $\oo$ is a one-form with values in the Lie-algebra of $SU(2)$. We find the relation
$$
(U_\oo e^X U_\oo^{-1})(\nabla) = e^X (\nabla + \oo)
$$
which is in fact an integrated version of the canonical commutation relations of quantum gravity formulated in terms of Ashtekar variables. Thus, we find that the algebra of holonomy-diffeomorphisms leads us directly into the realm of canonical quantum gravity.\\


Note that the algebra $\mathbf{HD}(M)$ is manifestly non-commutative. This means that an approach to quantum gravity based on this algebra also plays into the heartland of non-commutative geometry. \\

In order to approach the problem of finding Hilbert space representations of these algebraic structures we first introduce operators that correspond to infinitesimal translations on $\ca$. 
Using infinite sequences of lattice approximations we construct a Dirac-type operator $D$ from these infinitesimal translation operators and consider the algebra generated by $\mathbf{HD}(M)$ and commutators with $D$. This algebra, which also encodes the canonical commutation relations of quantum gravity, has a state that exist independently of the lattice approximations. We apply the GNS construction on this state to form a kinematical Hilbert space.

With a Dirac-type operator we then consider one-forms, which in the language of non-commutative geometry have the form $B= a[D,b]$, where $a$ and $b$ are elements of the $\mathbf{HD}(M)$ algebra. 
We also consider the corresponding curvature operators
$$
\cf_{B} = [D,B_{}] + \frac{1}{2}[B_{},B_{}]\;,
$$
which are curvature operators {\it over} the configuration space of Ashtekar connections, and find that the densitized Hamiltonian constraint of general relativity formulated in terms of Ashtekar variables emerges in a semi-classical limit from a scalar curvature operator built from operators $\cf_B$ in a way that involves the orientation of the manifold $M$.\\


The paper is organized as follows: In section \ref{firsttask} we first introduce the $\mathbf{HD}(M)$, $\mathbf{QHD}(M)$ and $\mathbf{dQHD}(M)$ algebras and show that the latter two encode the canonical commutation relations of canonical quantum gravity formulated in terms of Ashtekar variables. In section 3 we then introduce a lattice formulation of the holonomy-diffeomorphism algebra and consider states on $\mathbf{QHD}(M)$ and $\mathbf{dQHD}(M)$ and reach the conclusion that such are unlikely to exist. We define the $\mathbf{dQHD}^*(M)$ algebra via the Dirac type operator and show that semi-classical states exist hereon. Subsequently we obtain a kinematical Hilbert space via the GNS construction. 
In section 4 we show that a spatial Dirac operator as well as the principal part of the Dirac Hamiltonian emerges in a semi-classical limit from the expectation value of the Dirac type operator on certain states in the kinematical Hilbert space. We then consider the dynamics of general relativity in section 5 and show how a candidate for a Hamiltonian constraint operator is obtained from a scalar curvature operator. We then compute the constraint algebra. In section 6 we show that an almost-commutative algebra emerges from the $\mathbf{HD}(M)$ algebra in a semi-classical limit.  
The use of lattice approximations naturally entails the question of background independency, which is discuss in section 7, where we also work out an abstract formulation of the $\mathbf{dQHD}^*(M)$  algebra. 
Then we present in section 8 a tentative analysis of the overlap function, which suggest that it vanishes, and in section 9 we consider a complex Ashtekar connection. 
 Finally, in section 10 we show that the Dirac operator has an interpretation in terms of the quantized volume of the manifold M.
 We conclude in section 11 and add an appendix, which contains computations on the operator constraint algebra.

\section{The Quantum Holonomy-Diffeomorphism algebra}
\label{firsttask}

The first task is to introduce the Holonomy-Diffeomorphism algebra. This algebra was first described in \cite{Aastrup:2012vq, AGnew}, where its spectrum was analyzed. The extension to the Quantum Holonomy-Diffeomorphism algebra then follows canonically.\\

\subsection{The Holonomy-Diffeomorphism algebra}
\label{beent}

\begin{figure}[t]
\begin{center}
\resizebox{!}{4.5cm}{
 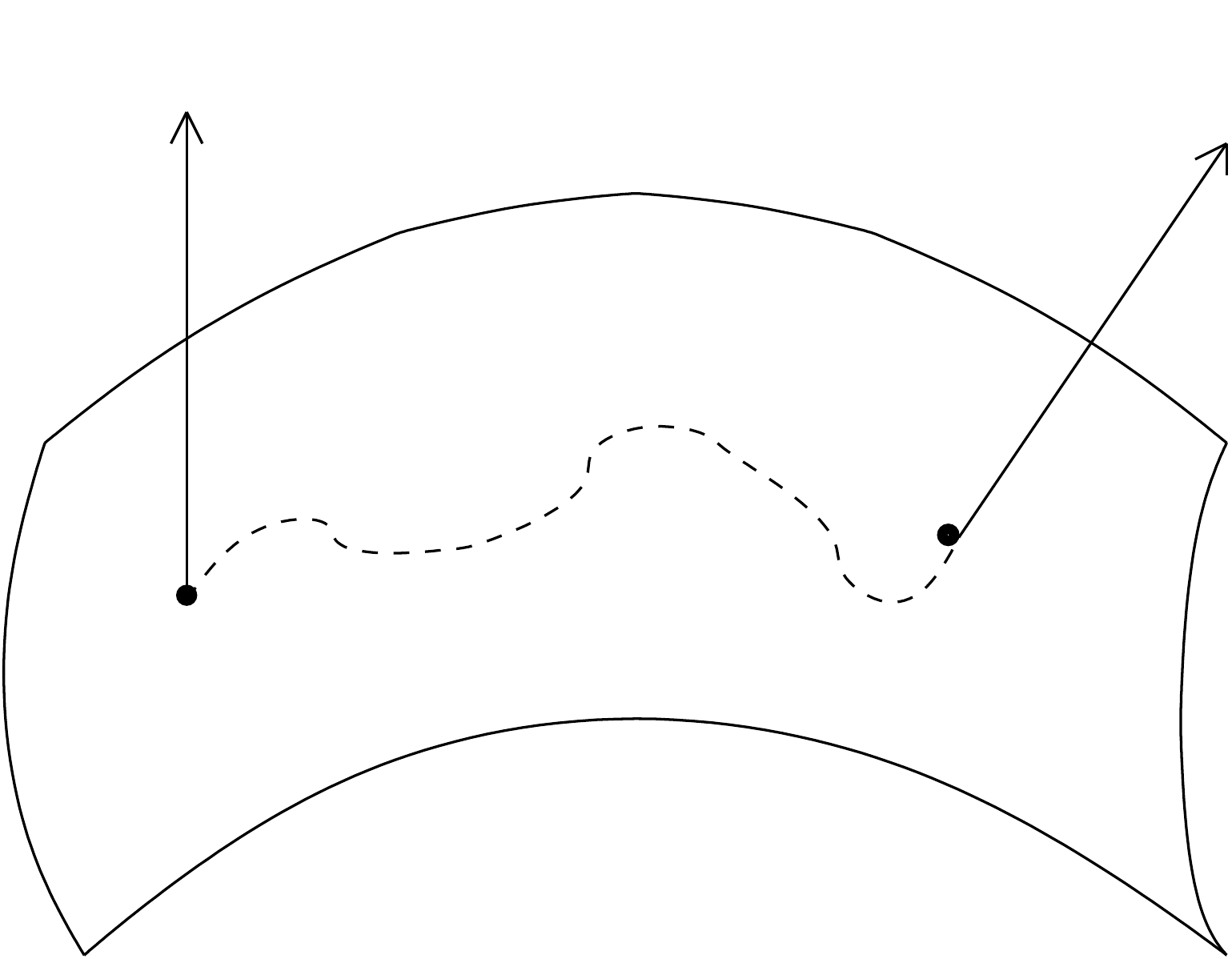}
\end{center}
\caption{An element in $\mathbf{H D}$ will parallel transport a vector on $M$ along the flow of a diffeomorphism.}
\label{lenin}
\end{figure}

Let $M$ be a connected $3$-dimensional manifold. We consider the two dimensional trivial vector bundle $S=M\times \C^2$ over $M$, and we consider the space of $SU(2)$ connections acting on the bundle. Given a metric $g$ on $M$ we get the Hilbert space $L^2(M,S,dg)$, where we equip $S$ with the standard inner produkt. Given a diffeomorphism $\phi:M\to M$ we get a unitary operator $\phi^*$ on  $L^2(M,S,dg)$ via
$$( \phi^* (\xi ))(\phi (m) )= (\Delta \phi )(m)  \xi (m) , $$
where  $\Delta \phi (m)$ is the volume of the volume element in $\phi (m)$ induced by a unit volume element in $ m$ under $\phi $.      

Let $X$ be a vectorfield on $M$, which can be exponentiated, and let $\nabla$ be a $SU(2)$-connection acting on $S$.  Denote by $t\to \exp_t(X)$ the corresponding flow. Given $m\in M$ let $\gamma$ be the curve  
$$\gamma (t)=\exp_{t} (X) (m) $$
running from $m$ to $\exp_1 (X)(m)$. We define the operator 
$$e^X_\nabla :L^2 (M , S, dg) \to L^2 (M ,  S , dg)$$
in the following way:
we consider an element $\xi \in L^2 (M ,  S, dg)$ as a $\C^2$-valued function, and define 
$$  (e^X_\nabla \xi )(\exp_1(X) (m))=  ((\Delta \exp_1) (m))  \hbox{Hol}(\gamma, \nabla) \xi (m)   .$$
Here  $\hbox{Hol}(\gamma, \nabla)$ denotes the holonomy of $\nabla$ along $\gamma$. Again, the factor $(\Delta \exp_1) (m)$ is accounting for the change in volumes, rendering $e^X_\nabla$ unitary.  \\

Let $\ca$ be the space of $SU(2)$-connections. We have an operator valued function on $\ca$ defined via 
$$\ca \ni \nabla \to e^X_\nabla  . $$
We denote this function $e^X$. 

Denote by $\cf (\ca , \cb (L^2(M, S,dg) ))$ the bounded operator valued functions over $\ca$. This forms a $C^*$-algebra with the norm
$$\| \Psi \| =  \sup_{\nabla \in \ca} \{\|  \Psi (\nabla )\| \}, \quad \Psi \in  \cf (\ca , \cb (L^2(M, S,dg )) ) $$

For a function $f\in C^\infty_c (M)$ we get another operator valued function $fe^X$ on $\ca$.

\begin{definition}
Let 
$$C =   \hbox{span} \{ fe^X |f\in C^\infty_c(M), \ X \hbox{ exponentiable vectorfield }\}  . $$
The holonomy-diffeomorphism algebra $\mathbf{H D} (M,S,\ca)   $ is defined to be the $C^*$-subalgebra of  $\cf (\ca , \cb (L^2(M,S,dg )) )$ generated by $C$.

We will often denote $\mathbf{H D} (M,S,\ca)   $ by $\mathbf{H D}   $ or $\mathbf{H D}  (M)$ when it is clear which $M$, $S$, $\ca$ is meant.
We will by $\ch \cd (M,S,\ca)   $ denote the  $*$-algebra generated by $C$.
\end{definition}

It was shown in \cite{AGnew} that  $\mathbf{H D} (M,S,\ca)   $ is independent of the chosen metric $g$.

\subsection{The Quantum Holonomy-Diffeomorphism algebra}

Let $\mathfrak{su}(2)$ be the Lie-algebra of $SU(2)$. It is well-known that two connections in $\mathcal{A}$ differs by an element in $\Omega^1(M,\mathfrak{su}(2))$, and that for $\nabla \in \mathcal{A}$ and $\omega \in \Omega^1(M,\mathfrak{su}(2))$, $\nabla +\omega$ defines a connection in $\mathcal{A}$.  Thus  a section $\omega \in \Omega^1(M,\mathfrak{su}(2))$ induces a transformation of $\ca$, and therefore an operator $U_\omega $ on $\mathcal{F}(\ca,  \cb(L^2 (M ,  S,g)))$ via   
$$U_\omega (\xi )(\nabla) = \xi (\nabla - \omega) .$$ 
%
Note that $U^{-1}_\omega=U_{-\omega}$. 

\begin{definition}

 $\;$ Let us denote by $\mathbf{QHD}(M,S,\ca)$ the sub-algebra of $\cf (\ca , \cb (L^2(M, S)) )$ generated by $\mathbf{HD}(M,S,\ca)$ and all the operators $U_{ \omega} $, $\omega \in \Omega^1(M,\mathfrak{su}(2))$. 
We will often denote $\mathbf{QHD}(M,S,\ca)$ by $\mathbf{QHD}$ or $\mathbf{QHD}(M)$ when it is clear, which $M$, $S$, $\ca$ is meant. 
We call $\mathbf{QHD}$ the Quantum-Flow algebra or the Quantum Holononomy-Diffeomorphism algebra.
\end{definition}

We note that we have the relation 
\begin{equation} \label{konj}
(U_{\omega}f e^X U_{ \omega}^{-1}) (\nabla) =f e^X (\nabla + \omega )  , 
\end{equation}
where $f\in C^ \infty_c(M)$. However $\mathbf{QHD}(M)$ is not a cross product of $\mathbf{HD}(M)$ with the additive group $\Omega^1 (M,\mathfrak{su}(2))$, since the function of operators given by $\nabla \to e^X_{\nabla +  \omega}$ need not be in $\mathbf{HD}(M)$.

\subsection{The infinitesimal $\mathbf{QHD}(M)$ algebra}

To get closer to the formulation of the holonomy-flux-algebra\footnote{By the holonomy-flux algebra we refer to the algebra used in loop quantum gravity, see \cite{AL1}.} and canonical quantization of gravity (see \cite{Aastrup:2012jj} for setup and notions) we need the infinitesimal version of $U_{t \omega}$. We simply do this by formally defining 
$$E_\omega  =\frac{d}{dt}U_{  t  \omega}|_{t=0} . $$
Due to the relation (\ref{konj}) we get 
\begin{equation} \label{flowkan}
[ E_\omega , e^X_\nabla ]= \frac{d}{dt}e^X_{\nabla +t\omega}|_{t=0}  . 
\end{equation}
Thus the infinitesimal version of the Quantum Holonomy-Diffeomorphism algebra is generated by the flows and the variables $\{ E_\omega\}_{\omega \in \Omega (M,T)}$. We denote this algebra by $\mathbf{dQHD}(M)$. \\

We note, that 
$$
E_{\omega_1+\omega_2}=E_{\omega_1}+E_{\omega_2\;.}
$$
This follows since the map $\Omega^1 (M,\mathfrak{su}(2))\ni \omega \to U_{ \omega}$ is a group homomorphism, i.e. $U_{(\omega_1+\omega_2 )}=U_{\omega_1}U_{ \omega_2}$. \\

\begin{figure}[t]
\begin{center}
\resizebox{!}{ 4 cm}{
 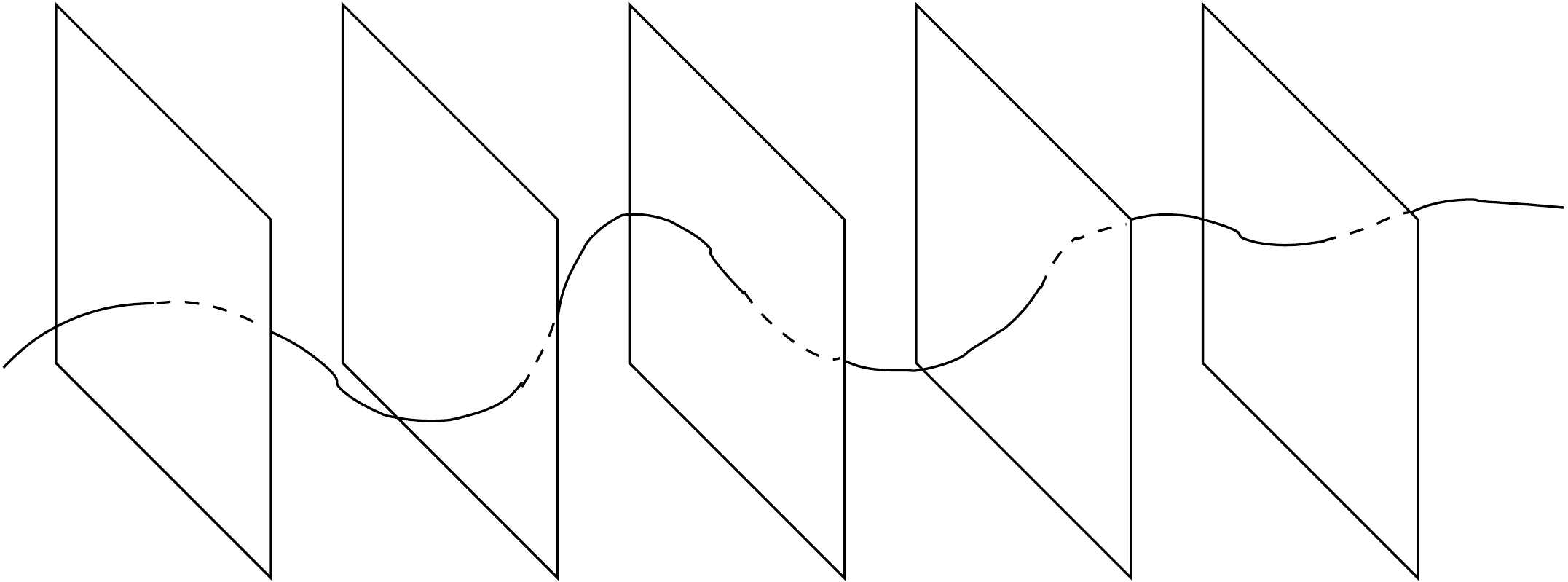}
\end{center}
\caption{\it The operator $E_{\sigma_i dx^\m}$ will, when it is commuted with a flow, insert Pauli matrices continuously along the course of the flow. This means that it acts as a sum of flux operators with surfaces, which intersect the flow at the points of insertion. }
 \label{heltvildtsvedig}
\end{figure}

To see the connection to the holonomy-flux algebra let us analyze the righthand side of (\ref{flowkan}). First we introduce local coordinates $(x_1 ,x_2,x_3)$. We decompose $\omega$: $\omega =\omega^i_\m\sigma_i dx^\m$. Due to the additive property of $E_\omega$ and that the action of $C^\infty_c(M)$ commutes with $U_{\omega }$ we only have to analyze an $\omega $ of the form $\sigma_i dx^\m$. For a given point $p\in M$ choose the points $$p_0=p,\quad p_1=e^{\frac{1}{n}X}(p),\ldots  ,\quad  p_n= e^{\frac{n}{n}X}(p)$$ 
on the path
$$t\to e^{tX}(p)  ,t\in [0,1].$$
We write the vectorfield $X=X^\n\partial_\n$. 
We have 
\begin{eqnarray} 
\lefteqn{e^X_{\nabla+t\omega}}\nn\\
& =&\lim_{n\to \infty } (1+\frac{1}{n}(A(X(p_0)+t\sigma_i X^\m(p_0) )) (1+\frac{1}{n}(A(X(p_1))+t\sigma_iX^\m(p_1)))\nn\\
&& \cdots (1+\frac{1}{n}(A(X(p_n)+t\sigma_i X^\m(p_n)) ,
\label{vlad}
\end{eqnarray}
where  $\nabla=d+A$, and therefore 
\begin{eqnarray}
 \lefteqn{\frac{d}{dt}e^X_{\nabla +t\omega}|_{t=0}}
 \nn
 \\
 &=& \lim_{n\to \infty }  \Big( \frac{1}{n} \sigma_iX^\m(p_0)  (1+\frac{1}{n}A(X(p_1)))\cdots (1+\frac{1}{n}A(X(p_n))) \nn\\
 &&+ (1+\frac{1}{n}A(X(p_0))) \frac{1}{n} \sigma_iX^\m(p_1)  (1+\frac{1}{n}A(X(p_2)))\cdots (1+\frac{1}{n}A(X(p_n))) \nn\\
 && + \quad\quad\quad\quad\quad\quad\quad\quad\quad\quad\quad \vdots \nn\\
 &&+ (1+\frac{1}{n}A(X(p_0))) )  (1+\frac{1}{n}A(X(p_2)))\cdots (1+\frac{1}{n}A(X(p_{n-1})) \frac{1}{n} \sigma_i X^\m(p_n) \Big)
\nn\\ \label{COOM}
\end{eqnarray}
We see that before taking the limit $\lim_{n\to \infty}$ this is just the commutator of the sum of the flux operators $\sum_k \frac{1}{n}X^\m (p_k) F^{S_k}_i $, where $S_k$ is the plane orthogonal to the $x_\m$-axis intersecting $p_k$, and the holonomy operator of the path $$t\to e^{tX}(p)  ,t\in [0,1],$$
 see figure \ref{heltvildtsvedig}.
 
It follows that $E_{\sigma_i dx^\m}$ is  a series of flux-operators $F^S_i$ sitting along the path $$t\to e^{tX}(p)  ,t\in [0,1],$$
where the surfaces $S$ are just the planes othogonal to the $x_\m$ direction.
But since there are infinitely many of them, they have been weighted with the infinitesimal length, i.e. with a $dx^\m$, see figure \ref{heltvildtsvedig}. We can formally write 
$$E_\omega = \int F^S_\m dx^\m  . $$   

Note that the phenomenon from the holonomy-flux-algebra, that a path $p$ running inside a surface $S$, has zero commutator with the corresponding flux operator is encoded in the quantum-flow-algebra, since the tangent vectors of $p$ will be annihilated by the differential form $dx^\m$.

  \subsection{The canonical commutation relations}

We can also make the holonomies infinitesimal in order to see the canonical commutation relations of general relativity. In the following we scale the translation operators $U_\oo\rightarrow U_{\kappa\oo}$, where $\kappa$ is the quantization parameter. Let us again first introduce a coordinate system $(x_1,x_2,x_3)$. We have the vectorfields $\partial_\m$, and we consider the operator
$$ \frac{d}{ds} e^{s\partial_\m}|_{s=0}  .$$ 
We note that if the local connection one form of $\nabla $ is $A_\m^i\sigma_i dx^\m$ we have 
 $$( \frac{d}{ds} e^{s\partial_\m}|_{s=0})(\xi (x,\nabla)) = \sigma_i A_\m^i(x) \xi (x,\nabla) . $$
We therefore consider the operator $\delta(x) \frac{d}{ds} e^{s\partial_\m}|_{s=0}$, where $\delta(x)$ is the delta function located in $x$, as the operator $\sigma_i A^i_\m(x)$ located in $x$.  
 
On the other hand consider $\omega =\sigma_ldx^\n$. We get 
\begin{eqnarray*}
\lefteqn{[E_{\sigma_ldx^\n},\sigma_i A^i_\m(x)] (\nabla )}\\
&=& \delta(x)  \frac{d}{ds} [E_{\sigma_ldx^\n},  e^{s\partial_\m}  ]|_{s=0}(\nabla)= \kappa \delta(x)  \frac{d}{ds} \frac{d}{dt} e^{s\partial_\m}_{\nabla+t\sigma_ldx^\n}|_{t=0}|_{s=0}\\
&=&  \kappa \delta(x)  \frac{d}{ds} \frac{d}{dt} (1+s(A_\m^i\sigma_i+t\sigma_l \delta_\m^\n  ))|_{t=0}|_{s=0}=\kappa \sigma_l \delta_\m^\n \delta(x) \;.
\end{eqnarray*}
If 
$$f_y(x)=\left\{ 
\begin{array}{cl}
1& x=y \\
0& x\not =y
\end{array}\right.$$
we can therefore consider the operator $f_y  E_{\sigma_ldx^\n}$ as $\hat{E}_l^\n (y)$ since then 
\begin{equation}
 [ \hat{E}_l^\n (y) , \sigma_i A^i_\m(x)] = \kappa \sigma_l \delta_\m^\n \delta (x-y),
\label{COOOM}
 \end{equation}
which is the quantized canonical commutation relation of general relativity formulated in terms as Ashtekar variables.

All together these results show that the algebra $\mathbf{QHD}(M)$ is intimately related to canonical quantum gravity since it is simply the algebra from which the infinitesimal operators forming the canonical commutation relations originate. 

Note, however, that with the choice of $SU(2)$ as the gauge group we are not, a priori, dealing with a construction based on the original Ashtekar connection, which takes values in the Lie-algebra of complexified $SU(2)$ -- the self-dual sector of $SL(2,\mathbb{C})$ --, but instead with an Ashtekar connection related to general relativity with a Euclidian signature\footnote{The issue is a little more complicated than this, since it depends on the form of the Hamiltionian as well. In the framework of this paper a real $SU(2)$ connection does imply a Euclidian signature since the Hamiltonian, which we eventually derive from our construction, has the simple algebraic structure '$EEF$', where $F$ is the field strength tensor of the Ashtekar connection. See for instance \cite{AL1}.}. We could, of course, try to work with $SL(2,\mathbb{C})$ instead of $SU(2)$, but it is our belief that the complexification, which takes us from $SU(2)$ to $SL(2,\mathbb{C})$ should arise naturally from the construction. See section \ref{REAL} and \ref{DIS}.

\subsection{The spectrum of $\mathbf{HD}(M) $}

In \cite{AGnew} we analyzed the spectrum of the algebra $\mathbf{HD}(M) $, which is defined as 
 the irreducible representations of $\mathbf{HD}(M) $ modulo unitary equivalence. There we obtained two main results concerning the spectrum of $\mathbf{HD}(M) $, which we will now review. Before we do so we first need to introduce the concepts of a measurable connection and of a generalized connection. For details and proofs of this section we refer to \cite{AGnew}.

\begin{definition}
Let $\cf$ be the group generated by flow operators $e^X$. A measurable $U(n)$-connection, $n=1,\ldots , \infty$, is a map $\nabla$ from $\cF$ to the group of measurable maps from $M$ to $U(n)$ satisfying
\begin{enumerate}
\item $\nabla (1)= 1$.
\item $\nabla (F_1 \circ F_2)(m)=\nabla (F_1) (m) \circ \nabla (F_2)(F_1^{-1}(m))$
\item If $F_1$ and $F_2$ are the same up to local reparametrization over some set $U\subset M$, then 
$$ \nabla ( F_1)_U= \nabla (F_2)_U  . $$ 
\end{enumerate} 
\end{definition}

Next, let $l$ be a piece-wise analytic path in $M$. We identify $l\cdot l^{-1}$ with the trivial path starting and ending at the start point of $l$. Furthermore we identify two paths that differ by a reparameterization. 
\begin{definition}
Let $G$ be a connected Lie-group. A generalized connection is an assignment $\nabla(l)\in G$ to each piece-wise analytic path $l$, such that
$$
\nabla(l_1\cdot l_2) = \nabla(l_1)\nabla(l_2)\;.
$$
\end{definition}

With this we can now state the two main results from \cite{AGnew}  on the spectrum of $\mathbf{HD}(M) $:
\begin{thm}
\label{TH1}
Any separable, irreducible representation of $\mathbf{HD}(M) $ is unitarily equivalent to a representation of the form $\varphi_\nabla$, where $\nabla$ is a measurable $U(2)$-connection\footnote{Here we disregard the special case where the representation decomposes into two $U(1)$ measurable connections.}.
\end{thm}

\begin{thm}
\label{TH2}
A generalized connection together with the counting measure on $M$ does not render a representation of $\mathbf{HD}(M) $.
\end{thm}

Theorem \ref{TH1} holds in more general settings -- manifolds of arbitrary dimensions and arbitrary vector bundles. The theorem is, however, particularly interesting in the case where $M$ is a three-dimensional manifold and $S$ is a two-spinor  bundle over $M$ with $SU(2)$ connections, since in this case one can interpret the spectrum as the completion of a configuration space of Ashtekar connections.

Theorem \ref{TH2} basically states that the bulk of the spectrum found in loop quantum gravity \cite{AL1}, which is given by generalized connections with support on finite graphs, is excluded from the spectrum of $\mathbf{HD}(M) $.

It is an open question what the non-separable part of the spectrum of $\mathbf{HD}(M) $ contains. The fact that we are for now unable to prove that the entire spectrum is given by measurable connections may indicate that we need to change the definition of $\mathbf{HD}(M) $. In particular, one may speculate that the topology of $\mathbf{HD}(M) $, which is the $C^*$-topology, is not the right one for our purpose. For further discussion of this point as well as other related issues see \cite{Aastrup:2012vq}.

Note finally that there is an open question as to how we get $SU(2)$ connections instead of $U(2)$ connections. Of course, we can simply put them in by hand -- which is what we do in this paper -- but a more natural solution would be to introduce a real structure and a conjugate action of the algebra, which would kill the $U(1)$ factor.

\section{Semi-classical states and a kinematical Hilbert space}

With the formulation of the Quantum Holonomy-Diffeomorphism algebra completed the first task is to determine whether states exist on this algebra.

\subsection{Lattice formulation}

The question concerning states on $\mathbf{QHD}(M)$ and $\mathbf{dQHD}(M)$ is best addressed in a setting where everything is reformulated in terms of lattice approximations. Aside from being a technical tool this also represents a coordinate dependent formulation since an infinite sequence of nested lattices corresponds to a coordinate system. The following is based on techniques introduced in \cite{Aastrup:2012vq}. 

Before we continue let is first note that at the present level of analysis we are unable to determine whether a lattice formulation exist for $\mathbf{QHD}(M)$ and $\mathbf{dQHD}(M)$ or if it is necessary to restrict $\mathbf{HD}(M)$ to analytic flows. We discuss this issue in section \ref{continuum} and in the final discussion in section \ref{DIS}. In the following we therefore leave open whether the lattice approximations of $\mathbf{HD}(M)$, $\mathbf{QHD}(M)$ and $\mathbf{dQHD}(M)$ are constructed from diffeomorphisms or analytic diffeomorphisms.    \\

\begin{figure}[t]
\begin{center}
\resizebox{!}{2.5cm}{
 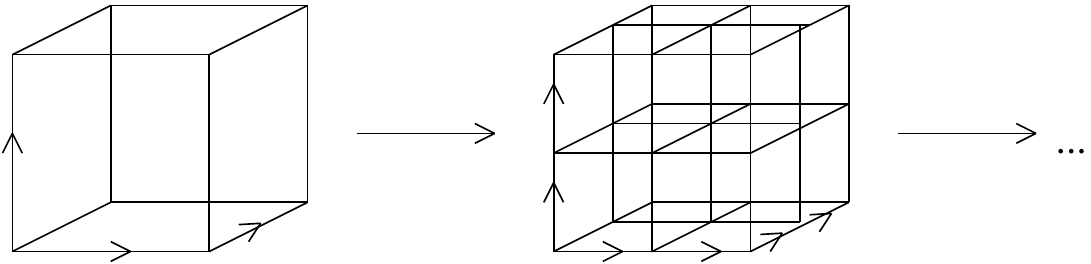}
\end{center}
\caption{\it $\OO$ is an infinite sequence of cubic lattices, where each lattice $\G_n$ in $\OO$ is a symmetric subdivision of the previous lattice $\G_{n-1}$ and where the union of all lattices $\G_n$ is dense in $M$. Thus $\OO$ represents a coordinate system in $M$.}
 \label{ronnie}
\end{figure}

In the lattice formulation an element in $\mathbf{HD}(M)$ is represented by an infinite family of operators acting in increasingly accurate lattice approximations associated to an infinite system $\{\G_n\}$ of nested cubic lattices, see figure \ref{ronnie}. 
Thus, we begin with a single finite cubic lattice $\G_n$ with vertices and edges denoted by $\{v_i\}$ and $\{l_j\}$. We assign to each edge $v_i$ a copy of $SU(2)$ 
\begin{equation}
\nabla(l_j) = g_j\in SU(2)
\label{MAAP}
\end{equation}
and obtain the space 
$$\ca_{n}= (SU(2))^{\vert{\bf l}_n\vert}$$
where $\vert{\bf l}_n\vert$ is the number of edges in $\G_n$. $\ca_{n}$ is an approximation of the space of $SU(2)$-connections $\ca$ and the map $\nabla$ should be understood as an approximation of a connection in $\ca$.


An element $f e^X$ in $\mathbf{HD}(M)$ is at the level of a lattice $\G_n$ approximated by a finite family of oriented, weighted paths in $\G_n$, denoted by  $\mathbbm{fF}_n$. 
Here $\mathbbm{F}_n$ denotes a family $\{ p_i\}$ of paths in $\G_n$, where each $p_i$ is a sequence of adjacent edges 
$$
p_i = \{l_{i_1}, \ldots,l_{i_n}\}\;,
$$
connecting two vertices in $\G_n$.
By $\mathbbm{f}$ we denote a corresponding set of weights assigned to each edge in $\mathbbm{F}_n$. There is a natural product between such lattice approximations given by composition of paths and a natural involution given by reversal of the paths (see \cite{Aastrup:2012vq} for details). The lattice approximation of $\mathbf{HD}(M)$ is denoted by $\mathbf{HD}_{n}$. We shall give a more precise definition of $\mathbf{HD}_{n}$ in the next section, where we describe the continuum limit.  

With the space $\ca_n$ we automatically have the Hilbert space $L^2(\ca_{n})$ via the Haar measure on $SU(2)$. We will, however, need more structure in order to construct a representation of $\mathbf{HD}_{n}$  and therefore introduce the Hilbert space
$$
\ch_{n} = L^2(\ca_{n}, M_2(\mathbb{C}))\times M_{\vert {\bf v}_n\vert}(\mathbb{C})
$$
where $\vert {\bf v}_n\vert$ is the number of vertices in $\G_n$. A representation of an element $\mathbbm{fF}_n$ in $\mathbf{HD}_{n}$ acts by multiplying the $M_2(\mathbb{C})$ factor in $\ch_n$ with the parallel transports 
$$
\nabla(p_i) = \nabla(l_{i_1})\cdot \ldots\nabla(l_{i_n})\;,\quad p_i \in \mathbbm{F}_n \;,
$$
and by acting on the $M_{\vert {\bf v}_n\vert}(\mathbb{C})$ factor with $\mathbbm{F}_n$ as an $\vert {\bf v}_n\vert \times \vert {\bf v}_n\vert$-matrix in the sense that each path in $\mathbbm{F}_n$ shifts lattice points according to its start and end-points. Thus, an example of a representation of an element $\mathbbm{fF}_n$ in $\mathbf{HD}_{n}$ as an operator in $\ch_n$ could be:
\begin{equation}
\left(
\begin{array}{ccccccccccc}
1 & \ldots & 0 &&&&&&&& \\
\vdots & \ddots & \vdots &&&&&&&&  \\
0 & \ldots & 1 &&&&&&&& \\
&&& \ddots &&&&&&&\\
&&&& 0 & \nabla(p_1) & \nabla(p_2) &&&& \\
&&&& 0 & 0 & \nabla(p_3) &&&& \\
&&&& 0 & 0 & 0 &&&& \\
&&&&&&& \ddots &&& \\
&&&&&&&& 1 & \ldots & 0 \\
&&&&&&&& \vdots & \ddots & \vdots \\
&&&&&&&& 0 & \ldots & 1
\end{array}
\right) 
\label{BigMatrix}
\end{equation}
where $\mathbbm{F}_n$ involves\footnote{In order to ease the notation we have not assigned weights to the paths in this example. If there had been weights these would have appeared as numbers multiplied the $\nabla(p_i)$'s.} the three paths $\{p_1,p_2,p_3\}$.
%

We also need to define the operators $U_\omega$ and $E_\omega$ on the lattice. We start with the latter since, if this is successfully constructed, a lattice approximation of the former is obtained via exponentiation hereof. 
Let $\mathfrak{su}_i(2)$ be the Lie algebra of the $i$'th copy of $SU(2)$ and choose an orthonormal basis $\{e_i^a\}$ thereon.  We also denote by $\{{\bf e}_i^a\}$ the corresponding right translated vector fields and by $\cl_{{\bf e}_i^a}$ the derivation with respect to the trivialization given by $\{e_i^a\}$. 
There is then a natural candidate for  a lattice approximation of $E_\oo$ given by 
$$
 E_{\oo,n}=  2^{-n}\kappa \sum_{i,a} \oo_i^a   \cl_{{\bf e}_i^a} \otimes \mathds{1}_{\vert {\bf v}_n\vert }
$$
where $\oo_i^a$ here denotes the value of $\oo$ evaluated at the start-point of the edge $l_i$ and where the sum runs over all edges in the lattice and over all Lie-algebra indices. Here we have introduced a quantization parameter $\kappa$, which corresponds to the transformation $U_\oo\rightarrow U_{\kappa \oo}$.

\subsection{The continuum limit}
\label{continuum}

\begin{figure}[t]
\begin{center}
\resizebox{!}{ 6 cm}{
 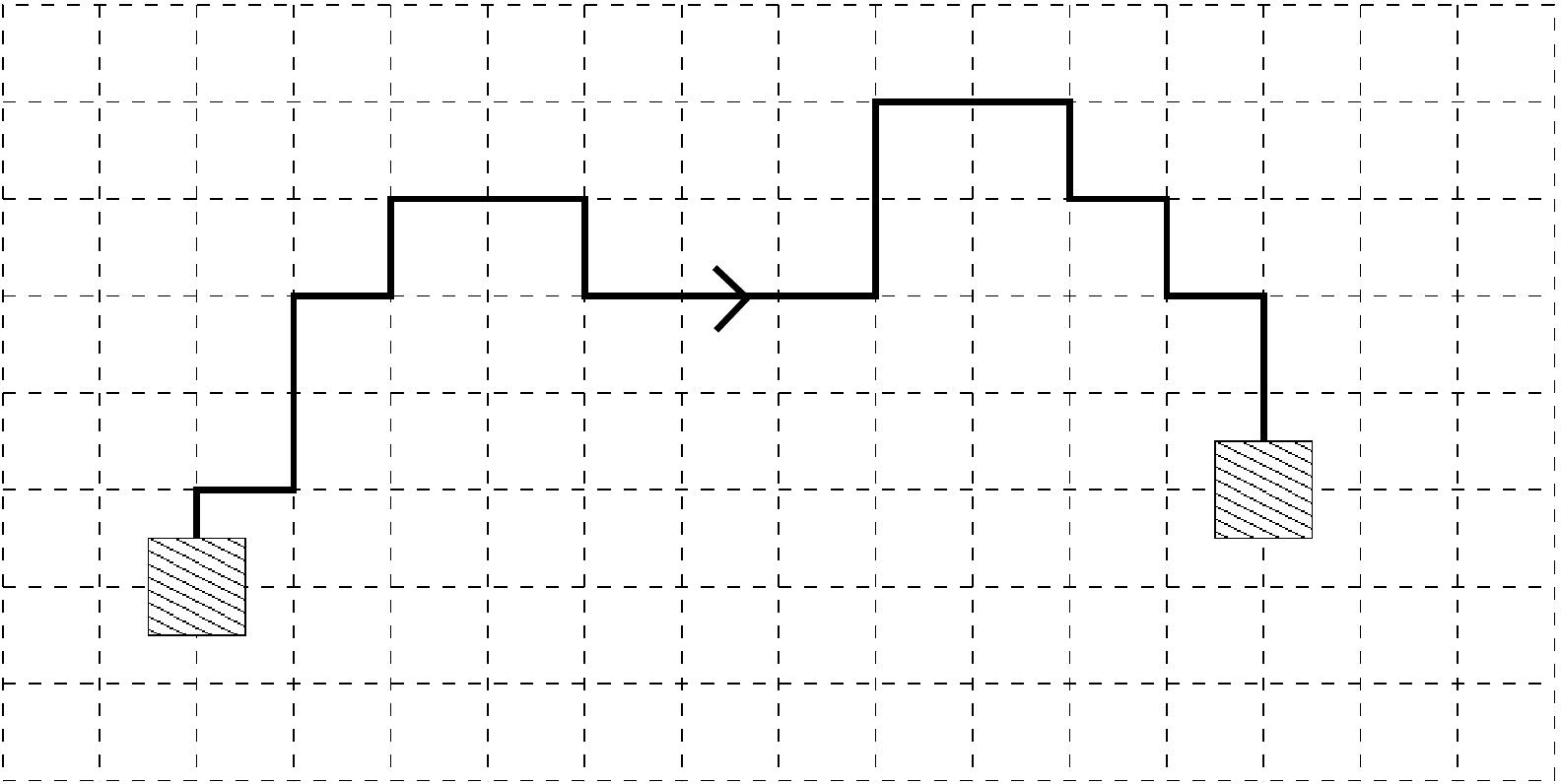}
\end{center}
\caption{\it The action of a path in lattice on $L^2(M,S)$. It transports the content of the first square to the content of the second square and multiplies with the holonomy of the connection $\nabla$ along the path.}
 \label{brian}
\end{figure}

Consider now a sequence $\{\mathbbm{fF}_n\}$ of lattice approximations as described above. In order to give meaning to the notion that the sequence approximates  an element $f e^X$ in $\mathbf{HD}(M)$  we need to define an action of $\mathbbm{fF}_n$ as an operator in $L^2(M,S)$. This is done in the following manner (see also section 5 in \cite{Aastrup:2012vq}). First, we subdivide $M$ into cubes $\{c_i\}$, where a cube $c_i$ is assigned to each vertex $v_i\in\G_n$, and in such a way that the cubes fill out $M$. Let us assume that $\mathbbm{F}_n$ includes a path $p$ which connects two vertices $v_i$ and $v_j$. Then $\mathbbm{F}_n$ acts on a spinor $\psi\in L^2(M,S)$ by shifting each value of $\psi$ in the cube $c_i$ to the same relative location in the cube $c_j$ while multiplying it with the holonomy transform $\nabla(p)$ of the connection along $p$, see figure \ref{brian}. We denote this representation by $\varphi(\mathbbm{F}_n)$.
Furthermore, $\mathbbm{fF}_n$ has the same representation, denoted $\varphi(\mathbbm{fF}_n)$, in $L^2(M,S)$ as $\mathbbm{F}_n$ except that each holonomy with respect to $\nabla$ of a paths in the family is also multiplied with the weight associated.

We say that a sequence $\{\mathbbm{fF}_n\}$ converges to $f e^X\in \mathbf{HD}(M) $ in the representation $\varphi$ if
\begin{equation}
\lim_{n\rightarrow\infty}\vert\vert (\varphi(\mathbbm{fF}_n) - \varphi(fe^X)\xi) \vert\vert = 0\;,
\label{xtxt}
\end{equation}
for all $\xi\in L^2(M,S)$ and for all smooth connections,
where $\varphi(fe^X)$ denotes the representation of $fe^X$ in $L^2(M,S)$. Keep in mind that the representation $\varphi$ depends on a chosen $SU(2)$ connection and that (\ref{xtxt}) must hold for any such choice.
There is a natural equivalence relation between the sequence $\{\mathbbm{fF}_n\}$ and $\{\mathbbm{fF}'_n\}$ given by
\begin{equation}
\{\mathbbm{fF}_n\}\sim \{\mathbbm{fF}_n'\}\quad iff \quad \lim_{n\rightarrow\infty}\vert\vert ({\varphi(\mathbbm{fF}_n)} - \varphi(\mathbbm{fF}_n'))\xi \vert\vert = 0
\label{xtxtx}
\end{equation}
for all $\xi\in L^2(M,S)$ and all smooth connections. Note that this equivalence relation is just saying that if $\{\mathbbm{fF}_n\}\sim \{\mathbbm{fF}_n'\}$ then $\{\mathbbm{fF}_n\}$ and $\{\mathbbm{fF}_n'\}$ converge to the same element in  $\mathbf{HD}(M)$, if they converge. 
If a family $\{\mathbbm{fF}_n\}$ converges to $fe^X$ in $\varphi$ we 
shall say that  $\varphi(\mathbbm{fF}_n)$ approximates $fe^X$.

This notion of convergence can also be defined for diffeomorphisms and volume-preserving diffeomorphisms, see \cite{Aastrup:2012vq}.

In a next step we need to describe how the algebra $\mathbf{HD}(M) $ is represented by sequences of lattice approximations. Let $a,b,c$ be elements in $\mathbf{HD}(M) $ and let $\{ a_n\}$, $\{ b_n\}$ and $\{ c_n\}$ be sequences of lattice approximations converging to $a,b,c$. $\{ a_n\}$, $\{ b_n\}$ and $\{ c_n\}$ are elements in equivalence classes according to (\ref{xtxtx}) and we choose these elements such that if
$
a = bc
$
then the sequences $\{  a_n \}$ and $\{b_n c_n\}$ are identical. We call this choice of lattice approximations of elements in $\mathbf{HD}(M) $ for a consistent set of lattice approximations. The lattice approximation of $\mathbf{HD}(M) $ that arises in this way at the level of $\G_n$ is denoted $\mathbf{HD}_n $. 

The reason why it is necessary to require a consistent set of lattice approximations is that this is the only way one can preserve the algebra structure in lattice approximations of the flow-dependent version of $\mathbf{dQHD}(M) $, which is the subject of the next section. It is, however, not clear to us if it is in fact possible to obtain a consistent set of lattice approximations of $\mathbf{HD}(M) $. One possibility is to restrict $\mathbf{HD}(M) $ to analytic flows, in which case the requirement can be met, but we do not know if this restriction is necessary. We shall return to this issue in the final discussion.

Finally, we also need a notion of convergence of states on $\mathbf{HD}(M) $ from states on $\mathbf{HD}_n $.
Let again $\{\mathbbm{fF}_n\}$ converge to $fe^X\in \mathbf{HD}(M) $ and let $\x$ be a state on $\mathbf{HD}(M) $, i.e. a positive linear functional on $\mathbf{HD}(M) $. Let $\{\x_n\}$ be a sequence of states with $\x_n\in H_n$. We will say that $\x_n$ approximates $\x$ if
$$
\lim_{n\rightarrow\infty}\left\langle \mathbbm{fF}_n \x_n\vert \x_n\right\rangle = \x(fe^X)
$$
for all elements in the equivalence class $\{\mathbbm{fF}_n\}$. We say that two sequences $\{\x_n\}$ and $\{\x_n'\}$ are equivalent if they both approximate the same state $\x$. We will, when discussing states and their approximations always assume that we are dealing with such equivalence classes.

\subsection{A semi-classical state on $\mathbf{HD}(M)$}


We are now going to write down necessary conditions for a state to exist on $\mathbf{HD}(M)$ and $\mathbf{dQHD}(M)$, respectively.
%
Let $\{\xi_{(n,i)}(g_i)\}$ be a set of functions on $SU(2)$ associated to edges in $\G_n$.
To define a candidate for a state on $\mathbf{HD}(M)$ and $\mathbf{dQHD}(M)$ we first write down the state $\xi_n$ in $\ch_n$ 
\begin{equation}
\xi_n =  c_n \P_{i=1}^{\vert {\bf l}\vert} \xi_{(n,i)}\otimes \mathds{1}_2\otimes \mathds{1}_{\vert {\bf v}_n\vert }\;,
\label{state-semi}
\end{equation}
where $c_n$ is a normalization constant, and consider the continuum limit $n\rightarrow\infty$ of $\xi_n$, which
we denote by $\xi_c$.

The necessary conditions for $\xi_c$ to be a state on $\mathbf{HD}(M)$ and $\mathbf{dQHD}(M)$, respectively, are the following\footnote{to be precise, it might be possible to exchange smoothness in condition 3) with some weaker requirement such as $C^1$.}:
\begin{eqnarray}
1) &&\left\langle \xi_{n} \vert \nabla(l_i) \vert \xi_{n}\right\rangle = \mathrm{1} - \co(dx) \hspace{5.4cm}
\nn\\
2) &&\left\langle \xi_{n} \vert \cl_{{\bf e}_{i_1}}\cl_{{\bf e}_{i_2}}\ldots \cl_{{\bf e}_{i_m}}  \vert \xi_{n}\right\rangle = \co(dx^{2m}),\; \forall m, i_n\in\{1,2,3\}
\label{conditionss}
\end{eqnarray}
where $dx=2^{-n}$, as well as:\\

\hspace{4mm}3)\vspace{-0.5cm}
\begin{addmargin}[2cm]{1cm}
the expectation values of $ \nabla(l_i) $ and the expectation values of all powers of the $\cl_{{\bf e}_{i}}$'s must converge to smooth functions on $M$ in the limit $n\rightarrow\infty$. i.e. the coefficients in front of $\co(dx)$ and $\co(dx^{2m})$ must depend smoothly on the points on the manifold.
\hspace{6cm}\eqnum\label{conditions}\\
\end{addmargin}

%
Conditions $1)$ and $3)$ ensure the the expectation value of a parallel transport $\nabla(p)$ will converge in the limit $n\rightarrow\infty$. These conditions are easily satisfied and we may therefore use the GNS construction to obtain the Hilbert space $\ch_{\xi_n}$ and its limit $\ch_{\xi_c}$. Furthermore, we may write
\begin{equation}
\left\langle \xi_{n} \vert \nabla(l_i) \vert \xi_{n}\right\rangle = \mbox{Tr}_{M_2}\left( \mathrm{1} + dx \left(A_{(n,i)} - t B_{(n,i)}\right)\right)
\label{Ash}
\end{equation}
where $A_{(n,i)}$ is the skew-self-adjoint component and where $B_{(n,i)}$ is the remainder, which is a strictly positive operator, and where $t$ is a formal parameter. If we let $t$ play the role of a quantization parameter we see that the states on $\mathbf{HD}(M)$ constructed this way will always have a semi-classical structure with $A$ being the classical point and $B$ the quantum correction. 

Condition $2)$ ensures that the expectation value of the translation operators $E_{\oo,n}$ and all polynomials thereof will converge in the limit $n\rightarrow\infty$. However, we strongly expect that this condition can never be satisfied simultaneous with condition $1)$, since this condition ensures that the state is peaked around the identity of the group, whereas the second condition more or less gives that the state is constant. These two requirements are mutually exclusive.

Finally, condition $3)$ also ensures that the commutator between $E_{\oo,n}$ and elements in $\mathbf{HD}_n$ exist in the large $n$ limit as well as the elements of $\mathbf{HD}(M)$ themselves. 
 \\

To sum up, we make the following conclusions:
\begin{enumerate}
\item
There is a state $\xi_c$ on $\mathbf{HD}(M)$ and by the GNS construction a Hilbert space $\ch_{(\xi_c,\mathbf{HD})}$. 
\item
Within the framework of lattice approximations it appears that no states on $\mathbf{dQHD}(M)$ exist. The infinitesimal translation operators $E_\oo$ are not well defined operators in $\ch_{(\xi_c,\mathbf{HD})}$.
\end{enumerate}


\subsection{A Dirac type operator and the graded $\mathbf{HD}(M)$ algebra}

Our failure to find a state on $\mathbf{dQHD}(M)$ leads us to consider an alternative approach, which involves a Dirac-type operator. At the end of our analysis we shall see that this Dirac-type operator is, at this point, a somewhat superfluous construction, which can be omitted. In the light of results of later sections in this paper, where this Dirac-type operator plays an important role, we choose however to include it in our analysis here too.  \\

We return to the lattice approximation $\G_n$ and the operator $E_{\oo,n}$. We first add an additional factor to the Hilbert space $\ch_n$
$$
\ch'_n  = L^2(\ca_{n}, Cl(T^*\ca_n)\times M_2(\mathbb{C}))\times M_{\vert {\bf v}_n\vert}(\mathbb{C})
$$
where $Cl(T^*\ca_n)$ is the Clifford algebra over the co-tangent space over $\ca_n$. 
This enables us to turn $E_{\oo,n}$ into a Dirac-type operator over the space $\ca_n$ by substituting $\oo$ in $E_{\oo,n}$ with an element of the Clifford algebra:
\begin{equation}
E_{\oo,n}  \stackrel{\oo\rightarrow{\bf e}}{\longrightarrow} D_n = 2^{-n}\kappa \sum_{i,a}  {\bf e}_i^a \cdot \cl_{{\bf e}_i^a} \otimes 1_{\vert {\bf v}_n\vert }
\label{Rothstein}
\end{equation}
where we by ${\bf e}_i^a$ also denote the element of $Cl(T^*\ca_n)$ that corresponds to the right-translated vector field ${\bf e}_i^a$ and where '$\cdot$' denotes Clifford multiplication. One could think of the Clifford element ${\bf e}_i^a$ as a lattice approximation of a Grassmann valued element of $\Omega^1(M,\mathfrak{su}(2))$. The operator $D_n$ is essentially the Dirac type operator, which we studied in the papers \cite{Aastrup:2012jj}-\cite{AGNP1} . We denote by $D$ the $n\rightarrow\infty$ limit of $D_n$.

Here, however,  we shall require the Clifford elements to form the non-standard Clifford algebra
\begin{equation}
\left\{ {\bf e}_i^a, {\bf e}_j^b   \right\} =  -2^{n}  \d^{ab}  \d_{ij}\;.
\label{Cliff}
\end{equation}
This means that the anti-commutator, in the large $n$ limit, will approach a one-dimensional delta function. Of course, if ${\bf f}_i^a$ denotes the standard generators, we get the above generators via $ {\bf e}_i^a=2^{\frac{n}{2}}{\bf f}_i^a$.

Next, define the operation $\d_n$, which consist of taking the commutator\footnote{All operator brackets in this paper are graded. We shall occasionally write '$\{\cdot,\cdot\}$' for the anti-commutator but it is understood that the bracket '$[\cdot,\cdot]$' is graded.} with $D_n$

$$
\d_n a_n := [a_n, D_n]\;,\quad a_n\in \cb(\ch'_n).
$$
One can check that for $a_n\in\mathbf{HD}_{n}$ $\d_n a_n$ will be a sum over insertions in $a_n$ of the form
$$
2^{-n} \kappa \sigma^a  {\bf e}^a_j  
$$
where $j$ now labels the point of insertion in $a_n$ and where $\sigma^a$ are the skew-adjoint Pauli matrices satisfying $[\sigma^a,\sigma^b]= -2\e^{abc}\sigma^c$.

\begin{definition}
We define the graded $\mathbf{HD}_n$ algebra, which we denote by $\mathbf{HD}^*_n$, as the algebra generated by elements of $\mathbf{HD}_n$ and elements  $\d_n a_n$, $a_n\in \mathbf{HD}_n$. Furthermore, we define the graded $\mathbf{HD}(M)$ algebra, which we denote by $\mathbf{HD}^*(M)$, as the $n\rightarrow\infty$ limit of $\mathbf{HD}^*_n$.
 \end{definition}
 The $n\rightarrow\infty$ limit is again as described in \cite{Aastrup:2012vq}. Also, we denote the $n\rightarrow\infty$ limit of $\d_n$ by $\d$.\\ 



Note that the emphasis put here and in the following on the commutator of the Dirac type operator $D$ instead of $D$ itself is quite natural since the central relation (\ref{konj}) describes the conjugation of a flow with the operator $U_\oo$, the infinitesimal of which is of course the commutator. Thus, it does not seem necessary that the operator $E_\oo$, nor its graded version $D$, should be well defined in the lattice formulation as long as its commutators are.\\

We shall now consider the GNS construction of the algebra $\mathbf{HD}^*_n$ on the state $\x_n$ which we wrote down in (\ref{state-semi}) and which we now require to satisfy condition 1) in (\ref{conditionss}). Since $\x_n$ does not take values in the Clifford algebra we immediately see that 
$$
\left\langle \xi_{n} \vert a_n (\d_n b_n) c_n \vert \xi_{n}\right\rangle = 0 \;,\quad \forall \{a_n, b_n, c_n\}\in \mathbf{HD}_n
$$
and we check that
\begin{equation}
\lim_{n\rightarrow\infty}\left\langle \xi_{n} \vert  (\d_n a_n)^* \d_n a_n \vert \xi_{n}\right\rangle < \infty \;,\quad  a_n\in \mathbf{HD}_n\;,
\label{finiteee}
\end{equation}
where the sequence $\{a_n\}$ is a representation of an element in $\mathbf{HD}(M)$. This relation holds since $\d_n a_n$ is a sum of insertions of the form $2^{-n} \kappa \sigma^a  {\bf e}^a_j  $ and since $\tr_{\mbox{\tiny Cl}} ({\bf e}^a_i {\bf e}^b_j )= 2^n \d^{ab}\d_{ij}$ collapses the two sums in (\ref{finiteee}) to a single sum with the right factor of $2^{-n}$. In fact, if we compute (\ref{finiteee}) for a single path the result is simply the length $L$ of that path, computed with respect to the metric provided by the lattice.

We readily generalize (\ref{finiteee}) to
\begin{equation}
\lim_{n\rightarrow\infty}\left\langle \xi_{n} \vert a_n (\d_n b_n)^* c_n (\d_n d_n) e_n \vert \xi_{n}\right\rangle < \infty \;,\quad  \{a_n,b_n,c_n,d_n,e_n\}\in \mathbf{HD}_n\;,
\nn
\end{equation}
where we again assume that the sequences $\{a_n\},\{b_n\},\{c_n\},\{d_n\}$ and $\{e_n\}$ are representations of elements in $\mathbf{HD}(M)$. Here, due to the Clifford algebra the result can only be non-zero if $\{b_n\}$ and $\{d_n\}$ are representations of the same element in $\mathbf{HD}(M)$. Finally, one checks that this result holds for all polynomials: 
\begin{equation}
\lim_{n\rightarrow\infty}\left\langle \xi_{n} \vert  P_m(\d_n a_{n_1},\ldots,\d_n a_{n_i}, a_{m_1},\ldots,a_{m_j}) \vert \xi_{n}\right\rangle < \infty \;,\quad  \{a_{n_i},a_{m_j}\}\in \mathbf{HD}_n\;,
\nn
\end{equation}
where $P_m$ is an arbitrary polynomial of degree $m$ and where
the sequences $\{a_{n_i}\},\{a_{m_j}\}$ represent elements in $\mathbf{HD}(M)$. We therefore conclude that the state $\xi_n$ provides us, in the limit $n\rightarrow\infty$, with a state on $\mathbf{HD^*}(M)$. We denote the corresponding Hilbert space, obtained from the GNS construction, by $\ch_{(\xi_c,\mathbf{HD^*})}$.\\


\subsection{Unbounded operators and the $\mathbf{dQHD}^*(M)$ algebra }

The next step is to analyze whether the operator $\d$ itself can be understood in terms of a GNS construction over a suitable algebra. To this end we start by introducing a graded and flow-dependent version of the $\mathbf{dQHD}_n$ algebras as well as the $\mathbf{dQHD}(M)$ algebra.

\begin{definition}
We define the flow-dependent $\mathbf{dQHD}_n$ algebra, which we denote by $\mathbf{dQHD}^*_n$, as the algebra generated by elements in $\mathbf{HD}_n$ and by elements of the form
$$
\left[D_n, \left[D_n, \ldots\left[ D_n, a\right]\ldots\right]\right]  \;,\quad a\in \mathbf{HD}_n.
$$
Furthermore, we define the flow-dependent $\mathbf{dQHD}(M)$ algebra, which we denote by $\mathbf{dQHD}^*(M)$, as the $n\rightarrow\infty$ limit of $\mathbf{dQHD}^{*}_n$.

 \end{definition}

We are now going to show that the state $\xi_c$ on $\mathbf{HD}^*(M)$ can be generalized to a state on $\mathbf{dQHD}^*(M)$ as well. To ease the notation we shall in the following write $\d_n$ as $\d$.\\

First, consider the $m$'th commutator, with $m$ even
\begin{equation}
\left\langle \xi_{n} \vert \d^m a_n \vert \xi_{n}\right\rangle ,\quad  a_n\in \mathbf{HD}_n\;.
\label{even}
\end{equation}
One checks that the $m$ sums in (\ref{even}) are collapsed by the trace over the Clifford elements to $m/2$ sums with the correct factor of $dx^{m/2}= 2^{-m/2}$, where each sum indexed by $i$ runs over an insertion of the form
\begin{equation}
2^{-n} \kappa^2 \sigma^a \cl_{{\bf e}_i^a} + 2^{-2n} \kappa^2  ( \sigma^a  {\bf e}^a_i)    ( \sigma^b  {\bf e}^b_i)  
\label{insertion}
\end{equation}
where sums over the Lie-algebra indices $a$ and $b$ are implied. Thus, in the limit this converges to $m/2$ line integrals. The convergence of (\ref{even}) in the limit $n\rightarrow\infty$ depends therefore entirely on how the expectation values of the $\cl_{{\bf e}_i^a}$'s behave. Therefore we now require the states $\xi_n$ to satisfy both condition 1) in (\ref{conditionss}) and condition 3) in (\ref{conditions}), where the latter here simply implies that the expectation values of powers of the vector-fields are required to depend smoothly on the points in $M$. We thus conclude that (\ref{even}) will converge and be finite:
\begin{equation}
\lim_{n\rightarrow\infty}\left\langle \xi_{n} \vert \d^m a_n \vert \xi_{n}\right\rangle < \infty ,\quad  a_n\in \mathbf{HD}_n, \; m \;\mbox{even}.
\label{even2}
\end{equation}

Next, consider also the the $m$'th commutator, with $m$ uneven. In this case the entire expression will be non-trivial with respect to the Clifford algebra and thus its expectation value vanish
\begin{equation}
\lim_{n\rightarrow\infty}\left\langle \xi_{n} \vert \d^m a_n \vert \xi_{n}\right\rangle =0 ,\quad  a_n\in \mathbf{HD}_n, \; m\; \mbox{uneven}.
\label{ueven}
\end{equation}
However, in the case where $m$ is uneven one may consider the expression 
$$
(\d^m a) (\d^k a^*)\;,
$$
where $k$ is uneven as well. This is an unbounded operator of order $(m+k)/2$, which involves $(m+k)/2$ sums over insertions of the form (\ref{insertion}). Again, since the expectation values of each $\cl_{{\bf e}_i^a}$ is a smooth function on $M$ we conclude that this expression will have a finite expectation value
\begin{equation}
\lim_{n\rightarrow\infty}\left\langle \xi_{n} \vert  (\d^m a) (\d^k a^*)    \vert \xi_{n}\right\rangle< \infty  ,\quad  a_n\in \mathbf{HD}_n, \; m,k\; \mbox{uneven}.
\label{ueven-even}
\end{equation}
Also, this expectation value will vanish if we instead choose $m +k$ to be uneven.

Next, we consider an expression of the form 
$$
(\d^{m_1} a_{1,n})(\d^{m_2} a_{2,n})\ldots (\d^{m_l} a_{l,n})\;, \quad a_{i,n}\in \mathbf{HD}_n
$$ 
If the sum $\sum_{i=1}^l m_i$ is uneven, then the expectation of this expression will again vanish since it is non-trivial with respect to the Clifford algebra. If the sum is even, then it depends whether each element $a_{i,n}$ in $ \mathbf{HD}_n$ equals the inverse of one of the other elements $a_{i,n}$ in $ \mathbf{HD}_n$ (or, possible, equals the inverse of a section of one of the other elements).  If this is not the case then the expression will again be non-trivial with respect to the Clifford algebra and the expectation value will vanish. If this is the case, then it will involve $(\sum_{i=1}^l m_i)/2$ sums over insertions (\ref{insertion}), which is again well defined since the right factor of $dx$ appears, and we conclude:
\begin{equation}
\lim_{n\rightarrow\infty}\left\langle \xi_{n} \vert   (\d^{m_1} a_{1,n})(\d^{m_2} a_{2,n})\ldots (\d^{m_l} a_{l,n})   \vert \xi_{n}\right\rangle< \infty  ,\quad  a_{i,n}\in \mathbf{HD}_n\;.
\label{multi-multi}
\end{equation}

Finally, none of the above results are changed by insertion of an element of $\mathbf{HD}_n$, for example $(\d^m a_n) b_n(\d^m c_n)$ with $ a_n, b_n, c_n \in \mathbf{HD}_n$, and we may therefore conclude that 
\begin{equation}
\lim_{n\rightarrow\infty}\left\langle \xi_{n} \vert  P_m(a_{1,n}, \d a_{2,n}, \d^2 a_{2,n},\ldots,  \d^m a_{m,n} )   \vert \xi_{n}\right\rangle< \infty  ,\quad  a_{i,n}\in \mathbf{HD}_n,
\label{polypop}
\end{equation}
where $P_m$ is a polynomial of order $m$. This, in turn, implies that $\xi_n$ converges to a state $\xi_c$ on $\mathbf{dQHD}^*(M)$ in the limit $n\rightarrow\infty$. 

Note that the operator $\d$ is not a well defined operator in the GNS construction around the state $\xi_{c} $ since the expectation value of $\d^2$ diverges on $\xi_{c} $. This is in accord with our general philosophy that only operators where $D$ is commuted with elements in $\mathbf{HD}(M)$ -- and thus not the vacuum state $\xi_{c} $ -- are well defined due to their one-dimensional smearing.

We denote by $\ch_{(\xi_n,\mathbf{dQHD}^*_n)}$ the Hilbert space emerging from this GNS construction over $\mathbf{dQHD}^*_n$ and by $\ch_{(\xi_c,\mathbf{dQHD}^*)}$ its $n\rightarrow\infty$ limit. The Hilbert space $\ch_{(\xi_c,\mathbf{dQHD}^*)}$ can -- and will in the following -- be understood as a kinematical Hilbert space for quantum gravity since the operators in $\mathbf{dQHD}^*(M)$ encode information about the canonical commutation relations of canonical quantum gravity formulated in terms of Ashtekar variables.
%
%
The state $\xi_c$ will play the role of a vacuum state in our construction. 

Note that we could have carried out the above construction without reference to a Dirac type operator, but simply by inserting the right-invariant vector fields into the flows by hand. As we pointed out above, we believe that the Dirac type operator represents a significant structure and therefore we have included it.

Note also that there exist an alternative approach, where instead of the algebra $\mathbf{dQHD}^*(M)$ we consider an algebra generated by elements of $\mathbf{HD}(M)$ and by second commutators between the Dirac type operator and elements of $\mathbf{HD}(M)$ {\it only}:
$$
a,\;\left\{ D, \left[ D, b \right]\right\}   \;\quad a,b\in \mathbf{HD}(M)\;.
$$
 This algebra will not involve polynomials of vector-fields associated to the same position and will therefore in some respect be better behaved. Since this algebra also encodes the kinematics of canonical quantum gravity it could be an interesting alternative to consider.

Let us finally point out that it is in fact an open technical question exactly what algebra the GNS construction is built over. The uncertainty has to do with the norm for the $C^*$-algebra that arises in the continuum limit. Since the vector fields in the $\mathbf{dQHD}^*(M)$ algebra may have a non-measurable effect on the spectrum of the holonomy-diffeomorphism algebra (see section \ref{ISIL} for a discussion hereof) it may be that the relevant algebra is built with respect to the counting measure instead of the Lebesque measure. We shall return to this question in the final discussion.

\subsection{Constructing the state}
\label{LarsL}

So far we have identified conditions for a state on $\mathbf{dQHD}^*(M)$ to exist. The conditions were that the expectation value of $\nabla(l_i)$ must be infinitesimally close to the identity and that expectation values of $\nabla(l_i)$ and of all powers of the vector fields should be smooth with respect to $M$, see (\ref{conditions}). 

Let us now for a moment concern ourselves with the actual construction of this state. In particular, let us consider a construction, which relies on coherent states on $SU(2)$ as have been considered by Hall \cite{H1,H2} and  others \cite{BT1,TW,BT2}. Thus, we pick a phase-space point $(A(x),E(x))$ of Ashtekar variables and consider first the function 
$$
\xi_{(n,i)}(g) = \phi^\kappa_{(A,E,i)}(g)
$$
where $\phi^\kappa_{(A,E,i)}(g)$ is the coherent state on a copy of $SU(2)$ assigned to the $i$'th edge in $\G_n$, where $\kappa$ is a quantization parameter and which satisfies the properties
\begin{equation}
 \lim_{\kappa \to 0}\left\langle   \phi^\kappa_{(A,E,i)} \vert \kappa \cl_{{\bf e}^a_i} \vert   \phi^\kappa_{(A,E,i)}  \right\rangle=   \mathrm{i}E_a^\m(\bar{l}_{i})\;,
 \label{JP1}
 \end{equation}
and
\begin{equation}
\lim_{\kappa \to 0}\left\langle  \phi^\kappa_{(A,E,i)}  \otimes v\vert \nabla(l_i) \vert \phi_{(A,E,i)}^\kappa\otimes v \right\rangle=(v,Hol(l_i,A)v)\;,
\label{JP2}
\end{equation}
where $v \in \bbC^2$, and $(,)$ denotes the inner product hereon. The index $\m$ denotes the spatial orientation of the ledge $l_i$.
The construction of the coherent state $\phi^\kappa_{(A,E,i)}$ is described in \cite{BT1,TW,BT2} and relies on a so-called complexifier, which determines the exact localization properties of $\phi^\kappa_{(A,E,i)}$. 

Consider first
$$
M^\kappa_{l_i}=\left\langle \phi_{(A,E,i)}^\kappa \vert \nabla(l_{i})\vert \phi_{(A,E,i)}^\kappa \right\rangle_{\mbox{\tiny SU(2)}} \;,
$$
which is the $SU(2)$-valued expectation value of a parallel transport on the edge $l_i$. $M^\kappa_{l_i}$ will be of the general form
\begin{equation}
M_{l_i}^\kappa = g_i (1 - \kappa (c_i + \sigma^j c_i^j))
\nn
\end{equation}
where $c_i,c_i^j$ are positive parameters, which depend on the exact form of the coherent state and on $\kappa$, and where $g_i=Hol(l_i,A)$. The matrix $(1 - \kappa(c_i + \sigma^j c_i^j))$ will have operator norm strictly smaller than 1, which implies that the first condition in (\ref{conditions}) cannot be satisfied.
If, however, we scale $\kappa$ in the coherent states $ \phi_{(A,E,i)}^\kappa$ with a factor $dx$
\begin{equation}
 \phi_{(A,E,i)}^\kappa \rightarrow  \phi_{(A,E,i)}^{ s } \;,\quad s = \kappa dx
 \label{scaling}
\end{equation}
where $dx$ in the lattice approximation equals $dx= 2^{-n}$, then we obtain
\begin{equation}
M_{l_i}^s = (1+ dx A) (1 - s (c_i + \sigma^j c_i^j)) := 1 + dx(A +\kappa B)
\label{passing}
\end{equation}
where $B$ is a quantity that depends on both $(A,E)$, $\kappa$ and the specific form of the coherent state $\phi^s_{(A,E,i)}$. Thus, $\phi^s_{(A,E,i)}$ satisfies the first condition in (\ref{conditions}).

In the passing we note that $B$ in (\ref{passing}) cannot be a connection since 
$$ M_{l_i}^{s} M_{l_i^{-1}}^{s}\not= 1\;. $$

The scaling of $\kappa$ has, however, implications for the peakedness of $\phi^s_{(A,E,i)}$ over the point $E_a^\m(\bar{l}_{i})$ in (\ref{JP1}) since this is now shifted with a factor of $dx^{-1}$. Thus, we must adjust this with a second scaling
$$
 \phi_{(A,E,i)}^{ s } \rightarrow  \phi_{(A, dx E,i)}^{ s }
$$
and we thus reach the result that the state built from
$$
\xi_{(n,i)}(g) = \phi^{s}_{(A, dx E,i)}(g)
$$
satisfies the requirements for the expectation values on it. Since $A$ and $E$ are smooth on $M$ all expectation values will be smooth as required.

Let us for later reference introduce the notation
$
 \xi^{\kappa}_{(A,E)}
$
for a state on $\mathbf{dQHD}^*(M)$, which is constructed from Hall's coherent states on $SU(2)$ in this way.\\


\subsection{The spectrum of the semi-classical states}
\label{ISIL}

We will now look at how the algebra affects the spectrum of the coherent state, i.e. which kind of translations on the space of connections that appears.

 In order to see more clearly what happens to the spectrum, we will in this section consider operators, which are a bit different from those in the previous sections. Also we will consider a slightly different state. At the end we will comment on the difference.

We first need a little bit of notation. The system af Graphs $\{ \Gamma_n \}$ induces a coordinate system. It therefore also induces  a one-norm
$$\| (x_1,x_2,x_3 )\|_1=|x_1|+|x_2|+|x_3|  ,$$
and a length function of path, i.e. if $\gamma: [a,b]\to M$ is a path, the length is defined via
$$ L(\gamma )=\int_a^b \|\dot{\gamma} (t)\|_1 dt  .$$ 
Let $p_n=\{ l_1, \cdots , l_k \}$ be a path in $\Gamma_n$, and let us assume that $p_n$ approximates $p$ when $n\to \infty$, i.e. where $p$ is a smooth path in $M$. Again we denote the operator associated to $p_n$ with $\nabla({p_n})=\nabla(l_1)\nabla(l_2)\ldots\nabla(l_k)$. The double commutator with the Dirac operator is given by
\begin{eqnarray} 
\label{formeldi}
 \{ D_n,[D_n, \nabla(p_n)] \} &=& 
 \nn\\&&\hspace{-2cm}
 2^{-n}\left( D_1\nabla({p_n}) +\nabla({l_1})D_2 \nabla({l_{2}}) \cdots \nabla({l_k})+\right.
 \nn\\&&\left.
\hspace{-0.5cm} \ldots+ \nabla({l_1}) \cdots \nabla({l_{k-1}}) D_k \nabla({l_k})    \right),
\end{eqnarray}
where $D_i$ denotes here the Dirac operator on the copy of $G$ associated to $l_i$, i.e. the operator $ \sigma^a\cl_{e^a} $ on $SU(2)$. 

Instead of coupling the $\cl_{e^a}$ to $\sigma^a$, we can couple it to an $\omega \in \Omega^1 (M, \mathfrak{su}(2))$ in the following way: 
we consider $l_i$ as a tangent vector in the starting point $p_i$ of $l_i$. In this way we can replace  $\sigma^a\cl_{e^a}$ on the $i$'th copy of $G$ with 
$$ \omega^a_{l_i} (p_i) \cl_{e^a_{i}} , $$
and insert this in (\ref{formeldi}) instead of $D_i$. Since the length of $l_i$ is $2^{-n}$ we propose the operator 
\begin{eqnarray}
E_{(\omega,p,n)}&=& \omega^a_{l_1} (p_1) \cl_{e^a_{1}} \nabla(p_n) +\nabla({l_1}) \omega^a_{l_2} (p_2) \cl_{e^a_{2}}\nabla({l_{2}})  \cdots \nabla({l_k})+
\nn\\&&
\ldots+ \nabla({l_1})\cdots \nabla({l_{k-1}}) \omega^a_{l_k} (p_k) \cl_{e^a_{k}} \nabla({l_k})  \nn  \;.
\end{eqnarray}
Note, that this is nothing but the derivative with respect to $t$ in $t=0$ of 
\begin{eqnarray} 
U_{(t \omega,p,n)} &=  &
\exp (t \omega^a_{l_1} (p_1) \cl_{e^a_{1}}) \nabla({l_1}) \exp (t\omega^a_{l_2} (p_2)  \cl_{e^a_{2}})\nabla({l_2})
\nn\\&&
\ldots 
 \exp (t\omega^a_{l_k} (p_k) \cl_{e^a_{k}} ) \nabla({l_k})    \; .
 \label{Yoda}
\end{eqnarray}
If we therefore compare to the formula (\ref{vlad}) and if we neglect the $\nabla({l_i})$ and the moving of the starting point of $p$ to the endpoint of $p$, then we see that in the continuum limit ($n\to \infty $) the operator $U_{(\omega , p, n)}  $ corresponds to a translation of $A$  over $p$ by $\omega$, i.e.
$$ (B_{(p,\omega)})_X(m)=\left\{  \begin{array}{cl}   
A_X(	m	)&, (m,X)\notin p \\ 
A_X(m)+\omega_X(m)&, (m,X) \in p
\end{array}  \right.  , $$
where $(m,X)\in TM$. Note that this  is a "singular" connection since it is given by a smooth connection which is deformed over the path $p$.

Let us now look at what happens if we act on a semi-classical state. We will here consider a semi-classical state, which is associated to a single vertex. Let $\psi$ be a spinor over $M$. With the notation of section 3.3 we consider 
$$
\xi_{(n,A,E)}(v)=\psi (v)\prod_{i=1}^{|l|}\xi_{(n,i)}\;,
$$
where $v$ denote a vertex in $\Gamma_n$ and where $(A,E)$ denotes a semi-classical phase-space point.  The state $\xi_{(A,E)}$ is the continuum limit of $\xi_{(n,A,E)}$. Furthermore we also consider 
$$
\eta_{(n,A,E)}=\prod_{i=1}^{|l|}\xi_{(n,i)}
$$ 
and $\eta_{(A,E)}$ as the continuum limit thereof.
Let $F= f e^X$ be a flow. This is approximated in the lattice $\Gamma_n$ by a family of paths. Let $F_m$ be the  path ending in $m$ generated by $F$, and let $U_{(\omega,F)}$ be the operator $\sum_{m\in M} U_{(\omega,F_m)}  $, where $U_{(\omega,F_m)} $ is the continuum limit of $U_{(\omega,F_m,n)} $.  From the above we see that\footnote{Here we only concern ourselves with the effect on the connection. We are not sure what happens to the field $E$.}  
$$   (U_{(\omega,F)}  (\xi_{(A,E)}))(m)= Hol (B_{(p,\omega)}, F_m)\psi (F^{-1}(m)) \eta_{(B_{(F_m,\omega)},E)} .  $$
If we forget the factor $Hol (B_{(p,\omega)}, F_m)\psi (F^{-1}(m))$ we see that each semi-classical state $\eta_{(B_{(F_m,\omega)},E)}$ is sitting in a point $m\in M$. Therefore the operator $U_{(\omega,F)}$ introduces discontinuities in the state, since for two neighboring points $m_1,m_2$ the states $\eta_{(B_{(F_{m_1},\omega)},E)}$ and $\eta_{(B_{(F_{m_2},\omega)},E)}$ are not necessarily close, because the singularities added to $A$ in the two states are sitting over different paths, namely over $F_{m_1}$ and $F_{m_2}$.

We can see the state $U_{(\omega,F)}\xi_{(A,E)}$ as a kind of integral over the $\eta_{(B_{(F_m,\omega)},E)}$, and therefore also as an integral over the representations of the holonomy-diffeomorphism algebra induced by the connections $B_{(F_m,\omega)}$. However, although the spectrum contains a lot of singular objects, and it does not appear to contain any transition between different smooth connections (see section \ref{overlap}), some of the expectation values in the GNS-construction bear resemblance to a transition between smooth connections. Let us for example consider a closed flow $F$. The expectation value
$$ \langle U_{(\omega,F)}\xi_{(A,E)} |F  | U_{(\omega,F)}\xi_{(A,E)} \rangle $$
will look like the expectation value of the connection $A+\omega' $, where $\omega'$ depends on $F$ and $\omega$, i.e. it will look like the expectation value of a smooth connection different from $A$.

Note that exponentiated vector-fields of the form (\ref{Yoda}) do in fact {\it not} directly arise in the $\mathbf{dQHD}^*(M)$ algebra. However the difference between the double commutators and the exponentiated vector-fields should be seen as the difference between the usual Dirac-operator on a manifold and the flows of vectorfields on the same manifold. The first does not directly moves point on the manifold, but contains all the infinitesimal translations. We therefore expect that the double commutators contain infinitesimal translations in the directions between $A$ and the $B_{(p,\omega)}$'s. More analysis is needed to determine exactly how the algebra $\mathbf{dQHD}^*(M)$ affects the spectrum of the semi-classical state.

\subsection{A bimodule over volume-preserving diffeomorphisms}
\label{lahar}

Let us end this section with a brief comment on the mathematical structure, which we have obtained so far\footnote{Here we shall assume that the issue with the invariance properties of $D$ raised in section \ref{mozart} has been resolved. This means that the factor $2^{-n}$ in $D$ should be viewed as a one-form.}.
 Before we do this we need however a few definitions and some notation. Also, we shall in this subsection use mathematical terms, which are explained in \cite{Aastrup:2012vq} and in the literature cited there. 

Consider again the representation of an element in $\mathbf{HD}_n$ in $\ch_n$ and $\ch_n'$ according to (\ref{BigMatrix}). We may, in a completely similar manner represent diffeomorphisms in $\ch_n$ and $\ch'_n$ as well as  in $\ch_{(\xi_n,\mathbf{dQHD}^*_n)}$, simply by replacing the holonomy transforms in matrices like (\ref{BigMatrix}) by the two-by-two identity matrix. We invite the reader to see \cite{Aastrup:2012vq} for details and shall here simply introduce the notation $\mbox{Diff}_n$ and $\mbox{Diff}(M)$ for the corresponding algebra of diffeomorphisms restricted to $\G_n$ and the algebra of diffeomorphisms on $M$, respectively. Furthermore, denote by $\mbox{Diff}_{vol,n}$ and $\mbox{Diff}_{vol}(M)$ the corresponding algebras of volume-preserving diffeomorphisms. At the level of a lattice $\G_n$ the volume-preserving diffeomorphisms are the diffeomorphisms, which are invertible.

Now, we first note that the state $\xi_c$ defines a Hilbert $\mbox{Diff}_{vol}(M)$-module with an action of both $\mathbf{HD}(M)$, $\mathbf{HD}^*(M)$ and $\mathbf{dQHD}^*(M)$. This arises by leaving out the trace over the ${\vert {\bf v}_n\vert \times \vert {\bf v}_n\vert}$ matrices in $\ch_{(\xi_n,\mathbf{dQHD}^*_n)}$. In this way the $\mathbf{dQHD}^*(M)$ algebra has a left-action and the $\mbox{Diff}_{vol}(M)$ algebra a right action. The reason why we restrict ourselves to volume-preserving diffeomorphisms -- and not simply all diffeomorphisms -- is given by our next observation: the operator $\d$ commutes with $\mbox{Diff}_{vol}(M)$. Thus, if $\d$ had been a Dirac-type operator we would have a Kasparov $(\mathbf{HD}(M),\mbox{Diff}_{vol}(M))$-bimodule.  Instead $\d$ is a commutator with a Dirac-type operator and thus the construction is of a somewhat different nature. 


\section{Emerging elements of fermionic QFT}

In this section we will analyze states in $\ch_{(\xi_c,\mathbf{dQHD}^*)}$ from which a link to fermionic QFT emerges in a semi-classical limit. We show that both a spatial Dirac operator and the principal part of the Dirac Hamiltonian emerges from the construction. This analysis built on the papers \cite{Aastrup:2009dy}-\cite{Aastrup:2011dt} and \cite{AGNP1}.


\subsection{Infinitesimals}

First we need to define infinitesimal elements in $\mathbf{HD}(M)$, where infinitesimal is understood with respect to the manifold $M$. These infinitesimal elements are again defined via the algebras $\mathbf{HD}_n$ and are given by an infinite sequence of lattice approximations, which we denote $\{da_n\}$, where $da_n\in\mathbf{HD}_n$ consist of matrix entries, which are only non-zero if they correspond to adjacent vertices:
\begin{equation}
(da_n)_{ij} = \left\{
\begin{array}{cl}
g_k\;\mbox{or}\; 0 & \mbox{if}\; v_i, v_j \;\mbox{are adjacent}\\
0     & \mbox{else}
\end{array}
\right.,\nn
\end{equation}
where $g_k$ is an element of a copy of $SU(2)$ assigned to the edge $l_k$ that connect adjacent vertices $v_i$ and $v_j$ in $\G_n$. 

Next, we define corresponding elements in $\mathbf{HD}^*_n$ and $\mathbf{HD}^*(M)$, which are both infinitesimal and non-trivial with respect to the Clifford algebra. The matrix entries  of these elements are of the form
\begin{equation}
(d\tilde{a}_n)_{ij}  = \left\{
\begin{array}{cl}
2^{-n}{\bf e}^a_k  \sigma^a g_k \;\mbox{or}\; 0&\mbox{if}\; v_i, v_j \;\mbox{are adjacent}\\
0     & \mbox{else}
\end{array}
\right.,
\label{MWind}
\end{equation}
where the configuration of vertices is as above. Note that 
$$ 
d\tilde{a}_n =  \frac{1}{\kappa}  \d_n d a_n\;.
$$
which means that $d\tilde{a}	$ is, in the language of non-commutative geometry, a one-form with respect to the space $\ca$ of connections.

\subsection{A two-point lattice state}

Let us for simplicity first consider a lattice $\G_n$ with only the two vertices $v_1$ and $v_2$ and an edge $l_k$ connecting them, and let $da_n$ and $d\tilde{a}_n$ denote the corresponding element in $\mathbf{HD}^*_n$ that connects them. Also, for the remainder of this section we shall ignore the second term in (\ref{insertion}) since this term does not play a role in the semi-classical analysis, which we are concerned with here. Although both terms in (\ref{insertion}) are proportional to $\kappa^2$ the first term is effectively of order $\kappa$ since the right-invariant vector field absorbs one order of $\kappa$ in the semi-classical limit (see for instance equation (\ref{JP1})).

Consider the state 
$$
\rho_n = (\psi(v_1)  +2^{2n} d\tilde{a} \psi(v_2)da^* )\xi_n
$$
where $\psi(v_i)$ is a $2\times 2$-matrix associated to $v_i$.  The double factor of $2^n$ corresponds to two one-dimensional delta-functions in the $n\rightarrow\infty$ limit, which means that this state does, strictly speaking, not correspond to a continuum state in the Hilbert space $\ch_{(\xi_c,\mathbf{dQHD}^*)}$. These delta-functions correspond to the localization of the infinitesimal element $da$ in $\mathbf{HD}(M)$ in the direction of its flow. Ignoring this issue we are going to compute the expectation value of $D_n$ to lowest order in $\kappa$ on this state. 
Due to the trace over the Clifford algebra we find
\begin{eqnarray}
\left\langle \rho_n \vert D_n \vert \rho_n\right\rangle &=&  \left\langle  \xi_n ( \psi(v_1) + 2^{2n} d\tilde{a}  \psi(v_2) da^*) \vert D_n \vert  ( \psi(v_1) + 2^{2n} d\tilde{a} \psi(v_2) da ^*)  \xi_n\right\rangle\nn\\
&=& 
2^{2n}\big(     \left\langle  \xi_n  \psi(v_1)  \vert D_n\vert  d\tilde{a}   \psi(v_2) da^*  \xi_n\right\rangle 
\nn\\&&
\quad\;+
\left\langle  \xi_n   d\tilde{a}  \psi(v_2) da^* \vert D_n\vert   \psi(v_1) \xi_n\right\rangle   \big)\;.
\nn
\end{eqnarray}
To proceed we need to denote the classical limit $\kappa\rightarrow 0$ of the basic operators on the state $\xi_{(n,k)}$. We do this in a manner parallel to (\ref{JP1}) and (\ref{JP2}):
\begin{eqnarray}
\lim_{\kappa\rightarrow 0} \left\langle \xi_{(n,k)} \vert     \kappa \cl_{{\bf e}_k^a}    \vert \xi_{(n,k)}\right\rangle &=& \mathrm{i}E^\m_a (x)
\label{clcondition}
\\
 \lim_{\kappa\rightarrow 0} \left\langle \xi_{(n,k)}\otimes v \vert    g_k   \vert \xi_{(n,k)}\otimes v \right\rangle &=&  ( v, Hol(l_k, A) v) 
\end{eqnarray}
 where $v\in \mathbb{C}^2$ and where $(,)$ denotes the inner product hereon. Also, the index $\m\in\{1,2,3\}$ corresponds to the spatial orientation of $l_k$ and '$x$' refers to the endpoint of $l_k$. We take the set $(E,A)$ to be a pair of conjugate Ashtekar variables -- $E^\m_a(x)$ is an inverse densitized triad field and $A_\m^a(x)$ an $SU(2)$ connection --  so that $Hol(l,A)$ denotes the holonomy of a connection $A$ along $l$. With this in place we find
\begin{eqnarray}
&&\hspace{-1cm}\lim_{\kappa\rightarrow 0} 2^{2n}    \left\langle  \xi_n  \psi(v_1) \vert D_n\vert   d\tilde{a} \psi(v_2)  da^*   \xi_n\right\rangle
\nn\\
&&\hspace{2cm}=
\lim_{\kappa\rightarrow 0} 2^{n}  \left\langle  \xi_n \psi^*(v_1)  \kappa \cl_{{\bf e}_k^a}\sigma^a g_k  \psi(v_2) g^*_k   \xi_n\right\rangle
\nn\\
&&\hspace{2cm}= 2^n  \mbox{Tr}_{M_2} (\psi^*(v_1)  \mathrm{i}E^\m_a \sigma^a Hol(l_k,A)  \psi(v_2) Hol(l_k,A)^* )\;.
\nn
\end{eqnarray}
We are now going to let the two vertices approach each other. With a deliberate misuse of notation, we call this the limit $n\rightarrow\infty$ despite the fact that we, for now, keep $n=2$ fixed. The point is that we shall, in a next step, consider a full lattice, for which the $n\rightarrow\infty$ limit implies that adjacent lattice points converge onto each other with respect to the lattice metric, i.e. $\lim_{n\rightarrow\infty} 2^{-n} =dx$.
Thus, keeping this in mind, we may write
$$
\lim_{n\rightarrow\infty}   Hol(l_k,A) = \mathds{1} + dx A_\m \;,\quad \lim_{n\rightarrow\infty} \psi(v_2) = \psi(v_1) + dx \pa_\m \psi(v_2)
$$
which implies
$$
\lim_{n\rightarrow\infty} Hol(l_k,A)  \psi(v_2) Hol(l_k,A)^* = (1+dx \nabla_\m)\psi(v_1)
$$
where $\nabla_\m\cdot = \pa_\m + [A_\m,\cdot]$ is the covariant derivative. Adding all this up we find
\begin{eqnarray}
&&\hspace{-1cm}\lim_{\kappa\rightarrow 0} \lim_{n\rightarrow\infty}  \left\langle \rho_n \vert D_n \vert \rho_n\right\rangle 
\nn\\
&&\hspace{1cm}=  \mbox{Tr}_{M_2} \left( \psi^*(x) E^\m_a  \mathrm{i} \sigma^a \nabla_\m  \psi(x)  -  ( \nabla_\m \psi(x) )^*  E^\m_a \mathrm{i}\sigma^a   \psi(x)  \right)
\nn\\
&&\hspace{1cm} =\mbox{Tr}_{M_2} \left(  \psi^*(x)  (   E^\m_a \mathrm{i}\sigma^a \nabla_\m + \nabla_\m  E^\m_a  \mathrm{i} \sigma^a ) \psi(x)   \right)
\nn\\
&&\hspace{1cm} =\mbox{Tr}_{M_2} \left(  \psi^*(x) \slashed{D} \psi(x)   \right)
\end{eqnarray}
where we wrote $\psi(v_1)$ as $\psi(x)$ and where we in the second line permitted ourselves a partial integration although we will not have this until we consider a full lattice and its continuum limit. Note that with our conventions $ \mathrm{i} \sigma^a$ equals the standard Pauli matrices.

To get this far we had to start with a state $\rho_n$ with a factor of $2^n$ in front of each infinitesimal element $da_n$, which means that this state does not correspond to a state in $\ch_{(\xi_c,\mathbf{dQHD}^*)}$ in the $n\rightarrow\infty$ limit. The issue is clear: when we act with $\d_n$ we get a factor of $2^{-n}$ from $D_n$, which means that $\d_n a_n$, with $a_n\in\mathbf{HD}^*_n$, has exactly the right factor to give a well-defined line-integral in the limit $n\rightarrow\infty$. This, however, becomes problematic when we have an infinitesimal element in $\mathbf{HD}^*_n$, since the corresponding line integral just gives a $dx$, which vanishes. This explains the factor of $2^n$, which effectively corresponds to a one-dimensional delta-function. Perhaps one can compare this to the situation in quantum mechanics, where the plane wave is not square integrable. It may be that the strict localization of $da$ in the direction of the flow is too idealized a point of view, albeit not of principal concern. 
Note also that $\d_n\rho_n$ does descent to a well-defined state in $\ch_{(\xi_c,\mathbf{dQHD}^*)}$.

\subsection{An emergent one-particle state}

Let us now consider a sequence of lattice approximations $\{\G_n\}$ and consider the state at the $n$'th level of approximation:
\begin{equation}
\rho_n = (\psi  +2^{2n}\sum_{m=1}^3  d\tilde{a}_n^{(m)} \psi da_n^{(m)*} )\xi_n 
\label{one-state}
\end{equation}
where $da_n^{(m)}$ and $d\tilde{a}_n^{(m)}$ are now infinitesimal elements of $\mathbf{HD}^*_n$, oriented in the $m$'th direction, $m\in\{1,2,3\}$, and otherwise as defined previously. By $\psi$ we now denote a $\vert {\bf v}_n\vert \times \vert {\bf v}_n\vert$ diagonal matrix, where each diagonal entry $(j,j)$ is a $M_2(\mathbb{C})$-valued field $\psi(v_j)$. In the continuum limit $\psi$ becomes a spinor $\psi(x)$.

It is now straight forward to repeat the computation of the previous section. We are now dealing with $\vert {\bf v}_n\vert \times \vert {\bf v}_n\vert$ matrices and their limit as $n$ grows infinitely large, where the trace over $M_{\vert {\bf v}_n\vert}(\mathbb{C}) $ converges onto a Riemann integral, but at the level of each matrix entry the situation is identical to the situation encountered above with the exception that we are now including all three spatial directions. Thus, we write
\begin{equation}
\lim_{\kappa\rightarrow 0} \lim_{n\rightarrow\infty}  \left\langle \rho_n \vert D_n \vert \rho_n\right\rangle 
= \lim_{\kappa\rightarrow 0}   \left\langle \rho \vert D \vert \rho\right\rangle 
 =\int_M d^3x \mbox{Tr}_{M_2} \left(  \psi^*(x) \slashed{D} \psi(x)   \right)
\label{LIMLIM}
\end{equation}
where the domain of the integral is determined by the spatial domain of the holonomy-diffeomorphism $da$.

Note that in the above derivation we did not pay particular attention to the order of the two limits, $\kappa\rightarrow0$ and $n\rightarrow\infty$, which is justified since these limits commute. This is one major difference between the computation given here and what we found in the papers \cite{Aastrup:2009dy}-\cite{Aastrup:2011dt} and \cite{AGNP1},
where the setup entailed that these limits did not commute: the semi-classical limit had to be taken before the continuum limit.

To obtain the Dirac Hamiltonian in a semi-classical limit we perform a transformation in the $M_2(\mathbb{C})$-factor in $\ch_{(\xi_n,\mathbf{dQHD}^*_n)}$ in order to include the lapse and shift fields. Thus, consider an $\vert {\bf v}_n\vert \times  \vert {\bf v}_n\vert$ matrix $M_n$ with diagonal entries given by two-by-two self-adjoint matrices $M(v_i)$, together with a sequence of such matrices assigned to lattice approximations $\G_n$. Consider the transformation
$$
\left\langle \rho_n \vert D_n \vert \rho_n\right\rangle \rightarrow \left\langle \rho_n \vert  M_n D_n \vert \rho_n\right\rangle 
$$
which gives
\begin{eqnarray}
\lim_{\kappa\rightarrow 0} \lim_{n\rightarrow\infty}  \left\langle \rho_n \vert M_n D_n \vert \rho_n\right\rangle 
\hspace{-2cm}&&
\nn\\
& =&  \int_M d^3x \mbox{Tr}_{M_2} \left(  \psi^*(x) N(x) \slashed{D} \psi(x)  +  \psi^*(x) N^\m(x) e \pa_\m \psi(x)  \right)
\nn\\
&&+\mbox{zero order terms}
\label{LIMLIM2}
\end{eqnarray}
where 
we wrote $M(v_i) := M(x) = N(x) + N^a(x)\sigma^a$ with $N(x)$ and $N^a(x)$ playing the role of the lapse and shift fields.

Thus, we find that the principal part of the Dirac Hamiltonian emerges in a semi-classical limit of the gravitational degrees of freedom from a natural class of states in $\ch_{(\xi_n,\mathbf{dQHD}^*_n)}$. We obtain the lapse and shift fields from a local transformation 
 in the spinor degrees of freedom represented by the $M_2(\mathbb{C})$-factor in $\ch_{(\xi_n,\mathbf{dQHD}^*_n)}$.
 This establishes a connection to fermionic quantum field theory and in particular it shows that the framework of quantum holonomy theory has the potential of producing canonical matter degrees of freedom coupled to quantum gravitational degrees of freedom.\\ 

 It is not clear what geometrical significance the structure of the one-particle state (\ref{one-state}) has. In terms of non-commutative geometry it involves a one-form, and if we consider the quantity $d\tilde{a} da^*$ alone, which we write (up to a constant) as
 $$
 [D,da] da^*\;,
 $$
then we recognize the quantity, which arises from a fluctuation of the Dirac type operator with by an infinitesimal inner automorphism:
$$
D\rightarrow  da D da^* = D - [D,da]da^*\;.
$$
It is, however, not clear to us if this is of significance. Alternatively one could reproduce the above results as the inner product between states, that involve the double commutator with $D_n$. This interpretation would be more in line with the emphasis put on the algebra $\mathbf{dQHD}^*_n$ and would circumvent the fact that the Dirac type operator is not a well defined operator in the Hilbert space $\ch_{(\xi_c,\mathbf{dQHD}^*)}$.

One might, however, speculate whether the algebra of differential forms generated by the Dirac type operator is related to a type of Fock-space structure, where one-forms corresponds to one-particle states and $n$-forms to $n$-particle states. This idea was put forward in \cite{Aastrup:2011dt}, but the results on $n$-particle states obtained there cannot be incorporated in the setting of this paper in a straight forward manner.\\

The introduction of the Dirac type operator and the Clifford algebra does entail the interesting concept of orthonormal diffeomorphisms, which might be of significance when one is to determine to what extend more elements of fermionic quantum field theory can be uncovered from quantum holonomy theory.
Therefore, let us end this section by developing the notion of orthogonality among diffeomorphisms. First, we define a class of elements in $\mathbf{HD}^*_n$, which are built from elements similar to the $d\tilde{a}_n$'s. A characteristic of the operator $d\tilde{a}_n$ is that it is not unitary. We can remedy this by changing its definition in (\ref{MWind}) to (see \cite{Aastrup:2011dt})
$$
(d\tilde{a}_n)_{ij}  =  2^{-(n+1)}({\bf e}^a_k  \sigma^a + \mathrm{i} {\bf e}^1_k{\bf e}^2_k{\bf e}^3_k ) g_k\;,    \qquad \mbox{(alternative definition)}
$$
and we check that $d\tilde{a}_n$ with this definition is unitary. In \cite{Aastrup:2011dt} we also found that
$$
\mbox{Tr}_{Cl} (  (d\tilde{a}_n)_{ab}  M (d\tilde{a}_n)^*_{ba}        ) = M^0 
$$
where $M$ is a two-by-two matrix, which we write as $M=M^a \sigma^a + M^0 \mathds{1}$, and where '$ab$' refers to a pair of adjacent vertices in $\G_n$. Thus, conjugation with $d\tilde{a}_n$ corresponds to taking the trace in the $M_2(\mathbb{C})$-factor of $\ch_{(\xi_n,\mathbf{dQHD}^*_n)}$. This fact played a crucial role in the analysis carried out in \cite{Aastrup:2011dt}.

We now define a class of elements in $\mathbf{HD}^*_n$ as products of these infinitesimals (we omit the subscript '$n$')
$$
\tilde a = d\tilde{a}_1 d\tilde{a}_2\ldots d\tilde{a}_k\;, \quad \mbox{with}\quad a = a_1 a_2\ldots a_k\;.
$$
Notice that these elements in $\mathbf{HD}^*_n$ are mutually orthogonal
$$
\left\langle \tilde{a}\vert \tilde{a}' \right\rangle =\d_{a,a'}
$$
where $\d_{a,a'}$ equals one if $a=a'$ and zero else. This, too, will apply in the continuum limit. Thus, this introduces a notion of orthogonality among holonomy-diffeomorphisms.



\section{On a dynamical principle}

We are now going to consider the dynamics of general relativity, which in the formulation in terms of Ashtekar variables is encoded in two constriants: the Hamilton and the diffeomorphism constraints. The Hamilton constraint has the form\footnote{Here we write the densitized Hamilton constraint \cite{Ashtekar:1986yd} in which the integral is not invariant but depends on scale due to an extra factor of $e$. Also, as pointed out elsewhere, this Hamiltonian corresponds either to gravity with Euclidian signature - if the connection takes values in $\mathfrak{su}(2)$, or to gravity with Lorentzian signature - if the connection takes values in the dual sector of $\mathfrak{sl}(2,\mathbb{C})$, see section \ref{REAL}. }
\begin{equation}
H(N)= \int_M d^3x N 
  \e^{ab}_{\;\;c} F^c_{\m\n} E^\m_a E^\n_b \;,
\label{DYBB¯L}
\end{equation}
where 
$N$ is the lapse field, $E_a^\m(x)$ the inverse densitized dreibein and $F_{\m\n}(x)$ the field-strength tensor of the Ashtekar connection $A_m(x)$.
The diffeomorphism constraint has the form 
\begin{equation}
D(\bar{N})= \int_M d^3x   N^\n
 F^c_{\m\n} E^\m_c \;,
\label{DYBB¯L2}
\end{equation}
where $N^\n$ is the shift field. 
The aim of this section is twofold: first to investigate how we can obtain these classical constraints in a semi-classical limit from operators acting in $\ch_{(\xi_c,\mathbf{dQHD}^*)}$, and second to check whether these constraint operators satisfy an algebra comparable to the classical constraint algebra. The latter issue is crucial since it determines to what extend general covariance is maintained in the quantized theory\footnote{see \cite{Nicolai:2005mc} for an interesting discussion on this issue in a framework comparable to the one presented here.}. The classical constraint algebra modulo the Gauss constraint has the form:
\begin{eqnarray}
\left\{  H(N), H(N') \right\}_{\mbox{\tiny PB}} &=& D(N\pa N' - N'\pa N)
\label{one}\\
\left\{ H(N),D(\bar{N})  \right\}_{\mbox{\tiny PB}} &=& H( N \pa \bar{N} -  \bar{N} \pa N)
\label{two}\\
\left\{ D(\bar{N}),D(\bar{N}')  \right\}_{\mbox{\tiny PB}} &=& D(\bar{N}\pa \bar{N}'  - \bar{N}'\pa \bar{N}  )
\label{three}
\end{eqnarray}

\subsection{The Hamilton constraint operator}


In the following we first provide a geometrical motivation for choosing our candidate for a Hamilton constraint operator. Then we compute the commutator between two Hamilton constraint operators -- which differ by choice of lapse fields -- and derive from that a candidate for a diffeomorphism constraint operator. With that we then compute the rest of the operator constraint algebra.\\

As already mentioned, a one-form in the vocabulary of noncommutative geometry is given by an operator of the form $ a [D,b]$ where $a,b$ are elements of the particular algebra at hand and $D$ is a Dirac operator. Here we need one-forms which are infinitesimal with respect to the manifold $M$ and which have a certain orientation. We denote by
$db_{n,i}$, $i\in\{1,2,3,4,5,6\}$ infinitesimal elements of $\mathbf{HD}_n$, where $db_{n,1}$ translates in the positive $x$-direction, $db_{n,2}$ in the negative $x$-direction, $db_{n,3}$ in the positive $y$-direction and so forth. We built two distinct one-forms by
\begin{eqnarray}
R_n = \sum_{i,j}^{\mbox{\tiny right}} db_{n,i} [D_n, db_{n,j}]      
&&
L_n =  \sum_{i,j}^{\mbox{\tiny left}}  db_{n,i} [D_n, db_{n,j}]    
\nn       
\end{eqnarray}
where the 'right' sum picks out combinations $(db_{n,i},db_{n,j})$ that form an angle in the lattice with a positive orientation and where the 'left' sum picks out terms with negative orientation. Recall that the brackets are graded. Note that the conjugation switches between the 'right' and 'left' sectors. 
%
We also write down the corresponding curvature operators
\begin{eqnarray}
\cf_{R_n}&=&  [D_n, R_n] + \frac{1}{2}[R_n,R_n] \;,
\nn\\ 
\cf_{L_n}&=& [D_n ,L_n] + \frac{1}{2}[L_n,L_n]  \;,
\nn
\end{eqnarray}
which can be understood as two curvature operators over the approximate space of connections $\ca_n$ and consequently in terms of curvature operators over $\ca$ in the $n\rightarrow\infty$ limit.

In the following we shall partly be concerned with the lowest orders in $\kappa$ in a semi-classical limit. Note that in a semi-classical limit one factor of $\kappa$ will be absorbed by a vector-field $\cl_{{\bf e}_j^a} $ due to (\ref{clcondition}) whereas a commutator $[D_n , a_n]$ with $a_n\in\mathbf{HD}_n$ will produce a factor of $\kappa$.
Furthermore, in line with the previous section, we shall equip each infinitesimal element $db$ with a factor of $2^n$. Again, this implies that these elements do in fact not correspond to operators in $\ch_{(\xi_c,\mathbf{dQHD}^*)}$, but we shall permit ourselves this discrepancy and refer to the comments of the previous section.

Let us now consider the matrix entries in $L_n$ and $R_n$. They are of the form
$$
\kappa 2^{n}  {\bf e}^a_j g_i  g_j\sigma^a \;,
$$
where $g_i$ and $g_j$ correspond to adjacent edges in $\G_n$, which do not have the same spatial direction. Correspondingly, matrix entries in $\cf_{R_n}$ and $\cf_{L_n}$ are of the form
\begin{equation}
2^n \kappa ( \kappa \cl_{{\bf e}_j^a} ) g_i    g_j  \sigma^a + \mbox{"three terms"} \;,
\label{strand}
\end{equation}
where "three terms" refer to three additional terms, which we shall for the moment ignore since they do not affect our analysis. We will return to these terms later in this section.

Let us consider the following operator:
\begin{eqnarray}
\mathds{h}_n &=&  \frac{\mathrm{i}}{2} \big( \cf_{R_n} (\cf_{L_n})^*  -  \cf_{L_n} (\cf_{R_n} )^*\big) \nn\\
&=& \mbox{Im}\big( \cf_{L_n} (\cf_{R_n} )^* \big) \;.
\label{cur}
\end{eqnarray}
We shall shortly explain why we choose this particular combination, but let us first write down the form of the matrix entries that arise.
Using (\ref{strand}) we find
\begin{equation}
 2^{2n}\kappa^2 (\kappa\cl_{{\bf e}_j^a})  \sigma^a  g_i g_j   g_k g_l \sigma^b  (\kappa \cl_{{\bf e}_k^b}) \;,
\label{Steinhard}
\end{equation}
where it is important to notice that the two derivatives in (\ref{Steinhard}) will be associated to adjacent edges due to the algebraic structure of (\ref{cur}).  Due to the particular choice of orientations we are certain that in those contributions, which are diagonal in the $M_{\vert {\bf v}_n\vert}(\mathbb{C})$ matrix, the four infinitesimal elements will form a closed loop, which gives the field strength tensor in the semi-classical limit:
\begin{eqnarray}
 \lim_{\kappa\rightarrow 0} \lim_{n\rightarrow\infty}\left\langle \xi_{n} \vert   g_i g_j   g_k g_l 
     \vert \xi_{n} \right\rangle = \mbox{Tr}_{M_2}
     \left(
  \mathds{1}_2 + dx^2 F_{\m\n} \right) + \co(dx^3)\;,
\label{knud}
\end{eqnarray}
where the indices $\m$ and $\n$ corresponds to the plane in which the loop sits.
The first term in (\ref{cur}) contribute with three loops in the three different lattice planes, all with the same orientation, whereas the second term in (\ref{cur}) contribute with the inverse of these same three loops. Thus, we have the algebraic structure 
$$\mbox{loop} - (\mbox{loop})^{-1}\;.$$
This is the reason why we choose the particular $L$-$R$ combination in (\ref{cur}), because with this algebraic structure, and keeping the first condition in (\ref{conditions}) in mind, we see that the identity terms of the loops as in (\ref{knud}) cancels out whereas the term, which gives the field strength tensor $F_{\m\n}$ in the semi-classical limit, add up. Also, the reason why we split the one-forms into "R" and "L" sectors is to avoid backtracking. 
With this we can now
 take the expectation value of (\ref{cur}) and find
\begin{equation}
\lim_{\kappa\rightarrow 0}\lim_{n\rightarrow\infty} \kappa^{-2}  \left\langle \xi_n \vert   \mathds{h}_n  \vert \xi_n \right\rangle = \int_M d^3x  \Big( E_a^\m E^\n_b F_{\m\n}^c \e^{ab}_{\;\;\;c}\Big)\;.
\label{dynamics}
\end{equation}
Comparing this to equation (\ref{DYBB¯L}) we see that the densitized Hamiltonian of general relativity emerges in the semi-classical limit. \\

The particular form of the operator (\ref{cur}) must be rooted in symmetry considerations. Without going into details let us here just note that it seems to be invariant under a symmetry that involves a change of orientation of $M$ as well as a gauge transformation of the "connections" $R_n$ and $L_n$.  It seems natural that the operator (\ref{cur}) should have the canonical form
$$
 \cf_{A_n} (\cf_{A_n})^* 
$$
where $A_n$ is a one-form that involves edges of both "right" and "left" orientation. Such a construction must, however, involve some sort of grading with respect to orientation, which we at the moment do not know how to implement. Thus, we work with the operator (\ref{cur}).
\\

Before we continue let us first deal with the "three terms" in equation (\ref{strand}). The first term is of the form
$$
\kappa^2 {\bf e}^a_i {\bf e}^b_j g_i \sigma^a g_j \sigma^b\;.
$$
This term can, however, due to the Clifford algebra elements only give something non-zero via a backtracking, which will not arise due to the argumentation given above. The second term is of the form
$$
\kappa^2 {\bf e}^a_j {\bf e}^b_j g_i g_j  \sigma^a\sigma^b\;,
$$
which can provide a non-trivial contribution. This, however, will be of higher orders in $\kappa$ and shall, therefore, not concern us here. The third term comes from the commutators $[L_n,L_n]$ or $[R_n,R_n]$, but again we find that this term will vanish since it is non-trivial with respect to the Clifford algebra.\\




The operator $\mathds{h}_n$ can, as it stands, not be interpreted as a Hamilton operator, since it does not involve a trace over $M_{\vert {\bf v}_n \vert}(\mathbb{C})$, which in the $n\rightarrow\infty $ limit gives an integral over $M$, nor does it involve a trace over $M_2(\mathbb{C})$. Without this trace the operator has no chance of satisfying the right operator constraint algebra. Therefore, we consider instead
\begin{equation}
\mathds{H}'_n = \mbox{Tr}_{\mbox{\tiny partial}} \left(\mathds{h}_n\right) := dx^3\sum_i \mathds{H}'_{v_i}
\label{rommel}
\end{equation}
where $\mathds{H}'_{v_i}$ denotes the $i$'th diagonal matrix entry of $\mathds{h}_n$ and where the partial trace is a normalized trace over the matrix factors $M_{\vert {\bf v}_n \vert}(\mathbb{C})$ and $M_2(\mathbb{C})$. Note that it is not certain that this partial trace exist in the $n\rightarrow\infty$; we here simply assume that it does. 

One possibility is now to define the Hamilton operator $\mathds{H}$ as the continuum limit of $\mathds{H}'_n $. Here, however, we run into the same problem that we encountered with the operator $E_{\oo}$ and with the Dirac type operator $D$, namely that there is a conflict between 1) the existence of the expectation value of the operator itself on the vacuum state and 2) its commutators with elements of $\mathbf{HD}(M)$ being non-zero. With the definition (\ref{rommel}) we will get the correct expectation value on the vacuum state in the $n\rightarrow\infty$ limit -- the classical Hamiltonian -- but commutators between $\mathds{H}$ and elements of $\mathbf{HD}(M)$ will all vanish.

The solution we choose to circumvent this discrepancy is to change definition (\ref{rommel}) to
\begin{equation}
\mathds{H}'_n := dx \sum_i \mathds{H}'_{v_i}\;,\quad\mbox{(new definition)}
 \label{rommel22}
\end{equation}
which gives an operator that is only well defined in terms of commutators with elements in $\mathbf{dQHD}_n$. 


\subsection{The operator constraint algebra}

With definition (\ref{rommel22}) we are in a position where we can start computing the operator constraint algebra and use the bracket $[\mathds{H}'_n(N),\mathds{H}'_n(N')]$ to derive the diffeomorphism constraint operator. Before we do that we shall, however, write down yet a new version of (\ref{rommel22}) where the vector-fields are ordered symmetrically and where we also include a lapse field. This definition of the Hamilton operator is a simplification of (\ref{rommel22}) since it involves three instead of twelve loops based at each vertex, which greatly simplifies the computations.
Thus we write
\begin{equation}
\mathds{H}_n(N)= dx \sum_{i} \mathds{H}_{v_i}(N_{v_i})
\label{AGW}
\end{equation}
with the vertex operators $\mathds{H}_{v_i}(N_{v_i})$ given by 
$$
 \mathds{H}_{v_i}(N_{v_i}) =     \mathds{H}^{(xy)}_{v_i}(N_{v_i}) + \mathds{H}^{(yz)}_{v_i}(N_{v_i}) + \mathds{H}^{(zx)}_{v_i}(N_{v_i})     
$$
and 
\begin{eqnarray}
 \mathds{H}^{(xy)}_{v_i}(N_{v_i}) 
 &=& 2^{2n-2}  N_{v_i}  \left\{ \cl^{v_i}_{{\bf e}^b_y} , \left\{ \cl^{v_i}_{{\bf e}^a_x}  , \mbox{Tr} \left( \sigma^a  \stackrel{\hspace{-3mm}\curvearrowleftright}{L^{v_i}_{xy}}       \sigma^b   \right) \right\} \right\}
 \nn\\
  \mathds{H}^{(yz)}_{v_i}(N_{v_i}) 
  &=& 2^{2n-2}  N_{v_i}  \left\{ \cl^{v_i}_{{\bf e}^b_z} , \left\{ \cl^{v_i}_{{\bf e}^a_y}  , \mbox{Tr} \left( \sigma^a  \stackrel{\hspace{-3mm}\curvearrowleftright}{L^{v_i}_{yz}}       \sigma^b   \right) \right\} \right\}
 \nn\\
  \mathds{H}^{(zx)}_{v_i}(N_{v_i}) 
  &=& 2^{2n-2}  N_{v_i}  \left\{ \cl^{v_i}_{{\bf e}^b_x} , \left\{ \cl^{v_i}_{{\bf e}^a_z}  , \mbox{Tr} \left( \sigma^a  \stackrel{\hspace{-3mm}\curvearrowleftright}{L^{v_i}_{zx}}       \sigma^b   \right) \right\} \right\}
 \label{symmm}
\end{eqnarray}
where $N$ is the element in $M_{\vert {\bf v}_n \vert}(\mathbb{C})\otimes  M_{2}(\mathbb{C})$, which is diagonal in the first factor and which gives rise to the lapse field. $N_{v_i}$ is the two-by-two diagonal matrix entry assigned to $v_i$.  We use the shorthand notation 
$$
\stackrel{\hspace{-3mm}\curvearrowleftright}{L^{v_i}_{\m \n}}  =  L^{v_i}_{\m \n}  - \left( L^{v_i}_{ \m \n} \right)^{-1} 
$$
where $L^{v_i}_{xy}$ is again the loop in the $xy$-plane based in $v_i$, where we write $L^{v_i}_{xy}=g_1 g_2 g^{-1}_3 g^{-1}_4$. Likewise for $L^{v_i}_{yz}$ and $L^{v_i}_{zx}$. Note again that we have here simplified the operator by only considering loops that start out in the positive $x$, $y$ and $z$ directions. The vector-fields are defined as
$$
\cl^{v_i}_{{\bf e}^a_x} =  \frac{1}{2}\left(  \cl^{v_i}_{{\bf e}^a_2} +\cl^{v_i}_{{\bf e}^a_4}    \right)   \;,\quad \cl^{v_i}_{{\bf e}^a_y} = \frac{1}{2}\left(\cl^{v_i}_{{\bf e}^a_1} +\cl^{v_i}_{{\bf e}^a_3}   \right)\;,
$$ 
where the right-invariant vector-field $ \cl^{v_i}_{{\bf e}^a_j}  $ is the vector-fields interacting with the $g_j$ element in the loop based in $v_i$. This means that the vector-fields in the operators (\ref{symmm}) appear in all possible adjacent pairs. The vector-fields appearing in the operators in the '$yz$' and '$zx$' planes are constructed in the same way. This notation is not completely unambiguous for the computational setup we shall encounter, but we trust that no confusion will arise.
In appendix \ref{appendixham} we show that
\begin{eqnarray}
\left[\mathds{H}_n(N), \mathds{H}_n(N')\right] &=& \mathds{D}_n (\bar{N}(N,N')) + \Xi_n(N,N')
\label{pochahontas}
\end{eqnarray}
where
\begin{eqnarray}
\mathds{D}_n (\bar{N}) = dx \sum_{v_i} \frac{1}{2}\left(  \left\{\bar{N}_{v_i}^x , \mathds{D}^{v_i}_x\right\}   + \left\{ \bar{N}_{v_i}^y,\mathds{D}^{v_i}_y\right\}  +\left\{ \bar{N}_{v_i}^z, \mathds{D}^{v_i}_z \right\}   \right)
\nn
\end{eqnarray}
and
\begin{eqnarray}
\mathds{D}^{v_i}_x &=& 2^{2n} \left( \mathds{D}^{v_i(xy)}_x -  \mathds{D}^{v_i(zx)}_x \right)
\nn\\
\mathds{D}^{v_i}_y &=& 2^{2n} \left( \mathds{D}^{v_i(yz)}_y -  \mathds{D}^{v_i(xy)}_y \right)
\nn\\
\mathds{D}^{v_i}_z &=&2^{2n} \left(  \mathds{D}^{v_i(zx)}_z -  \mathds{D}^{v_i(yz)}_z \right)
\nn
\end{eqnarray}
where
\begin{eqnarray}
\mathds{D}^{v_i(xy) }_{y} = \frac{1}{2}\left\{ \cl^{v_i}_{{\bf e}^a_x}   
 ,  \mbox{Tr} \left( \sigma^a  \stackrel{\hspace{-3mm}\curvearrowleftright}{L^{v_i}_{xy}}        \right)   \right\} &,&
 \mathds{D}^{v_i(xy)}_{x} = \frac{1}{2}\left\{ \cl^{v_i}_{{\bf e}^a_y}   
 ,  \mbox{Tr} \left( \sigma^a  \stackrel{\hspace{-3mm}\curvearrowleftright}{L^{v_i}_{xy}}        \right)   \right\}\;,
 \nn\\
\mathds{D}^{v_i(yz)}_{z} = \frac{1}{2}\left\{     \cl^{v_i}_{{\bf e}^a_y} 
 , \mbox{Tr} \left( \sigma^a  \stackrel{\hspace{-3mm}\curvearrowleftright}{L^{v_i}_{yz}}         \right) \right\}
  &,&
\mathds{D}^{v_i(yz)}_{y}  =\frac{1}{2} \left\{ \cl^{v_i}_{{\bf e}^a_z}  , \mbox{Tr} \left( \sigma^a  \stackrel{\hspace{-3mm}\curvearrowleftright}{L^{v_i}_{yz}}        \right) \right\}\;,
\nn\\
\mathds{D}^{v_i(zx)}_{x} = \frac{1}{2}\left\{     \cl^{v_i}_{{\bf e}^a_z} 
 , \mbox{Tr} \left( \sigma^a  \stackrel{\hspace{-3mm}\curvearrowleftright}{L^{v_i}_{zx}}         \right) \right\}
 &,&
\mathds{D}^{v_i(zx)}_{z}  =\frac{1}{2} \left\{ \cl^{v_i}_{{\bf e}^a_x}  , \mbox{Tr} \left( \sigma^a  \stackrel{\hspace{-3mm}\curvearrowleftright}{L^{v_i}_{zx}}        \right) \right\}\;,
\label{faces}
\end{eqnarray}
and with
$$
\bar{N}_\m(N,N') = \sum_{v_i} \sum_{\n=x,y,z} \left(N\pa_\n N'-N'\pa_\n N\right)\frac{1}{4}\left\{  \cl^{v_i}_{{\bf e}^a_\n} ,  \left\{  \cl^{v_i}_{{\bf e}^a_\m} , \circ  \right\}\right\}
$$
where we wrote $(N\pa_\n N'-N'\pa_\n N)$ instead of its lattice approximation involving factors like $2^n (N_{v_i} N'_{v_{i+1}} - N'_{v_i} N_{v_{i+1}} )$ and thereby anticipating its continuum limit. We also wrote a '$\circ$' indicating that $\bar{N}(N,N') $ appears in (\ref{pochahontas}) in terms of anti-commutators with the $\mathds{D}_{v_i}$'s.

Thus, we find that the commutator (\ref{pochahontas}) successfully reproduces the structure of the classical Poisson bracket (\ref{one}) up to a potentially anomalous term $\Xi_n(N,N')$. The operator $\Xi_n(N,N')$ consist of terms, which involve a factor $(A_{v_{i+1}}-A_{v_i})$ where $A_{v_i}$ is either a vector-field or a loop operator or a combination of both. In order to rule out anomalies in this sector of the constraint algebra we need to show that all commutators with $\Xi_n(N,N')$ vanishes in a non-trivial sector of $\ch_{(\xi_n,\mathbf{dQHD}^*_n)}$ in the continuum limit $n\rightarrow\infty$. This will happen
if the factor $(A_{v_{i+1}}-A_{v_i})$ produces a factor of $dx$ in this limit, which is the case at least on $\ch_{(\xi_n,\mathbf{HD}_n)}$ due to the requirement that expectation values of all powers of the vector-fields be smooth with respect to $M$.

The question is whether the factors $(A_{v_{i+1}}-A_{v_i})$ produces factors of $dx$ on all of $\ch_{(\xi_n,\mathbf{dQHD}^*_n)}$ in the continuum limit. In fact, we expect that this will not be the case since exponentiated vector-fields will produce finite translations on the configuration space of connections (see section \ref{ISIL}), which may cause the factor $(A_{v_{i+1}}-A_{v_i})$ to remain finite in the continuum limit. Put differently, in section \ref{actdiff} we show that the action of the diffeomorphism group on holonomy-diffeomorphisms is not strongly continuous when we also include the vector-fields -- i.e. on $\mathbf{dQHD}^*(M)$ -- and therefore there will be no infinitesimal generators of diffeomorphisms. Thus, one cannot expect the constraint algebra to close on $\ch_{(\xi_c,\mathbf{dQHD}^*)}$  and in fact the meaning of diffeomorphisms on the entirety of $\ch_{(\xi_c,\mathbf{dQHD}^*)}$ may be questionable. Since the holonomy-diffeomorphisms are a part of the conjugate variables, it seems plausible that diffeomorphism invariance and covariance, which, after all, must be regarded as classical concepts, cannot be maintained in the full quantum theory.

This line of reasoning only holds, however, whenever we are dealing with operators, which entail finite translations on the configuration space of Ashtekar connections. If we restrict ourselves to finite orders in $\kappa$, which means finite polynomials of vector-fields and thus infinitesimal translations only, then we can be certain that a factor $(A_{v_{i+1}}-A_{v_i})$ will produce a factor of $dx$ in the continuum limit. This means that the bracket (\ref{pochahontas}) will at least close off-shell at all {\it finite} orders in $\kappa$ without anomalies.

Thus, we arrive at the conclusion that the 'Hamilton-Hamilton' sector of the quantum constraint algebra closes off-shell without anomalies at all finite orders in perturbation theory. \\

Notice again that the operators on the right hand side of  Eq. (\ref{pochahontas}) are only well defined in terms of commutators with operators on $\ch_{(\xi_n,\mathbf{dQHD}^*_n)}$. In fact, had we instead provides the Hamilton constraint operator in Eq. (\ref{AGW}) with a factor of $dx^3$, so that its vacuum expectation value existed, then the commutator (\ref{pochahontas}) would vanish.\\

The result (\ref{pochahontas}) provides us with a candidate for a diffeomorphism constraint operator, which is $\mathds{D}_n (\bar{N}) $. If we multiply this operator with a $dx^2$ we can check that for $n\rightarrow\infty$ its semi-classical limit in fact coincides with the classical diffeomorphism constraint.

The next step is to compute the commutator $[\mathds{D}_n (\bar{N}) ,\mathds{D}_n (\bar{N}') ]$. This is done in appendix \ref{appendixhamm}. Here we find, however, that the operator $\mathds{D}_n (\bar{N}) $ does not entail commutators that matches the structure of the classical Poisson brackets (\ref{two}) and (\ref{three}). Furthermore, the anomalous terms are of a nature, which suggest that our choice of diffeomorphism constraint operator must be flawed and thus that our initial ansatz for a Hamilton constraint operator is in need of modification. 

To be specific, we find that terms emerge, which are divergent because they involve non-vanishing contributions from vertex operators associated to the same vertex. We also find terms, which correspond to classical terms, that involve a derivative of a triad field. Such terms also appear in the computations of the classical Poisson bracket except that here they fail to be covariant. Had they been covariant they could either be attributed to on-shell closure or to an anomaly involving a torsion operator, but with ordinary derivatives appearing and with divergent terms we believe this shows that our initial ansatz simply cannot be correct. 

The commutator $[\mathds{H}_n (N) ,\mathds{D}_n (\bar{N}) ]$ produces similar anomalous terms and we do not write it down.

It is possible to choose a diffeomorphism constraint operator from (\ref{pochahontas}), which is quadratic in the vector-fields. Classically this amounts to a diffeomorphism constraint, which takes values in the Lie-algebra and thus couples to a lapse field with an $\mathfrak{su}(2)$ index. This choice of diffeomorphism constraint operator, which may be attractive for other reasons that we comment on shortly, does, however, not solve the problems with anomalies. 

Let us consider what goes wrong (see appendix \ref{appendixhamm} for a more detailed discussion). When we computed the 'Hamilton-Hamilton' sector of the constraint algebra we found that certain commutators have the general algebraic structure 
\begin{equation}
A_{v_1} B_{v_2} + A_{v_2} B_{v_1}
\label{Indie}
\end{equation}
with an overall factor given by the lapse fields. This structure, which secures that no misplaced derivatives survives, is absent in the 'diffeomorphisms-diffeomorphisms' sector of the constraint algebra due to the richer index structures given by the lapse fields. We suspect that with an initial ansatz for a Hamilton constraint operator that is more symmetrical one might be able to obtain the algebraic structure (\ref{Indie}) and thereby avoid the problematic derivative terms and also the divergent terms.
Indeed, there exist a number of different operators within our framework, which all have the classical constraints as their semi-classical limit and the operator (\ref{AGW}) is not the most general candidate possible. Let us therefore provide a list of possible modifications to (\ref{AGW}), which we think could improve the situation:
\begin{enumerate}
\item
First of all, the operator $\mathds{H}_n(N)$ only involves three loops based in each vertex, whereas twelve are possible. The curvature operator $\mathds{h}_n$ in (\ref{cur}) in fact involves these twelve loops.
\item
The symmetrization of the vector-fields leading to (\ref{AGW}) does {\it not} respect the algebraic structure given by the operator $\mathds{h}_n$, that dictates that a loop based in $v_i$ has vector-fields associated to adjacent edges, which are not directly connected to $v_i$.
\item
The operator $\mathds{H}_n(N)$ only involves right-invariant vector-fields; none left-invariant.
\end{enumerate}
 A significant computational effort is required to explore the space of possible Hamilton constraint operators. We would like to stress, however, that we see no fundamental reason why such an endeavor could not succeed.\\

Let us end this section with three remarks: first, we suspect that the number of consistent quantum theories of gravity is very limited, probably equal to one. Thus, the operator constraint algebra will only close once we have all components of the theory in place. In this light it is perhaps no surprise that the constraint algebra, which we compute, is anomalous, since there is still a number of issues, which we have not yet dealt with. Rather, we see it as a great encouragement that the commutator between two Hamilton operators in fact has the right off-shell structure.

Second, we believe that the closure of the constraint algebra must be dictated by a powerful principle of symmetry. Our partial derivation of the Hamilton constraint operator from a curvature operator involving the Dirac type operator is an attempt to formulate such a principle. From a more mathematical vantage point we think it would be interesting to use the modular operator from Tomita-Takesaki theory as a candidate for a Hamilton operator. 

In fact, if we choose a diffeomorphism constraint operator, which is quadratic in the vector-fields then the three commutators in the constraint algebra will all have the same general algebraic structure and we would be able to combine the Hamilton constraint and the diffeomorphism constraint operators in a single operator as
\begin{equation}
\bar{N}^a_{v_i} \mathds{D}^a_{v_i}  + N_{v_i}  \mathds{H}_{v_i} := \mbox{Tr}  \left( M_{v_i} \mathds{M}_{v_i}   \right)
\label{FISH}
\end{equation}
where $M_{v_i}= N_{v_i} + \bar{N}^a_{v_i} \sigma^a$ is a self-adjoint two-by-two matrix and where $\mathds{M}_{v_i} $ would be a vertex operator quadratic in the vector fields. The constraint algebra would be of the form
$$
[\mathds{M}(M), \mathds{M}(M')] = \mathds{M}( M \pa M' - M'\pa M)
$$
where $\mathds{M}(M)= \sum_{v_i} \mathds{M}_{v_i}(M_{v_i})$.
It is an operator like (\ref{FISH}) that we speculate might be related to a modular operator coming from Tomita-Takesaki theory. 

Thirdly, our computations on the constraint algebra in the 'diffeomorhism-diffeomorhism' sector indicated that an anomaly proportional to a torsion operator could arise. Despite the fact that these computations showed that our diffeomorphism constraint operator cannot be correct, we think that the general computational structure leading to this potential anomaly could carry over to a computation involving a more realistic diffeomorphism constraint operator.

It may not be a surprise if an anomaly involving torsion shows up in a setting involving Ashtekar variables, since the Ashtekar connection only becomes Levi-Civita via the equations of motion. We need more analysis, however, before the existence of such an anomaly can be confirmed.

\section{Emergence of an almost-commutative algebra}

The algebra $\mathbf{HD}(M)$ can be formulated as the closure of the semi-direct product
\begin{equation}
\mathbf{HD}(M)  = C_c^\infty(M)  \rtimes  \cF  / I\;.
\label{alg1}
\end{equation}
in the norms described in \cite{AGnew}, where $\cf$ is the group generated by flow operators $e^X$, where $I$ is an ideal given by certain reparametrizations of flows (see \cite{AGnew} for details) and where $C^\infty_c (M) $ is the algebra of smooth functions with compact support. The semi-direct product comes with the multiplication relation
$$f_1F_1 f_2 F_2=f_1 F_1 (f_2) F_1 F_2 \;,  $$
where $F_1,F_2\in\cf$.


In \cite{Aastrup:2012vq} we observed that $\mathbf{HD}(M)$ reduces in the semi-classical limit to the algebra
\begin{equation}
 \left(C_c^\infty(M)\otimes M_2(\mathbb{C})\right) \rtimes \mbox{Diff}(M)\;,
\label{almost}
\end{equation}
where $\mbox{Diff}(M)$ is the group of diffeomorphisms on $M$. This is so because the holonomies on a fixed classical geometry generate a two-by-two matrix algebra\footnote{We assume we are considering a semi-classical analysis around a irreducible connection.}.
Thus we find the almost commutative algebra $C_c^\infty(M)\otimes M_2(\mathbb{C}) $ as a sub-algebra. 

This is interesting because the non-commutative geometrical formulation of the standard model of particle physics coupled to general relativity is based on an almost-commutative geometry, where the matrix factor is related to the gauge sector of the standard model. The algebra (\ref{almost}) suggest that the almost-commutative algebra on which the standard model is based might have its origin in a purely quantum gravitational setting.
Of course, there is some way to go between the algebra $C_c^\infty(M)\otimes M_2(\mathbb{C})$ and the almost-commutative algebra of standard model, but one has to bear in mind that there are several issues in the present construction, which we have not yet addressed. Let us therefore provide a list of issues, which we think are relevant here:
\begin{enumerate}
\item
so far we have dealt with a real Ashtekar connection, not the self-dual $SL(2,\mathbb{C})$ connection, that corresponds to a Lorentzian signature. A construction based on the latter connection may give rise to a richer structure than (\ref{almost}) in a semi-classical limit.
\item
the formulation of the Standard Model in terms of non-commutative geometry is in a Lagrangian setting whereas the analysis presented here plays into a Hamiltonian setting. Thus, a straight forward comparison is not possible.
\item
in the non-commutative formulation of the Standard Model, the finite algebra is represented on a corresponding finite-dimensional Hilbert space. In the present case, the finite sector of the algebra, which emerges in the semi-classical limit, acts directly on the spinors. 
\item
several natural components from the toolbox of non-commutative geometry has not been introduced in our construction. For instance the real structure.
\end{enumerate}

In fact, the original motivation for studying non-commutative algebras of holonomies was the hope that such algebras might reduce to an almost-commutative algebra in a semi-classical limit, see \cite{Aastrup:2005yk}. 

\section{Background independency
and action of the diffeomorphism group
}
\label{strong-cont}

In this section we address the question of background and lattice independency. In contrast to the algebra $\mathbf{QHD}(M)$, which is manifestly background independent, the algebra $\mathbf{dQHD}^*(M)$ has been defined via lattice approximations and is therefore not necessarily background independent. The same applies to the Dirac type operator $D$ and the state $\xi_c$. We therefore need to assess to what extend these objects depend on the lattice approximations and the coordinate system, which these represent. First we shall address this question within the lattice formulation and next  commence the construction of a lattice-independent formalism.


\subsection{Invariance properties of $D$ and $\mathbf{dQHD}^*(M)$}
\label{mozart}


First we want to look at the invariance properties of the Dirac type operator. In order to do this we will look at the classical value of the double commutator of the Dirac type operator with a flow, or rather we will only look at what happens for a path.

Thus, let $p$ be a path, and let $p$ be parametrized by $t\to p(t)$ by arc-length in $\| \cdot \|_1$.  
Furthermore let $p_{<x}$ be the path $[0,x] \ni t\to p(t)$ and let $p_{>x}$ be the path $[x,L(p)] \ni t\to p(t)$. 
We put 
$$
A(x)=Hol (A,p_{<x})D\omega (\dot{p}(x))Hol (A,p_{>x})\;,
$$ 
where $D\omega$ is the $M_2$ valued object
\begin{equation}
(D\omega)^\mu =E_a^\mu \sigma^ a  \;,
\label{NASER}
\end{equation}
with $E$ being a classical field.
If we look at the formula (\ref{formeldi}) we see that the classical expectation value of $\{ D,[D,\nabla(p)]\}$ is
\begin{equation}
\int_{0}^{L(p)}  ( \psi(p(L(p))),A(x)\psi (p(0)) dx .  
\label{P1}
\end{equation}
Now, it is important to note that with the present definition of the Dirac type operator formula (\ref{P1}) will depend on the path being parametrized by arc-length in the $1$-metric. This indicates that the Dirac operator is only invariant under coordinate changes which preserves the $1$-metric, which in $3$-dimensions consists of $48$ coordinate changes. 

This is of course not an acceptable state of affairs, since we want the algebra $\mathbf{dQHD}^*(M)$ to be background independent. What is missing from our construction is that the insertion (\ref{NASER}) together with the factor $2^{-n}$ should be a one-form, which would improve the invariance properties of formula (\ref{P1}). One way to obtain this is to ignore the Dirac type operator and simply promote by hand the factor $2^{-n}$ to a one-form $dx^\mu$, which then turns the insertion (\ref{NASER}) into a one-form. Using conditions (\ref{conditions}) we can then in fact secure covariance for the $\mathbf{dQHD}^*(M)$ algebra at least within the GNS construction of $\xi_c$ and to all finite orders in perturbation theory.

If, however, we wish to keep the Dirac type operator we need to change the construction so that the factor $2^{-n}$ in the definition of the Dirac type operator (\ref{Rothstein}) corresponds to a one-form. This change would also have to involve the Clifford algebra and in particular the relation (\ref{Cliff}), which comes into play when one computes the double commutator (\ref{P1}). We see no obstacles for such a modification but shall not work out the details here. 

\subsection{Unitary equivalence}
We now turn to the question of indepence of the Hilbert space construction of the chosen lattice. We will argue that at the level of the holonomy-diffeomorphism algebra the state is independent of the chosen lattice in the sense that the GNS-constructions for the holonomy-diffeomeorphism algebra for different lattices are unitarily equivalent.  

Let us first look at what the semi-classical state looks like on a path $p$. Let $p$ be approximated by a family of paths $\{p_n \}$, $p_n\in \Gamma_n$. From the  formula (\ref{Ash})  we see that the expectation value of $p_n=\{l_1,\ldots. l_k \}$ is approximated by 
$$ (\psi (e(p_n)), Hol(l_1,A)e^{- 2^{-n}B(l_1)}\cdots Hol(l_k,A)e^{- 2^{-n}B(l_k)}\psi (s(p_n)))  ,
\label{HSc}$$
where $e(p_n)$ and $s(p_n)$ denotes the end- and start-point of $p_n$ and where $B(l_k)$ denotes the value of $B_{(n,i)}$ at the endpoint of $l_k$.
Since $B_{(n,i)}$ is stricktly positive we get factors $c,C>0$ satisfying the estimate
\begin{eqnarray*}
&&ce^{-L(p_n)}| (\psi (e(p_n)),Hol (p_n,A)\psi (s(p_n)))|\\
&& \leq |(\psi (e(p_n)), Hol(l_1,A)e^{- 2^{-n}B(l_1)}\cdots Hol(l_k,A)e^{- 2^{-n}B(l_k)}\psi (s(p_n)))| \\ && \leq   Ce^{-L(p_n)}| (\psi (e(p_n)),Hol (p_n,A)\psi (s(p_n)))| .
\end{eqnarray*}
Therefore in the continuum limit we get the estimate 
\begin{eqnarray*}
\lefteqn{ce^{-L(p)}| (\psi (e(p)),Hol (p,A)\psi (s(p)))|} \\
&\leq |\langle \xi_{(A,E)}  ,\nabla(p)   \xi_{(A,E)}\rangle |  \leq   Ce^{-L(p)}| (\psi (e(p)),Hol (p,A)\psi (s(p)))| .
\end{eqnarray*}
Similarly the expectation value on a flow is just integration over the paths in the flow, and hence we get similar estimates in this case as well.

If we have two lattices they induce two different $1$-metrics and define two different coherent states $\xi_{(A,E,1)} $ and $\xi_{(A,E,2)}$. These coherent states give rise to states $\rho_1, \rho_2$ on the flow algebra via
$$ \rho_i (F)=\langle \xi_{(A,E,i)} | F|\xi_{(A,E,i)}  \rangle =\lim_{n\to \infty}  \langle \xi_{(n,A,E,i)} | F_n|\xi_{(n,A,E,i)}  \rangle , \quad i\in \{1,2 \},$$
where $F_n$ is again a lattice approximation to $F$. 
We see from the form of the expectation value, that their absolute value on a path $p$ differ by a factor $$ \exp(D_1L_1(p)-D_2L_2(p))\;,$$ where $L_1$ and $L_2$ denote the $1$-lengths induced by the two lattices, and $D_1,D_2>0$. It hence follows that as long as we only consider flows of vector-fields with compact support then for each element $F$ in the $\mathbf{HD}(M)$ algebra there exist $c_1,c_2>0$ with 
$$ \rho_1(F)\leq c_2\rho_2 (F), \quad  \rho_2(F)\leq c_1\rho_1 (F). $$   
We let $(\ch_1,\langle \cdot , \cdot \rangle_1 )$ and $(\ch_2,\langle \cdot , \cdot \rangle_2 )$ be the Hilbert spaces of the GNS representations induced by the $\mathbf{HD}(M)$ algebra and the states, i.e. $\ch_i$ consists of the closure of the elements in the $\mathbf{HD}(M)$ algebra with the inner product
$$\langle F_1|F_2 \rangle_i= \rho_i (F_1^*F_2) .$$
The bilinear form 
$$ \ch_1\times \ch_1 \ni (F_1,F_2)\to \rho_2(F_1^*F_2)  $$ 
is densely defined and positive, and gives therefore rise to a positive densely defined operator $V$ on $\ch_1$ with $\langle F_1|F_2\rangle_2=\langle F_1|VF_2\rangle_1 $. Furthermore $V$ commutes with the action of the $\mathbf{HD}(M)$ algebra on $\ch_1$ since 
$$ \langle F_1|F_3VF_2 \rangle_1=\langle F_3^*F_1|VF_2 \rangle_1= \langle F_3^*F_1|F_2 \rangle_2 =\langle F_1|F_3F_2 \rangle_2=\langle F_1|VF_3F_2 \rangle_1 .$$
 The square root $U=V^{\frac12}$ fulfills 
$$\langle F_1|F_2\rangle_2=\langle U F_1|U F_2\rangle_1 ,$$
and hence $U$ defines an isometry $U:\ch_2\to \ch_1$ via
$$\ch_2\ni F \to UF\in \ch_1  $$
which commutes with the action of the $\mathbf{HD}(M)$ algebra. 

Like we have constructed $U$ we can likewise construct an isometry $U':\ch_1\to \ch_2$ which commute with the action of the $\mathbf{HD}(M)$ algebra and fulfills $UU'=id_{\ch_1}$, $U'U=id_{\ch_2}$. 

We conclude that the two GNS-representations of the $\mathbf{HD}(M)$ algebra are unitarily equivalent and that the equivalence is implemented by $U$.

The above argument requires a more detailed analysis, especially since $V$ is unbounded. We will not give the analysis here. The argument is taken from \cite{murphybook}, where the case of two states, which are uniformly bounded by each other is treated.

Also further analysis is needed to extend the above argument to the $\mathbf{dQHD}^*(M)$ algebra. Especially one needs to deal with the transformation properties of the Dirac operator, see section \ref{mozart}.


\subsection{Action of the diffeomorphism group}  
\label{actdiff}

Since we have assumed that the $\mathbb{C}^2$-bundle over $M$ is trivial, the group of diffeomorphisms acts naturally on the Holonomy-Diffeomorpism algebra. We will now look at what continuity properties this action has, in particular we will look at the continuity properties of the action on the GNS-construction of a semi-classical state.  

In order to discuss the continuity properties of the action we need a topology on the group of diffeomorphisms. We will consider the weak topology. This is defined in the following manner: Choose a Riemmannian metric on $M$. We consider the metrics
$$d_{K,r}(\Psi,\Phi)= \sup_{m\in K} d(\Psi (m),\Phi (m))+\sum_{n=1}^r \|D^n\Psi(m)-D^n\Phi (m) \|,$$ 
where $ K\subset M $ is  compact.
The weak topology is the topology induced by the metrics 
$$\{ d_{K,r}\}_{r\in \mathbb{N},K\subset M \hbox{ compact} }.$$

The action of the diffeomorphism group on the Holonomy-Diffeomorphism algebra is not continuous, if we consider the Holonomy-Diffeomorphism algebra with the $C^*$-norm defined in section 2.1. This follows since the action of $\bbR$ on $L^2(\bbR)$ is not norm continuous.

On the other hand, we can also consider  $\mathbf{HD}(M,S,\ca )$ with the strong topology induced by the GNS-construction of a semi-classical state. With this topology the action of the diffeomorphism group will be continuous. This follows because the expectation value of a the semi-classical state on a paths depends smoothly on the start and endpoint on the path, see the continuum limit of formula (\ref{HSc}).

However if we consider the algebra $\mathbf{dQHD}^*(M)$ with the weak topology induced by the GNS-construction, the action of the diffeomorpism group will no longer be continuous. This stems from the following fact: Consider two paths $p_1$ and $p_2$ that can be composed. The double commutator with $D$ consists of inserting vector-fields along the paths. If $p_1$ does not overlap with $p_2$, in the product $\{ D,[D,p_1]\} \{ D,[D,p_2]\}  $ there will be no interference between the vector-fields in $\{ D,[D,p_1]\}$ and the holonomy part of $\{ D,[D,p_2]\}$, and vice versa. If however $p_1$ and $p_2$ do overlap, there will be an extra quantum correction stemming from the interference of the vectorfields in $p_1$ and the holonomy part of $\{ D,[D,p_2]\}$, and vice versa. Therefore if we consider a family of diffeomorpisms $\Phi_t$ such the $\Phi_t (p_1)$ can be composed with $p_2$, and such that $\Phi_t (p_1)$ has no overlap with $p_2$ when $t\not=0$, but do have an overlap when $t=0$, we see
\begin{eqnarray*}
&& \lim_{t\to 0} \langle\xi_c | \Phi_t (\{ D,[D,p_1]\})| \{ D,[D,p_2]\}\xi_c \rangle \\
 &&\not= \langle \xi_c | \Phi_0 (\{ D,[D,p_1]\})| \{ D,[D,p_2]\}\xi_c \rangle ,
\end{eqnarray*}
which shows that the action of the diffeomorphisms is not weakly continuous, and therefore also not strongly continuous.

\subsection{A lattice-independent formulation of $D$ and $\mathbf{dQHD}^*(M)$ }

We will in this section try to formulate the $\mathbf{dQHD}^*(M)$ algebra and the Dirac type operator $D$ without the use of lattice approximations.  We will mainly describe what happens on paths, since flows are just families of paths. 

Let $p$ be a path, and let $p:[a,b]\to M$ be a parametrization of the path. 
Consider a coordinate system $(x_1,x_2,x_3)$ and the infinitesimal translation operator $\hat{E}^\a_a(x) = f_x E_{\sigma^a dx_\a}$, see equation (\ref{COOOM}). We will consider the operator $E_{p}$ of the following form: for $t\in [a,b]$ we set 
$$
p_t=\nabla ({p_{<t} }) \sigma^a \dot{p}_\a(t)   \hat{E}^\a_a (p(t))\nabla( {p_{>t} })  
$$
and define
$$E_{p} =\int_{a}^b {p}_t\  dt .$$
One checks that 
$$
{E}_{p} {E}_{q} = {E}_{pq}.  
$$
With $F$ being an element of $\mathbf{HD}(M)$ we then also obtain the operator $E_F$, which is the flow consisting of a family of operators $E_p$ and which then satisfy
$$
E_{F_1}E_{F_2} = E_{F_1 F_2}.
$$
Next we define 
\begin{eqnarray}
{p}^{(2)}_{t}&=& E_{(p_{<t} )} \sigma^{a} \dot{p}_\a(t)     \hat{E}^{\a}_{a} (p(t))  \nabla( {p_{>t}} ) 
  + \nabla(p_{<t} )  \sigma^{a} \dot{p}_\a(t)     \hat{E}^{\a}_{a} (p(t))  E_{(p_{>t})}  
\nn
\end{eqnarray}
as well as
$${E}^{(2)}_{p} =\int_{a}^b {p}^{(2)}_{t}  dt ,$$
which then also gives us $E^{(2)}_F$ for $F\in \mathbf{HD}(M)$.

In a straightforward generalization we also define $E_F^{(n)}$, which involves $n$ insertions of vector-fields in the flow $F$.

In a next step we introduce a grading to $\mathbf{HD}(M)$ and let $\tilde{F}$ be the odd-graded element that corresponds to the even-graded flow $F\in\mathbf{HD}(M)$. Thus, $\tilde{F}$ satisfies the same commutator identities as $F$, just with a graded bracket. Likewise we introduce graded elements $\tilde{E}_F^{(n)}$, which are odd-graded elements that corresponds to ${E}_F^{(n)}$.

We define an abstract Dirac type operator $D$ via its commutator relations with elements in $\mathbf{HD}(M)$ and with elements ${E}_F^{(n)}$ and $\tilde{E}_F^{(n)}$. These are as follows:
$$
\left[ D  ,  E^{(n)}_F  \right] = \tilde{E}_F^{(n)}
\quad,\;
\left[ D  ,  \tilde{E}_F^{(n)}  \right] = E_{F}^{(n+1)}
\nn
$$
 where we define $E^{(0)}_F= F$ and $\tilde{E}_F^{(0)}=\tilde{F}$. The commutators 
 $$[ E^{(n)}_{F_1} ,  E^{(m)}_{F_2} ]\;,\quad [ \tilde{E}^{(n)}_{F_1} ,  E^{(m)}_{F_2} ]\;,\quad [ \tilde{E}^{(n)}_{F_1} ,  \tilde{E}^{(m)}_{F_2} ]$$ are non-trivial and shall not be worked out here. In general we can say that a non-trivial commutator with a vector-field $E^{(n)}_{F} $ will insert a Pauli matrix in the corresponding location in a flow. This will result in two Pauli matrices with contracted indices, located at the same point but inserted in different flows.

 The lattice-independent $\mathbf{dQHD}^*(M)$ algebra is then the $\star$-algebra generated by $\mathbf{HD}(M)$ and all commutators between $D$ and elements in $\mathbf{HD}(M)$.

Note that we have not provided a proof that the abstract $\mathbf{dQHD}^*(M)$ algebra defined here is in fact identical to the $\mathbf{dQHD}^*(M)$ algebra defined previously with the aid of lattice approximations.

\section{The overlap function}
\label{overlap}


In section \ref{LarsL} we argued that the states $\xi_c$ on $\mathbf{dQHD}^*(M)$ will always be associated to a phase-space point of Ashtekar variables and we explicitely constructed states, which are peaked over such a phase-space point. The question therefore arises what the overlap function between two states associated to different phase-space points amounts to. This is the subject of this section. 

To ease the analysis we shall restrict ourselves to the case where the state $\xi_c$ is constructed from coherent states on $SU(2)$, $\xi_c=\xi^{\kappa}_{(A,E)}$, as we did in section (\ref{LarsL}), and we shall consider two phase-space points, which only differ in the Ashtekar connection.
Thus, the overlap function we are interested in is this
$$
\OO^{\kappa}(A,A',E)  = \left\langle \xi^{\kappa}_{(A,E)} \vert \xi^{\kappa }_{(A',E)}  \right\rangle \;.
$$
We first define the overlap function for a single edge $l_i$ 
$$
\oo^{\kappa}_i(A,A',E) = \left\langle \phi^s_{(A,dxE,i)} \vert \phi^s_{(A,dxE,i)}  \right\rangle \;,
$$
where we recall that $s= \kappa dx$ and that $\phi^s_{(A,dxE,i)} $ is Hall's coherent state on one copy of $SU(2)$.
To find $\OO(A,A',E) $ one must work out the continuum limit
$$
\OO^\kappa(A,A',E)  = \lim_{n\rightarrow \infty} \P_{i=1}^{\vert {\bf l}_n \vert} \oo^\kappa_i(A,A',E)
$$
where the index $i$ runs over all edges in $\G_n$. 
We write
\begin{equation}
\OO^{\kappa}(A,A',E)  =  \lim_{n\rightarrow \infty} \exp\left( - \sum_{i=1}^{\vert {\bf l}_n \vert}  \ln \left(\oo^\kappa_i(A,A',E)^{-1} \right)    \right) 
\label{frank}
\end{equation}
and since $\oo^\kappa_i(A,A',E)\leq 1$ we know that $\ln \left(\oo^\kappa_i(A,A',E)^{-1} \right) $ is positive.
Next, since the sum $\sum_{i=1}^{\vert {\bf l}_n \vert}$ in (\ref{frank}) becomes an integral over the manifold $M$ in the large $n$ limit
$$
\sum_{i=1}^{\vert \bf{l}_n \vert} \stackrel{n}{\longrightarrow} \int_M
$$
it is important to count the powers of $dx=2^{-n}$ in order to check whether the measure that arises, matches the dimensions of the integral. The manifold $M$ is 3-dimensional and therefore we need a factor $2^{-3n}$ in order to obtain a volume form. In this way we can estimate whether and to what the expression in (\ref{frank}) converges by simple power-counting. 

The infinitesimals come from two sources: from the parameter $s$ in the coherent state $\phi^s_{i(A,E)}$ and from the connection $A$. Further, if we assume that we have coherent states based on a Laplace operator complexifier, then the relevant parameter is the difference $z= (A-A')$. 
Since
$$
\lim_{s\rightarrow 0}\OO^\kappa(A,A',E) \propto \d(A,A') \;,
$$ 
we assume that at least one term in $\ln \left(\oo^s_i(A,A',E) \right) $ comes with a factor $s^{-1}$.
We do not have a general proof for the validity of this assumption, but in the special case with a Laplace operator complexifier we know it holds true (see \cite{Thiemann:2000ca}).

Furthermore, since $\ln \left(\oo^s_i(A,A',E)^{-1} \right) $ is positive its Taylor expansion in $(A-A')$ can only involve even powers and since $\OO^s(A,A,E)=1$ its lowest possible power is quadratic.
$$
\ln \left(\oo^s_i(A,A',E)^{-1} \right)  = a_1 z^2 + a_2 z^4 + \ldots \;. 
$$
On the other hand, for the sum in (\ref{frank}) to converge we need {\it four} powers of $dx$ to arise from $(A-A')$ in $\oo^s_i(A,A',E) $ since $s$ may provide one negative power due to assumption made above. Thus, if there are terms in the Taylor expansion of $ \oo^s_i(A,A',E) $, which provides fewer powers than four, then the sum will diverge and the overlap function will vanish. Also, if there are only terms strictly higher than four, then the sum will vanish and the overlap function will equal $1$. 

Thus, in order for the overlap function in (\ref{frank}) to obtain a nonzero value less than one we must require that the lowest power of $(A-A')$ in the Taylor expansion of $ \ln (\oo^s_i(A,A',E)^{-1}) $  will be exactly four and that this term also involves a $1/s$ factor. In particular, this means that the quadratic term is required to vanish.

This seems questionable. One can check that this condition is not satisfied for the special case where the complexifier is the Laplace operator on $SU(2)$ (use formula 4.55 in \cite{Thiemann:2000ca}), and we are inclined to think that it will not be satisfied for any choice of complexifier coherent state, which also satisfies standard requirements for semi-classical states (minimizing the uncertainly conditions, for instance). We do, however, not have a general proof. Whether it holds for any state on $\mathbf{HD}(M)$ is also unknown.

Thus, for the special class of semi-classical states with a Laplace-operator complexifier, we conclude that the overlap function between two different semi-classical approximations vanishes
$$
\OO^\kappa(A,A',E) = 0\;.
$$

The overlap function is important since it is the expectation value of $U_\oo$
$$
 \left\langle \xi^\kappa_{(E,A)} \vert \xi^\kappa_{(E,A')}  \right\rangle  =  \left\langle \xi^\kappa_{(E,A)} \vert U_{\oo} \vert\xi^\kappa_{(E,A)}  \right\rangle\;, \quad A' = A + \oo\;.
$$
and therefore tells us about the possibilities of shifting between classical geometries. If our reasoning presented here holds in general, then it means that each classical geometry gives rise to a vacuum state and a GNS construction over it, which are all isolated from each other. 
This does, however, not necessarily imply that there cannot be finite changes in geometries caused by quantum effects, as our analysis in subsection \ref{ISIL} shows. But it implies that there cannot be quantum interference between semi-classical approximations associated to different classical geometries. We find this to be an interesting possibility.

\section{The complex Ashtekar connection}
\label{REAL}

So far we have based our construction on the group $SU(2)$. We choose $SU(2)$ because it plays into the framework of canonical quantum gravity formulated in terms of Ashtekar variables. The original Ashtekar connection, however, is a complex connection, that takes values in the self-dual sector of $\mathfrak{sl}(2,\mathbb{C})$, the Lie-algebra of $SL(2,\mathbb{C})$. The $SU(2)$ connection corresponds to a metric with Euclidian signature. Since $SU(2)$ is a compact group it is simpler to start with this type of connection. In order to encompass also Lorentzian signatures it is, however, necessary to consider how a complex connection can be built into or emerge from our construction. 

The Lie group $SL(2,\mathbb{C})$ is generated by the six generators $\{\frac{1}{\sqrt{2}}\sigma_i,\frac{\mathrm{i}}{\sqrt{2}}\sigma_i\}$, $i\in\{1,2,3\}$, and the self-dual sector is characterized by invariance under the exchange of generators:
$$
\frac{1}{\sqrt{2}}\sigma_i \longleftrightarrow \frac{\mathrm{i}}{\sqrt{2}}\sigma_i
$$
which corresponds to interchanging rotations with Lorentz boosts. Thus, in terms of degrees of freedom, the self-dual sector of $SL(2,\mathbb{C})$ matches those of $SU(2)$. In terms of algebras generated by holonomies, however, the two cases are distinctly different.

There are two natural strategies to encompass the complex Ashtekar connection. Either we repeat the entire construction of the algebra generated by holonomy-diffeomorphisms and semi-classical states thereon with $SU(2)$ replaced by $SL(2,\mathbb{C})$, or, alternatively, we try to obtain a complexification of $SU(2)$ by doubling the Hilbert space.

Let us briefly discuss the first option.
The key difference between holonomy-diffeomorphisms of $SU(2)$ and $SL(2,\mathbb{C})$ connections is that the latter does not correspond to bounded operators. The norm given in section \ref{beent} diverges for $\mathfrak{sl}(2,\mathbb{C})$ connections
$$
 \sup_{\nabla\in\ca_{\tiny \mathfrak{sl}}} \| e^X_{\nabla}\| =\infty\;,\quad \ca_{\tiny \mathfrak{sl}} = \mbox{space of $\mathfrak{sl}(2,\mathbb{C})$ connections. }
$$
which means that we do not have a $C^*$-algebra. It is therefore not possible to define $\mathbf{HD}(M,S,\ca_{\tiny \mathfrak{sl}})$. There is, however, still a $*$-algebra and thus $\ch\cd(M,S,\ca_{\tiny \mathfrak{sl}})$ is available for a Hilbert space representation.

The conditions for a state to exist on $\ch\cd(M,S,\ca_{\tiny \mathfrak{sl}})$ are identical to the $SU(2)$ case and we strongly expect states on $\ch\cd(M,S,\ca_{\tiny \mathfrak{sl}})$ as well as on $d\cq\ch\cd(M,S,\ca_{\tiny \mathfrak{sl}})$, which is the $*$-algebra version of $\mathbf{dQHD}(M,S,\ca_{\tiny \mathfrak{sl}})$, to exist. A GNS construction around such a state would then involve only unbounded operators.

Once a Hilbert space representation of $d\cq\ch\cd(M,S,\ca_{\tiny \mathfrak{sl}})$ is obtained we can proceed to construct the Hamilton operator in the same manner as we did for $SU(2)$ and project into the self-dual sector. This will give the complex Hamiltonian that corresponds to the Lorentzian signature.

It is beyond the scope of this paper to work out the details for a construction based on $SL(2,\mathbb{C})$, but we would like to stress that we do not see any major obstacles for such a construction to exist.

\section{Interpreting $D$ in terms of the volume of $M$}

\begin{figure}[t]
\begin{center}
\resizebox{!}{4.2cm}{
 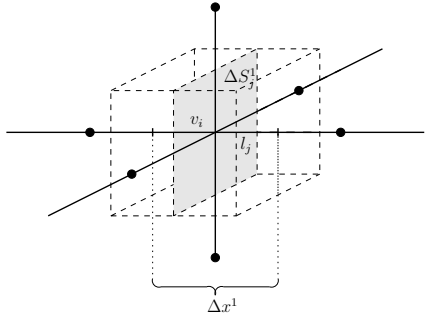}
\end{center}
\caption{The surfaces $\Delta S^1_j$, which are orthogonal to the line interval $\Delta x^1$ corresponding to a subdivision of $M$ in cubic lattices, are associated to flux variables located at the edges $l_j$.}
\label{leninnnn}
\end{figure}

The Dirac type operator $D$ has an interesting interpretation in terms of a quantization of the volume of the manifold $M$. To see this we first write the volume of $M$
$$
\mbox{Vol}(M) = \int_M e \e_{\m\n\r} dx^\m dx^\n dx^\r
$$
where $e=\mbox{det}(e_\m^a)$, and rewrite it in terms of a triad field
\begin{eqnarray}
\mbox{Vol}(M) &=& \int_M dx^a   e e^\m_a   \e_{\m\n\r}  dx^\n dx^\r
\nn\\
&=&  \int_M dx^a   dF_a
\label{windyt}
\end{eqnarray}
where $dF_a= e e^\m_a   \e_{\m\n\r}  dx^\n dx^\r$ is a sum over the three infinitesimal flux variable in the $x^\m$'s directions, which is conjugate to the holonomy of the Ashtekar connection, see \cite{AL1}.
If we rewrite the Riemann integral in (\ref{windyt}) as a limit of lattices we then we find
\begin{equation}
\mbox{Vol}(M) = \lim_{n \rightarrow \infty} \sum_{i,a} (\Delta x^a)_i F_{\Delta S^a_i}
\nn
\end{equation}
where the sum runs over two indices: the index $i$ over all edges\footnote{In fact, this sum could equally well run over vertices. The point is that the '$dx^a$' does not depend on the three directions in the lattice and thus only 'sees' the vertices.} in the lattice approximation and the "flat" index $a\in\{1,2,3\}$. Here $F_{\Delta S_i}$ is a lattice approximation of one of the three $x^\m$-components in $dF^a$ where the surface $\Delta S_i$ is perpendicular to the direction of $x^\m$. Thus, here the sum runs over all surfaces $S_i$ in all three directions. In the quantization procedure, which involves holonomies of Ashtekar connections, this flux variable corresponds to a right-invariant vector-field $\cl_{{\bf e}_i^a}$ on a copy of $SU(2)$ associated to the edge $l_i$, see figure \ref{leninnnn}, which gives us:
\begin{equation}
\mbox{Vol}(M)\stackrel{\mbox{\tiny quant.}}{\longrightarrow} \lim_{n \rightarrow \infty} \sum_{i,a} (\Delta x^a)_i \cl_{{\bf e}_i^a}\;,
\label{2070}
\end{equation}
There is a striking similarity between (\ref{2070}) and the Dirac operator $D_n$ in (\ref{Rothstein}), where the Clifford elements play the role of the infinitesimal element $dx^a$ or its lattice approximation $\Delta x^a$. Note too that this is in sync with the our conclusions in section \ref{mozart}, where we found that the Clifford algebra must be related to the exterior algebra of $M$.


\section{Summary and discussion}
\label{DIS}

In this paper we have presented quantum holonomy theory, which is a non-perturbative and background independent theory of quantum gravity coupled to quantized degrees of matter.

Four central objectives are met in this paper. The first is the formulation of a {\it first principle} -- namely the $\mathbf{QHD}(M) $ algebra --, which serves as the foundation for this approach to a theory of quantum gravity.
 It is the conceptual simplicity of the $\mathbf{QHD}(M) $ algebra, which makes it attractive. What could be more natural, more {\it poetic}, as a foundation for a theory of quantum gravity than an algebra that simply encodes how tensorial degrees of freedom -- i.e. {\it stuff} -- are moved in space? We find it surprising that this algebra, which encodes the mathematical setup of canonical quantum gravity, has not, to the best of our knowledge, been studied before.   


The second central objective met in this paper is the finding that semi-classical states exist on the algebra $\mathbf{dQHD}^*(M) $, which is the algebra obtained from $\mathbf{QHD}(M) $ by forming a canonical Dirac type operator and considering its commutators with the algebra of holonomy-diffeomorphisms. 
%
A state gives us a kinematical Hilbert space and a stage for a semi-classical analysis. It is remarkable that where other non-perturbative approaches to quantum gravity are challenged by the necessity of producing semi-classical states, we find that semi-classical states appear extremely natural in our approach. Also and perhaps no less surprising is the evidence we find that the overlap function between two different semi-classical states might vanish. 

The third central objective met in this paper is the formulation of a geometrical principle, which provides us with a Hamilton constraint operator from which the classical Hamilton constraint emerges in a semi-classical limit. This means that we do obtain general relativity in a semi-classical limit from our construction.

The key step to obtain the Hamilton constraint operator is again the construction of the Dirac type operator. Since this operator provides us with a canonical metric structure on the spectrum of the $\mathbf{HD}(M) $ algebra it gives us access to geometrical notions such as curvature. Indeed, it is a scalar curvature operator that provides us with a candidate for a Hamilton constraint operator. The construction of this scalar curvature operator does, however, involve a certain measure of ad-hoc reasoning, which is likely to be the cause of the failure of the operator constraint algebra to close. Nevertheless, the fact that our provisional candidate for a Hamilton constraint operator entails off-shell closure in the 'Hamilton-Hamilton' sector of the operator constraint algebra is an encouraging result, which should motivate further analysis in this direction.

The fourth central objective met in this paper is the identification of elements and mechanisms of unification. This point is essentially an adaptation of results already published. 
First of all, the  $\mathbf{HD}(M) $ algebra is inherently non-commutative, which immediately places our construction well within the domain of non-commutative geometry with its toolbox of unifying mechanisms. 
The fact that the  $\mathbf{HD}(M) $ algebra produces an almost-commutative algebra in the semi-classical limit 
suggest a link to the basic mathematical setup behind the non-commutative formulation of the standard model. 
Secondly, we find states in the kinematical Hilbert space from which the expectation value of the Dirac type operator gives a spatial Dirac operator in a semi-classical limit and from which we are also able to obtain the Dirac Hamiltonian in the same limit. 
This provides a link to fermionic quantum field theory.\\

The results presented in this paper raises a number of both conceptual and technical questions. If we start in the visionary end of the spectrum, then one may speculate whether there exist a link between the constraint operators and Tomita-Takesaki theory. For the operator constraint algebra to close off-shell in a non-trivial sector of the Hilbert space there must be a powerful symmetry principle that dictates such closure. We suspect this to be Tomita-Takesaki theory, which under certain conditions prescribe the existence of a one-parameter group of automorphisms, that is unique up to inner automorphisms. Indeed, with a state on the $\mathbf{dQHD}^*(M) $ algebra we are in a position where we can attempt to derive the one-parameter group and the modular operator. We speculate that the latter is related to the constraint operators and that the eventual off-shell closure of the operator constraint algebra will turn out to be a consequence of this theory. 

Note again that because the action of the diffeomorphism group on the $\mathbf{dQHD}^*(M) $ algebra is not strongly continuous and thus will not have infinitesimal generators one cannot expect the constraint algebra to close on the entire kinematical Hilbert space but merely in a sector defined by finite perturbation theory in the quantization parameter. This observation appears to be generic and may reflect a deeper conceptual issue. Indeed, since the diffeomorphisms themselves are quantized -- the holonomy-diffeomorphisms are part of the conjugate variables --, one may question to what extent it makes sense to talk about diffeomorphisms and diffeomorphism invariance outside the perturbative regime.

Concerning the constraint operators it is clearly possible to push the analysis commenced in this paper much further, i.e. by brute force calculations of the constraint algebra, to determine which candidates for a Hamilton constraint operator are physically realistic. As a first step it would be interesting (and rather demanding) to compute the operator constraint algebra with the new candidate for a Hamilton operator, that we obtain in the appendix and which involves twelve loops in each vertex operator. 
 It would also be favorable to tighten the geometrical argument, that we use to derive our candidate for a Hamilton constraint operator from a curvature operator on the configuration space of Ashtekar connections. To do this one would have to consolidate the non-commutative geometrical notions, which we have already built, i.e. move further towards a spectral geometry over a space of spatial connections.


It is an open question what support the measure on the configuration space of Ashtekar connections, which the semi-classical state provides, has. Our analysis on the overlap function suggest that the measure could be localized over a single classical point and that this point is the only measurable one in its support. If this result holds in general it would render a theory of quantum gravity, 
which has a measure markedly different from the Ashtekar-Lewandowski measure used in loop quantum gravity. There may, however, be finite geometrical changes caused by transitions generated by the $\mathbf{dQHD}^*(M) $ algebra, as our analysis of how the algebra affects the spectrum of the states suggest, but there will not be quantum interference between vacuum states associated to different classical geometries. On the other hand it is also possible that states exist, which render a non-zero overlap function. If this should be the case one would have to work out whether both types of states are physically feasible.

The fact that we use lattice approximations for much of our analysis raises the question wether or not the construction is lattice and background independent. The algebra $\mathbf{QHD}(M) $ is manifestly background independent, but the bulk of our analysis is based on a version of $\mathbf{dQHD}^*(M) $, which is constructed using lattice approximations. We have a formulation of $\mathbf{dQHD}^*(M) $ and a Dirac type operator, which is independent of the lattices, but we have not provided a proof that these objects are in fact identical to the ones on which our analysis is based. Also, the construction of the state has so far only been carried out using lattice approximations. It seems, however, very plausible that also the state -- and in fact the entire construction presented in this paper -- can be constructed independently of lattice approximations. To do this is a primary task.

So, can this approach to quantum gravity be said to be background independent? The fact that the kinematical Hilbert space is provided by a GNS construction over a semi-classical state of course implies that the Hilbert space always depends on a classical "background" metric. But this is not the background dependency, that is usually meant by the term. The background dependency encountered here comes naturally out of a manifestly background independent construction  -- the $\mathbf{QHD}(M) $ algebra -- as a consequence of representation theory. 

Concerning the lattice approximations, it is in fact not clear 
 whether they approximate diffeomorphisms or merely analytic diffeomorphisms. Note that if the latter should be the case this does not necessarily imply that this framework only harbors analytic diffeomorphisms. It would only imply that the lattices approximations do. Also, since exponentiated vector-fields have a non-measurable effect on the spectrum of the holonomy-diffeomorphism algebra, it is not clear 
 with respect to what norm the $C^*$-algebra is built. The $\mathbf{dQHD}^*(M) $ algebra does not directly involve exponentiated vector-fields, but such non-measurable effects could still occur and could indicate that the holonomy-diffeomorphism algebra should be build over the counting measure instead of a measure coming from a Riemannian metric. On the other hand  the algebra should tie the various non-measurable effects together in a measurable way, so that we in the end should end up with a Riemannian metric.

Another question is to determine the nature of the yet tentative connections to fermionic quantum field theory and the non-commutative formulation of the standard model. Overall, we believe that this construction begs for a deeper application of non-commutative geometry.  For instance, is the emergence of elements of fermionic QFT related to the algebra of differential forms generated by the Dirac type operator? And what is the spectral action? One serious obstacle for a direct comparison to the non-commutative formulation of the standard model is that we are operating with a Hamiltonian setup whereas the former is in the Lagrangian formalism.


Let us end by commenting on the issue of the complex Ashtekar connection. Within the framework presented in this paper it is entirely plausible that a formulation set with $SL(2,\mathbb{C})$ exist. The way the complex Ashtekar connection takes values in the self-dual sector of $\mathfrak{sl}(2,\mathbb{C})$ seems, however, to suggest that it should be obtained in a more intrinsic manner, probably by doubling the Hilbert space and using methods of non-commutative geometry. It would then be natural that the real structure should be used to define a reality condition. 

\vspace{1cm}
\noindent{\bf Acknowledgements}\\
We would like to express our thanks to Mario Paschke for helpful comments and suggestions.\\

\begin{appendix}

\section{The operator constraint algebra}

\subsection{The commutator $[\mathds{H}(N),\mathds{H}(N')]$}
\label{appendixham}

To compute the commutator 
\begin{equation}
\left[\mathds{H}_n(N), \mathds{H}_n(N')\right]
\label{maaan}
\end{equation}
we first compute all possible commutators between vertex operators $\mathds{H}^{\bullet\bullet}_{v_i}$, which will contribute with a factor $N(x)\pa_x N'(x)$ or $N'(x)\pa_x N(x)$ in the continuum limit. Once we have this we can add up the three spatial directions.

\begin{figure}[t]
\begin{center}
\resizebox{!}{6cm}{
 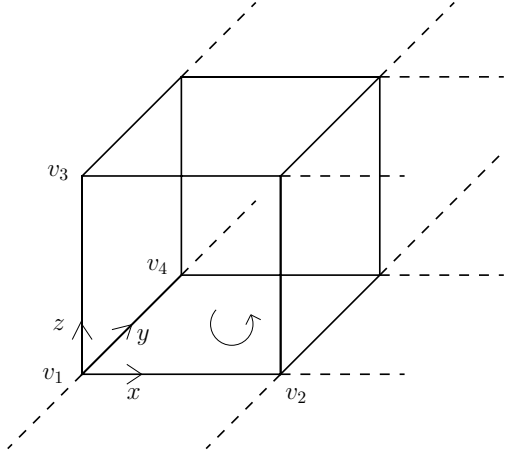}
\end{center}
\caption{The location of the vertices $\{v_1,v_2,v_3,v_4\}$ in a lattice plaquette. The circular arrow indicates the orientation of the "$xy$" loop.}
\label{leningrad}
\end{figure}

Thus, we start by computing the commutator between the two '$(xy)$'-operators at two adjacent vertices $v_1$ and $v_2$, see figure \ref{leningrad}
\begin{eqnarray}
\left[\mathds{H}^{(xy)}_{v_1}(N_{v_1}),\mathds{H}^{(xy)}_{v_2}(N'_{v_2})\right] &&\nn\\
&&\hspace{-5cm} = 2^{4n-4}N_{v_1}N'_{v_2} 
\left[
\left\{ \cl^{v_1}_{{\bf e}^b_y} , \left\{ \cl^{v_1}_{{\bf e}^a_x}  , \mbox{Tr} \left( \sigma^a  \stackrel{\hspace{-3mm}\curvearrowleftright}{L^{v_1}_{xy}}       \sigma^b   \right) \right\} \right\}
,
\left\{ \cl^{v_2}_{{\bf e}^d_y} , \left\{ \cl^{v_2}_{{\bf e}^c_x}  , \mbox{Tr} \left( \sigma^c  \stackrel{\hspace{-3mm}\curvearrowleftright}{L^{v_2}_{xy}}       \sigma^d   \right) \right\} \right\}
\right]
\nn\\
&&\hspace{-5cm} = 2^{4n-4}N_{v_1}N'_{v_2} 
\left\{ \left\{ \cl^{v_1}_{{\bf e}^b_y} , \left\{ \cl^{v_1}_{{\bf e}^a_x}  , \left[ \mbox{Tr} \left( \sigma^a  \stackrel{\hspace{-3mm}\curvearrowleftright}{L^{v_1}_{xy}}       \sigma^b   \right), \cl^{v_2}_{{\bf e}^d_y} \right] \right\} \right\}
 , \left\{ \cl^{v_2}_{{\bf e}^c_x}  , \mbox{Tr} \left( \sigma^c  \stackrel{\hspace{-3mm}\curvearrowleftright}{L^{v_2}_{xy}}       \sigma^d   \right) \right\} \right\}
\nn\\
&&\hspace{-5cm} + 2^{4n-4}N_{v_1}N'_{v_2} 
\left\{ \cl^{v_2}_{{\bf e}^d_y} , \left\{ \cl^{v_2}_{{\bf e}^c_x}  , \left\{ \left[ \cl^{v_1}_{{\bf e}^b_y} ,\mbox{Tr} \left( \sigma^c  \stackrel{\hspace{-3mm}\curvearrowleftright}{L^{v_2}_{xy}}       \sigma^d   \right) \right]   , \left\{ \cl^{v_1}_{{\bf e}^a_x}  , \mbox{Tr} \left( \sigma^a  \stackrel{\hspace{-3mm}\curvearrowleftright}{L^{v_1}_{xy}}       \sigma^b   \right) \right\} \right\}
\right\}\right\} \;.
\nn
%
%
%
\end{eqnarray}
In a next step we evaluate the commutator between loops and vector-fields and use (\ref{conditions}) to obtain 
\begin{equation}
 \left[  \cl^{v_i}_{{\bf e}^b_x} ,  \mbox{Tr} (  \sigma^a    \stackrel{\hspace{-3mm}\curvearrowleftright}{L^{v_{i+1}}_{xy}}       \sigma^c   )     \right]  
 = \pm \frac{1}{2} \mbox{Tr} ( \sigma^a \sigma^b \sigma^c ) + \co\left(dx^2\right) 
 \;,
\label{dannelse}
\end{equation}
where $ \cl^{v_i}_{{\bf e}^b_x} $ may be interchanged with $ \cl^{v_i}_{{\bf e}^b_y} $ and
where the sign depends on which part of the loop the vector-field interacts with. If the vector-field interacts with either $g_1$ or $g_2$ in the loop, then the sign is plus; else minus. In obtaining relation (\ref{dannelse}) we are explicitly using that we are operating within a GNS construction of the semi-classical state $\xi_n$ and its limit.  Since we are only interested in the continuum limit we may ignore all terms except those at lowest orders of $dx$. 

We continue the computation and find:
\begin{eqnarray}
\left[\mathds{H}^{(xy)}_{v_1}(N_{v_1}),\mathds{H}^{(xy)}_{v_2}(N'_{v_2})\right] &&\nn\\
&&\hspace{-5cm} =- 2^{4n-4}N_{v_1}N'_{v_2} 
\left(
\left\{ \cl^{v_2}_{{\bf e}^b_x}  , \left\{  \cl^{v_1}_{{\bf e}^b_y} , \left\{ \cl^{v_1}_{{\bf e}^a_x}   
 ,  \mbox{Tr} \left( \sigma^a  \stackrel{\hspace{-3mm}\curvearrowleftright}{L^{v_2}_{xy}}        \right) \right\}\right\} \right\}
\right.\nn\\
&&\hspace{-2.4cm}\left.
-
\left\{ \cl^{v_1}_{{\bf e}^a_x}   
 , \left\{ \cl^{v_2}_{{\bf e}^a_x}  ,  \left\{  \cl^{v_1}_{{\bf e}^b_y} , \mbox{Tr} \left( \sigma^b  \stackrel{\hspace{-3mm}\curvearrowleftright}{L^{v_2}_{xy}}        \right) \right\}\right\} \right\}
\right.
\nn\\
&&\hspace{-2.4cm} 
\left.
+\left\{ \cl^{v_2}_{{\bf e}^a_y} ,  \left\{ \cl^{v_1}_{{\bf e}^a_x}  ,\left\{ \cl^{v_2}_{{\bf e}^c_x}  , \mbox{Tr} \left( \sigma^c  \stackrel{\hspace{-3mm}\curvearrowleftright}{L^{v_1}_{xy}}         \right)  \right\}
\right\}\right\} 
\right.\nn\\
&&\hspace{-2.4cm}\left.
-
 \left\{ \cl^{v_2}_{{\bf e}^a_x}  , \left\{ \cl^{v_1}_{{\bf e}^a_x}  ,\left\{ \cl^{v_2}_{{\bf e}^d_y} , \mbox{Tr} \left( \sigma^d  \stackrel{\hspace{-3mm}\curvearrowleftright}{L^{v_1}_{xy}}         \right)  \right\}
\right\}\right\} 
\right)\;.
\label{koleskab}
\end{eqnarray}
Here the operator in the bracket has the general structure
$$
A_{v_1} B_{v_2} + A_{v_2} B_{v_1}
$$
which we rewrite as
$$
A_{v_1}B_{v_1} + A_{v_2}B_{v_2} + (A_{v_2}- A_{v_1}) (B_{v_1}-B_{v_2})\;.
$$
We use this, together with definition (\ref{faces}), to rewrite (\ref{koleskab}) as
\begin{eqnarray}
\left[\mathds{H}^{(xy)}_{v_1}(N_{v_1}),\mathds{H}^{(xy)}_{v_2}(N'_{v_2})\right] &&\nn\\
&&\hspace{-5cm} =- 2^{4n-3}N_{v_1}N'_{v_2} 
\left(
\left\{ \cl^{v_2}_{{\bf e}^b_x}  , \left\{  \cl^{v_2}_{{\bf e}^b_y} ,  \mathds{D}^{v_2(xy) }_{y}    \right\} \right\}
-
\left\{ \cl^{v_2}_{{\bf e}^a_x}   
 , \left\{ \cl^{v_2}_{{\bf e}^a_x}  ,  \mathds{D}^{v_2(xy)}_{x}   \right\} \right\}
\right.
\nn\\
&&\hspace{-2.5cm}
\left.
+\left\{ \cl^{v_1}_{{\bf e}^a_y} ,  \left\{ \cl^{v_1}_{{\bf e}^a_x}  ,  \mathds{D}^{v_1(xy) }_{y} 
\right\}\right\} 
-
 \left\{ \cl^{v_1}_{{\bf e}^a_x}  , \left\{ \cl^{v_1}_{{\bf e}^a_x}  ,\mathds{D}^{v_1(xy) }_{x} 
\right\}\right\} 
\right)
\nn\\
&&\hspace{-2.5cm}
+ \Xi^{(xy,xy)}_{v_1v_2}(N,N')
\label{likudo}
\end{eqnarray}
where $\mathds{D}^{v_i(\m\n) }_{\m}$ are defined in (\ref{faces}) and where $\Xi^{(xy,xy)}_{v_1v_2}(N,N')$ is an operator that consist of terms of the form
\begin{equation}
2^{4n-3} N_{v_1}N'_{v_2} (A_{v_2}- A_{v_1}) (B_{v_1}-B_{v_2})\;.
\label{likud}
\end{equation}
We see that the general structure of the classical Poisson bracket (\ref{one}) emerges in equation (\ref{likudo}) with $\Xi^{(xy,xy)}_{v_1v_2}(N,N')$ as a possible candidate for an anomaly. 
Note also that there is an additional factor $\tfrac{1}{2}$ in equation (\ref{likudo}) due to the fact that contributions at both vertices $v_1$ and $v_2$ show up. When we eventually add all contributions to the commutator (\ref{maaan}) this factor will be needed.

We obtain the opposite commutator
$$
\left[\mathds{H}^{(xy)}_{v_2}(N_{v_2}),\mathds{H}^{(xy)}_{v_1}(N'_{v_1})\right] 
$$
simply by interchanging $N$ and $N'$ in equation (\ref{likudo}) and adding a sign. Adding these two commutators will produce the factor $\left(N_{v_1}N'_{v_2} -N'_{v_1}N_{v_2}   \right)$, which in the continuum limit converges to $dx(N(x)\pa_x N'(x) - N'(x)\pa_x N(x)) $. Thus, in the following computations we shall not explicitly write down the opposite commutators but simply write $dx(N(x)\pa_x N'(x) - N'(x)\pa_x N(x) )$ in the end.

Next we work out the commutator between two vertex operators involving perpendicular loops. We first compute
\begin{eqnarray}
\left[\mathds{H}^{(xy)}_{v_1}(N_{v_1}),\mathds{H}^{(yz)}_{v_2}(N'_{v_2})\right] &&\nn\\
&&\hspace{-4,5cm} = -2^{4n-4}N_{v_1}N'_{v_2} \e^{acb}
 \left\{ \cl^{v_2}_{{\bf e}^d_z} ,  \left\{  \left\{  \cl^{v_1}_{{\bf e}^b_y} , \cl^{v_1}_{{\bf e}^a_x}     \right\}
 , \mbox{Tr} \left( \sigma^c  \stackrel{\hspace{-3mm}\curvearrowleftright}{L^{v_2}_{yz}}       \sigma^d   \right) \right\} \right\}
\nn\\
&&\hspace{-4,5cm} + 2^{4n-4}N_{v_1}N'_{v_2} \e^{cbd}
 \left\{ \cl^{v_2}_{{\bf e}^d_z} , \left\{ \cl^{v_2}_{{\bf e}^c_y}  ,  \left\{ \cl^{v_1}_{{\bf e}^a_x}  , \mbox{Tr} \left( \sigma^a  \stackrel{\hspace{-3mm}\curvearrowleftright}{L^{v_1}_{xy}}       \sigma^b   \right)  \right\}
\right\} \right\}
\label{edderkop!}
\end{eqnarray}
as well as
\begin{eqnarray}
\left[\mathds{H}^{(zx)}_{v_1}(N_{v_1}),\mathds{H}^{(yz)}_{v_2}(N'_{v_2})\right] &&\nn\\
&&\hspace{-4,5cm} = 2^{4n-4}N_{v_1}N'_{v_2} \e^{adb}
 \left\{ \left\{ \cl^{v_1}_{{\bf e}^b_x} ,  \cl^{v_1}_{{\bf e}^a_z}      \right\}
, \left\{ \cl^{v_2}_{{\bf e}^c_y}  , \mbox{Tr} \left( \sigma^c  \stackrel{\hspace{-3mm}\curvearrowleftright}{L^{v_2}_{yz}}       \sigma^d   \right) \right\} \right\}
\nn\\
&&\hspace{-4,5cm} - 2^{4n-4}N_{v_1}N'_{v_2} \e^{cad}
 \left\{ \cl^{v_2}_{{\bf e}^d_z} , \left\{ \cl^{v_2}_{{\bf e}^c_y}  , \left\{ \cl^{v_1}_{{\bf e}^b_x} ,   \mbox{Tr} \left( \sigma^a  \stackrel{\hspace{-3mm}\curvearrowleftright}{L^{v_1}_{zx}}       \sigma^b   \right) \right\}
 \right\} \right\}
\label{edderkopper!}
\end{eqnarray}
The second term in  (\ref{edderkop!}) appears to be a problem since it has a spatial index structure that cannot correspond to the diffeomorphism constraint. There will, however, be a term with the same index structure coming from the commutator
\begin{eqnarray}
\left[\mathds{H}^{(xy)}_{v_3}(N_{v_3}),\mathds{H}^{(yz)}_{v_2}(N'_{v_2})\right] &&\label{haandvaark}
\end{eqnarray}
where $v_3$ is the vertex neighboring $v_1$ in the $z$-direction, see figure \ref{leningrad}. This term will have the opposite sign since the intersection $L^{v_2}_{yz}  \cap L^{v_3}_{xy}$ has the opposite orientation as that of $L^{v_2}_{yz} \cap L^{v_1}_{xy}$. Also, this term will be proportional to $N_{v_3}N'_{v_2}$, which in the $n\rightarrow\infty$ limit gives a $N (x) (\pa_x + \pa_y) N'(x)$. The $N(x)\pa_x N'(x)$-term will cancel the first term in  (\ref{edderkop!}) and the $N(x)\pa_y N'(x)$-term will contribute to the computation in the $y$-direction. This term coming from the commutator (\ref{haandvaark}) will of course only be identical to the second term in (\ref{edderkop!}) up to another term, which involves a factor 
\begin{equation}
\left( \left\{ \cl^{v_1}_{{\bf e}^a_x}  , \mbox{Tr} \left( \sigma^a  \stackrel{\hspace{-3mm}\curvearrowleftright}{L^{v_1}_{xy}}       \sigma^b   \right) \right\} -   \left\{ \cl^{v_3}_{{\bf e}^a_x}  , \mbox{Tr} \left( \sigma^a  \stackrel{\hspace{-3mm}\curvearrowleftright}{L^{v_1}_{xy}}       \sigma^b   \right)\right\} \right)
 \label{Toronto}
 \end{equation}
which we must keep in mind in the end of this computation.

In a completely parallel manner the second term in (\ref{edderkopper!}) also has a wrong index structure and again we find that this term will cancel -- up to a term similar to (\ref{Toronto}) -- with a term coming from a commutator 
$$
\left[\mathds{H}^{(zx)}_{v_4}(N_{v_4}),\mathds{H}^{(yz)}_{v_2}(N'_{v_2})\right] 
$$
where $v_4$ is the vertex neighboring $v_1$ in the $y$-direction, see figure \ref{leningrad}.

Notice that two identical terms come from the commutator involving $v_3$ and $v_2$ and the commutator involving $v_4$ and $v_2$. 

Let us now continue our analysis of the first term in both (\ref{edderkop!}) and (\ref{edderkopper!}). We find
\begin{eqnarray}
\mbox{(\ref{edderkop!}) and (\ref{edderkopper!})}\Big\vert_{\mbox{\tiny first terms}} \hspace{1cm} &&\nn\\
&&\hspace{-4cm}= 2^{4n-3}N_{v_1}N'_{v_2}
\left(
- \left\{ \cl^{v_2}_{{\bf e}^a_z} ,  \left\{  \cl^{v_1}_{{\bf e}^a_x}     , \left\{     \cl^{v_1}_{{\bf e}^b_y} 
 , \mbox{Tr} \left( \sigma^b  \stackrel{\hspace{-3mm}\curvearrowleftright}{L^{v_2}_{yz}}         \right) \right\} \right\} \right\}\right.
\nn\\&&\hspace{-1.35cm}
 +
  \left\{ \cl^{v_2}_{{\bf e}^b_z} ,  \left\{   \cl^{v_1}_{{\bf e}^b_y}  , \left\{  \cl^{v_1}_{{\bf e}^a_x}    
 , \mbox{Tr} \left( \sigma^a  \stackrel{\hspace{-3mm}\curvearrowleftright}{L^{v_2}_{yz}}         \right) \right\} \right\} \right\}
\nn\\
&&\hspace{-1.35cm}+
 \left\{ \cl^{v_1}_{{\bf e}^b_x} ,  \left\{ \cl^{v_2}_{{\bf e}^b_y}       
, \left\{ \cl^{v_1}_{{\bf e}^a_z}  , \mbox{Tr} \left( \sigma^a  \stackrel{\hspace{-3mm}\curvearrowleftright}{L^{v_2}_{yz}}        \right) \right\} \right\} \right\}
\nn\\&&\hspace{-1.35cm}
-
\left. \left\{ \cl^{v_1}_{{\bf e}^b_x} ,  \left\{ \cl^{v_1}_{{\bf e}^a_z}     
, \left\{ \cl^{v_2}_{{\bf e}^a_y}  , \mbox{Tr} \left( \sigma^b  \stackrel{\hspace{-3mm}\curvearrowleftright}{L^{v_2}_{yz}}        \right) \right\} \right\} \right\}
\right)
%
\label{ukrainee}
\end{eqnarray}
Here the $\cl^{v_i}_{{\bf e}^a_z}$ and $\cl^{v_i}_{{\bf e}^a_y}$ vector-fields display the structure
$$
A_{v_1} B_{v_2} + A_{v_2} B_{v_1}
$$
which again permits us to rewrite (\ref{ukrainee}) as
\begin{eqnarray}
\mbox{(\ref{edderkop!}) and (\ref{edderkopper!})}\Big\vert_{\mbox{\tiny first terms}} =\hspace{0cm} &&\nn\\
&&\hspace{-4cm}= 2^{4n-3}N_{v_1}N'_{v_2}
\left(
 -\left\{  \cl^{v_2}_{{\bf e}^a_z} ,  \left\{  \cl^{v_2}_{{\bf e}^a_x}     , \mathds{D}^{v_2(yz)}_{z}  \right\} \right\}
+
 \left\{ \cl^{v_2}_{{\bf e}^a_x} ,  \left\{ \cl^{v_2}_{{\bf e}^a_y}       
, \mathds{D}^{v_2(yz)}_{y}  \right\} \right\}
\right) 
\nn\\&&\hspace{-4cm}
+\Xi^{(xy,yz)}_{v_1v_2}(N,N')
+\Xi^{(zx,yz)}_{v_1v_2}(N,N')%
\label{komisch}
\end{eqnarray}
where $\Xi^{(xy)(yz)}(N,N') + \Xi^{(zx)(yz)}(N,N')$ is an operator similar to (\ref{likud}), that involves a factor of the form $(A_{v_2}- A_{v_1}) (B_{v_1}-B_{v_2})$.

We do not write down the '$zx$-$zx$' commutators since they are identical to the '$xy$-$xy$' commutators just with '$zx$' replacing '$xy$'.

Thus, we have now identified and analyzed all parts of the commutator (\ref{maaan}), which involves a derivative of the lapse fields $N$ and $N'$ in the $x$-direction. Adding all this up and also adding the $y$- and $z$-directions we find
\begin{eqnarray}
\left[\mathds{H}_n(N), \mathds{H}_n(N')\right] &=& \mathds{D}_n (\bar{N}(N,N')) + \Xi_n(N,N')
\nn
\end{eqnarray}
where $\Xi_n(N,N')$ is a potentially anomalous term that reads
\begin{equation}
\Xi_n(N,N') = \sum_{v_i, v_j} \Xi_{v_iv_j}(N,N')
\label{kreditkarte}
\end{equation}
where the sum runs over adjacent vertices $v_i$ and $v_j$ and where $\Xi_{v_iv_j}(N,N')$ is given by the sum of all additional contributions like (\ref{likud}), which consist of terms that involve an operator of the form $(\cl^{v_i}_{{\bf e}^a_\m} -\cl^{v_j}_{{\bf e}^a_\m} )$ and else are identical to terms contributing to $\mathds{D}_n $ in (\ref{pochahontas}). In $\Xi_n(N,N')$ are also the additional terms involving factor like (\ref{Toronto}), which arise when terms from commutators involving the pairs of vertices $(v_1,v_2)$ cancel with commutators involving $(v_3,v_2)$ and $(v_4,v_2)$ - as described above.


\subsection{The commutator $[\mathds{D}(\bar{N}),\mathds{D}(\bar{N}')]$}
\label{appendixhamm}

We continue the computation of the operator constraint algebra. The task now is to compute the commutator between two diffeomorphism constraint operators
\begin{equation}
[\mathds{D}_n(\bar{N}),\mathds{D}_n(\bar{N}')]\;.
\label{REGNvejr}
\end{equation}
Since we now have three different vertex operators, one for each spatial direction, we first need to check whether we must also include commutators between different vertex operators associated to the same vertex. The relation
\begin{equation}
 \left[  \cl^{v_i}_{{\bf e}^b_\m} ,  \mbox{Tr} (  \sigma^a    \stackrel{\hspace{-3mm}\curvearrowleftright}{L^{v_{i}}_{xy}}       \sigma^c   )     \right]  
 =  \co\left(dx^2\right) \;,
\nn
\end{equation}
implies that we do not need to worry about vertex operators, that involve the same loop, since such contributions will be of such orders in $dx$ that converges to zero in the continuum limit. There will, however, be a contribution from vertex operators, which are associated to the same vertex and which involve perpendicular loops. Such contributions will be problematic since they do not have the right orders of $dx$, i.e. they will be divergent. Thus, already here do we find that our computation of the commutator (\ref{REGNvejr}) runs into severe difficulties.

Nevertheless, let us for now ignore this problem and continue the computation in order to assess how close we come to the wished for result. Thus, we compute the commutator between to vertex operators locates at adjacent vertices and which both involve loops in the '$xy$'-plane. We find
\begin{eqnarray}
&&\hspace{-0.8cm}\left[
\bar{N}_{v_1}^x \mathds{D}_x^{v_1(xy)} -  \bar{N}_{v_1}^y \mathds{D}_y^{v_1(xy)} , \bar{N'}_{v_2}^{x} \mathds{D}_x^{v_2(xy)} -  \bar{N'}_{v_2}^y \mathds{D}_y^{v_2(xy)} 
 \right]  \nn\\
 &&=-2^{4n-2}
 \bar{N}_{v_1}^x  \bar{N'}_{v_2}^x \left( \left\{      \cl^{v_1}_{{\bf e}^a_y}   , \mbox{Tr} \left( \sigma^a  \stackrel{\hspace{-3mm}\curvearrowleftright}{L^{v_2}_{xy}} \right) \right\}
 + \left\{     \cl^{v_2}_{{\bf e}^a_y}   ,   \mbox{Tr} \left( \sigma^a  \stackrel{\hspace{-3mm}\curvearrowleftright}{L^{v_1}_{xy}} \right) \right\}  \right)
\nn\\
&&+2^{4n-2}
   \left(\bar{N}_{v_1}^x  \bar{N'}_{v_2}^y  \left\{     \cl^{v_2}_{{\bf e}^a_x}   ,  \mbox{Tr} \left( \sigma^a  \stackrel{\hspace{-3mm}\curvearrowleftright}{L^{v_1}_{xy}} \right) \right\} 
+ \bar{N}_{v_1}^y  \bar{N'}_{v_2}^x \left\{     \cl^{v_1}_{{\bf e}^a_x}   , \mbox{Tr} \left( \sigma^a  \stackrel{\hspace{-3mm}\curvearrowleftright}{L^{v_2}_{xy}} \right) \right\}\right)
\nn\\ \label{LARMLARM}
\end{eqnarray}
where the loops are again based in $v_1$ and $v_2$, see figure \ref{leningrad}. To obtain (\ref{LARMLARM}) we used
$$
\left[   \cl^{v_i}_{{\bf e}^a_y}  ,\mbox{Tr} \left( \sigma^b  \stackrel{\hspace{-3mm}\curvearrowleftright}{L^{v_{i+1}}_{xy}} \right) \right] =\pm \frac{1}{2}\d^{ab} + \co\left( dx^2 \right)
$$
where the sign again depends on the interaction between the vector-field and the loop, see equation (\ref{dannelse}).

Here again  do we encounter difficulties. The commutator (\ref{LARMLARM}) is not -- as we found in the previous section when we computed the $[\mathds{H}_{v_1}(N),\mathds{H}_{v_2}(N')]$ commutator -- of the general form
$
A_{v_1}B_{v_2} + A_{v_2}B_{v_1}
$ multiplied with a factor involving the shift fields. This means that what amounts in the semi-classical limit to a derivative of an inverse, densitized triad field or a field strength tensor will arise that must be taken into account. This also happens in the computation of the classical Poisson bracket (\ref{two}) but there the derivative is covariant, which is not the case here. \\

Nevertheless, we continue the computation to see how close we get to the wished for result and in doing so simply ignore any further problems with non-covariant derivatives. Thus, we first write
\begin{eqnarray}
&&\hspace{-0.8cm}\left[
\bar{N}_{v_1}^x \mathds{D}_x^{v_1(xy)} -  \bar{N}_{v_1}^y \mathds{D}_y^{v_1(xy)} , \bar{N'}_{v_2}^{x} \mathds{D}_x^{v_2(xy)} -  \bar{N'}_{v_2}^y \mathds{D}_y^{v_2(xy)} 
 \right]  \nn\\
&&=-2^{4n-1}
 \bar{N}_{v_1}^x  \bar{N'}_{v_2}^x \left( \mathds{D}_x^{v_2(xy)}
 + \mathds{D}_x^{v_1(xy)} \right)
\nn\\
&&
+2^{4n-1}
    \left( \bar{N}_{v_1}^x  \bar{N'}_{v_2}^y\mathds{D}_y^{v_1(xy)}
+\bar{N}_{v_1}^y  \bar{N'}_{v_2}^x \mathds{D}_y^{v_2(xy)}  \right)
 +\Upsilon_{v_1v_2}^{(xy,xy)} (\bar{N},\bar{N'}) \;,
 \label{TORBEN}
\end{eqnarray}
with
\begin{eqnarray}
\Upsilon_{v_1v_2}^{(xy,xy)} (\bar{N},\bar{N'}) 
&=&-2^{4n-2}
 \bar{N}_{v_1}^x  \bar{N'}_{v_2}^x \left( \left\{    \left(  \cl^{v_1}_{{\bf e}^a_y} -  \cl^{v_2}_{{\bf e}^a_y}\right)  , \mbox{Tr} \left( \sigma^a  \stackrel{\hspace{-3mm}\curvearrowleftright}{L^{v_2}_{xy}} \right) \right\}
\right.\nn\\
&&\hspace{2.5cm}\left.
 + \left\{   \left(  \cl^{v_2}_{{\bf e}^a_y} -  \cl^{v_1}_{{\bf e}^a_y} \right) ,   \mbox{Tr} \left( \sigma^a  \stackrel{\hspace{-3mm}\curvearrowleftright}{L^{v_1}_{xy}} \right) \right\}  \right)
\nn\\
&&+2^{4n-2}
  \left(  \bar{N}_{v_1}^x  \bar{N'}_{v_2}^y  \left\{  \left(   \cl^{v_2}_{{\bf e}^a_x} -  \cl^{v_1}_{{\bf e}^a_x}  \right)  ,  \mbox{Tr} \left( \sigma^a  \stackrel{\hspace{-3mm}\curvearrowleftright}{L^{v_1}_{xy}} \right) \right\} 
\right.\nn\\
&&\hspace{1.2cm}\left.
+  \bar{N}_{v_1}^y  \bar{N'}_{v_2}^x  \left\{   \left(  \cl^{v_1}_{{\bf e}^a_x}  -  \cl^{v_2}_{{\bf e}^a_x}   \right) , \mbox{Tr} \left( \sigma^a  \stackrel{\hspace{-3mm}\curvearrowleftright}{L^{v_2}_{xy}} \right) \right\}\right)
\label{facebook}
 \end{eqnarray}

Next we consider again the commutator involving two loops perpendicular to each other
\begin{eqnarray}
&&\left[
\bar{N}_{v_1}^x \mathds{D}_x^{v_1(xy)} -  \bar{N}_{v_1}^y \mathds{D}_y^{v_1(xy)} , \bar{N'}_{v_2}^{y} \mathds{D}_y^{v_2(yz)} -  \bar{N'}_{v_2}^z \mathds{D}_z^{v_2(yz)} 
 \right]  \nn\\
&&\hspace{5cm} =
2^{4n-2}  \bar{N}_{v_1}^x \bar{N'}_{v_2}^y \left\{     \cl^{v_2}_{{\bf e}^a_z}    , \mbox{Tr} \left( \sigma^a  \stackrel{\hspace{-3mm}\curvearrowleftright}{L^{v_1}_{xy}} \right)  \right\}
\nn\\
&&\hspace{5cm} +
2^{4n-2}  \bar{N}_{v_1}^x \bar{N'}_{v_2}^z \left\{     \cl^{v_1}_{{\bf e}^a_y}    , \mbox{Tr} \left( \sigma^a  \stackrel{\hspace{-3mm}\curvearrowleftright}{L^{v_2}_{yz}} \right) \right\}
\nn\\
&&\hspace{5cm} -
2^{4n-2}  \bar{N}_{v_1}^x \bar{N'}_{v_2}^z \left\{     \cl^{v_2}_{{\bf e}^a_y}     , \mbox{Tr} \left( \sigma^a  \stackrel{\hspace{-3mm}\curvearrowleftright}{L^{v_1}_{xy}} \right) \right\} 
\nn\\
&&\hspace{5cm} -
2^{4n-2}  \bar{N}_{v_1}^y   \bar{N'}_{v_2}^z \left\{       \cl^{v_1}_{{\bf e}^a_x}   , \mbox{Tr} \left( \sigma^a  \stackrel{\hspace{-3mm}\curvearrowleftright}{L^{v_2}_{yz}} \right) \right\}
\label{nisse}
\end{eqnarray}
Similar to what happened when we computed the $[\mathds{H}_n(N),\mathds{H}_n(N')]$ commutator we here see that the first and third term in (\ref{nisse}) match similar terms coming from the commutator
$$
\left[ \bar{N}^x_{v_3} \mathds{D}^{v_3(xy)}_x -\bar{N}^y_{v_3} \mathds{D}^{v_3(xy)}_y   ,    \bar{N'}^y_{v_2} \mathds{D}^{v_2(yz)}_{y} -   \bar{N'}^z_{v_2} \mathds{D}^{v_2(yz)}_{z}\   \right] \;.
$$
We continue with
\begin{eqnarray}
&&\left[
\bar{N}_{v_1}^z \mathds{D}_z^{v_1(zx)} -  \bar{N}_{v_1}^x \mathds{D}_x^{v_1(zx)} , \bar{N'}_{v_2}^{y} \mathds{D}_y^{v_2(yz)} -  \bar{N'}_{v_2}^z \mathds{D}_z^{v_2(yz)} 
 \right]  \nn\\
&&\hspace{4cm} =
 2^{4n-2} \bar{N}_{v_1}^z  \bar{N'}_{v_2}^y \left\{        \cl^{v_1}_{{\bf e}^a_x}    , \mbox{Tr} \left( \sigma^a  \stackrel{\hspace{-3mm}\curvearrowleftright}{L^{v_2}_{yz}} \right) \right\}
 \nn\\
 &&\hspace{4cm}
-2^{4n-2} \bar{N}_{v_1}^x \bar{N'}_{v_2}^y \left\{        \cl^{v_1}_{{\bf e}^a_z}    , \mbox{Tr} \left( \sigma^a  \stackrel{\hspace{-3mm}\curvearrowleftright}{L^{v_2}_{yz}} \right) \right\}
 \nn\\
 &&\hspace{4cm}
+2^{4n-2} \bar{N}_{v_1}^x \bar{N'}_{v_2}^y \left\{     \cl^{v_2}_{{\bf e}^a_z}    , \mbox{Tr} \left( \sigma^a  \stackrel{\hspace{-3mm}\curvearrowleftright}{L^{v_1}_{zx}} \right)   \right\}
 \nn\\
 &&\hspace{4cm}
- 2^{4n-2} \bar{N}_{v_1}^x\bar{N'}_{v_2}^z \left\{     \cl^{v_2}_{{\bf e}^a_y}    , \mbox{Tr} \left( \sigma^a  \stackrel{\hspace{-3mm}\curvearrowleftright}{L^{v_1}_{zx}} \right) \right\}
\label{edd}
\end{eqnarray}
where we again note that the last two terms match similar terms coming from the commutator
$$
\left[ \bar{N}^z_{v_4} \mathds{D}^{v_4(zx)}_z -\bar{N}^x_{v_4} \mathds{D}^{v_4(zx)}_x   ,    \bar{N'}^y_{v_2} \mathds{D}^{v_2(yz)}_{y} -   \bar{N'}^z_{v_2} \mathds{D}^{v_2(yz)}_{z}\   \right] \;.
$$
Thus, we continue with
\begin{eqnarray}
\mbox{(\ref{nisse}) and (\ref{edd})}\Big\vert_{\mbox{\tiny relevant terms}} =
&&
2^{4n-2}  \bar{N}_{v_1}^x \bar{N'}_{v_2}^z \left\{     \cl^{v_1}_{{\bf e}^a_y}    , \mbox{Tr} \left( \sigma^a  \stackrel{\hspace{-3mm}\curvearrowleftright}{L^{v_2}_{yz}} \right) \right\}
\nn\\
&&
-2^{4n-2}  \bar{N}_{v_1}^y   \bar{N'}_{v_2}^z \left\{       \cl^{v_1}_{{\bf e}^a_x}   , \mbox{Tr} \left( \sigma^a  \stackrel{\hspace{-3mm}\curvearrowleftright}{L^{v_2}_{yz}} \right) \right\}
\nn\\
&&
+ 2^{4n-2} \bar{N}_{v_1}^z  \bar{N'}_{v_2}^y \left\{        \cl^{v_1}_{{\bf e}^a_x}    , \mbox{Tr} \left( \sigma^a  \stackrel{\hspace{-3mm}\curvearrowleftright}{L^{v_2}_{yz}} \right) \right\}
 \nn\\
 &&
-2^{4n-2} \bar{N}_{v_1}^x \bar{N'}_{v_2}^y \left\{        \cl^{v_1}_{{\bf e}^a_z}    , \mbox{Tr} \left( \sigma^a  \stackrel{\hspace{-3mm}\curvearrowleftright}{L^{v_2}_{yz}} \right) \right\}
\label{tordenvejr}
\end{eqnarray}
Here we immediately notice the two middle terms, which do not appear to have the index structure required for a diffeomorphism constraint operator. If we look at the front factors to these terms and also include those of the opposite commutators, where $\bar{N}$ and $\bar{N'}$ are interchanged and a sign is added, and if we consider the continuum limit of these factors, then we find a total derivative:
\begin{eqnarray}
&&-\bar{N}_{v_1}^y   \bar{N'}_{v_2}^z +  \bar{N}_{v_1}^z  \bar{N'}_{v_2}^y 
+\bar{N'}_{v_1}^y   \bar{N}_{v_2}^z -  \bar{N'}_{v_1}^z  \bar{N}_{v_2}^y 
\nn\\
&\stackrel{\mbox{\tiny continuum}}{\longrightarrow}&
\pa_x\bar{N}^y   \bar{N'}^z -  \pa_x \bar{N}^z  \bar{N'}^y 
-\pa_x \bar{N'}^y   \bar{N}^z +\pa_x  \bar{N'}^z  \bar{N}^y 
\nn\\
&=& \pa_x \left(   \bar{N}^y  \bar{N'}^z- \bar{N}^z   \bar{N'}^y    \right)\;.
\label{Henkel}
\end{eqnarray}
Here we ignored that there is also a contribution in (\ref{Henkel}) without a derivative, which will be divergent.
Further, if we consider also the classical correspondent to the operator $\tfrac{1}{2}\left\{        \cl^{v_1}_{{\bf e}^a_x}    , \mbox{Tr} \left( \sigma^a  \stackrel{\hspace{-3mm}\curvearrowleftright}{L^{v_2}_{yz}} \right) \right\}$, which is $E^x_a F^a_{yz}=\mbox{Tr}\left( E^x F_{yz}\right)$, then we find
\begin{eqnarray}
&& \mbox{Tr} \left( D_x \left(\left( \bar{N}^y  \bar{N'}^z   - \bar{N}^z   \bar{N'}^y \right)  E^x \right) F_{yz} \right)
\nn\\
&=&-\left( \bar{N}^y  \bar{N'}^z   - \bar{N}^z   \bar{N'}^y    \right)\mbox{Tr}\left(  E^x  D_x F_{yz}     \right)
\end{eqnarray}
where we used partial integration and interchanged $\pa_x$ with the covariant derivative $D_x$. Now, we use the Bianchi identity to proceed
\begin{eqnarray}
&&-\left( \bar{N}^y  \bar{N'}^z   - \bar{N}^z   \bar{N'}^y    \right) \mbox{Tr}\left( E^x  D_x F_{yz}  \right)
\nn\\
&=&- \left( \bar{N}^y  \bar{N'}^z   - \bar{N}^z   \bar{N'}^y    \right) \mbox{Tr}\left( \left(  D_y E^x\right)  F_{zx} \right)  - \pa_y \left( \bar{N}^y  \bar{N'}^z   - \bar{N}^z   \bar{N'}^y    \right)\mbox{Tr}\left( E^x  F_{zx}     \right)
\nn\\
&& -\left( \bar{N}^y  \bar{N'}^z   - \bar{N}^z   \bar{N'}^y    \right) \mbox{Tr}\left(\left( D_z E^x\right)   F_{xy}      \right) -  \pa_z\left( \bar{N}^y  \bar{N'}^z   - \bar{N}^z   \bar{N'}^y    \right)\mbox{Tr}\left( E^x  F_{xy}   \right) 
\nn
\end{eqnarray}
If we add the classical correspondents to the remaining two terms in equation (\ref{tordenvejr}) and the classical correspondents to (\ref{TORBEN}) as well as the '$zx-zx$' commutators, which we again do not write down, then we obtain the classical expression
\begin{eqnarray}
-\left(  \bar{N}^\a \pa_\a \bar{N'}^\m -    \bar{N'}^\a \pa_\a \bar{N}^\m   \right) E^\n_a F^a_{\m\n} + \mbox{"e.o.m. terms"}
\end{eqnarray}
where "e.o.m. terms" are terms which involve a $D_\m E^\n_a$. Thus, here too do we encounter derivatives of the triad fields, but now they are, as they should be, covariant. In the classical computation such derivative terms cancel out and leave only a Gauss constraint. Here they remain and form the torsion tensor coupled to the field strength tensor and the lapse fields.

Of course, this final 'ad-hoc' analysis of equation (\ref{tordenvejr}) is classical and should be elevated to the level of operators. This means formulating partial integration and the Bianchi identity for this setting.

We see, however, that we come close to reproducing the structure of the classical constraint algebra. Apart from terms like the one we found in (\ref{facebook}), which fail to give at the operator level what corresponds to covariant derivatives, and apart from terms, which are simply divergent, we discovered the possibility that an anomaly proportional to a torsion operator might emerge. Despite the fact that the computational setup that leads to this potential anomaly is flawed, we think that the overall structure of this computation might reflect something more general, that could carry over into a setting, that involves a more realistic candidate for a diffeomorphism constraint operator.

Our analysis of the commutator (\ref{REGNvejr}) can also be used to shed light on what a more realistic candidate for a Hamilton constraint operator should look like. First of all, in order to avoid the initial problem that vertex operators associated to the same vertex do not commute, we think one needs to built a Hamilton constraint operator, which fully incorporates the algebraic structure dictated by the operator $\mathds{h}_n$ in (\ref{cur}). This means that vector-fields should only be associated to edges, which do not have the base-point of the loop as a start or endpoint. Secondly, the problem with the non-covariant derivatives of the triad fields might be avoided if one is able to obtain the algebraic structure $A_{v_1}B_{v_2} + A_{v_2}B_{v_1}$ multiplied with a factor involving the lapse fields. To obtain this structure from the commutator of two constraint operators one needs to include vertex operators, which involve loops in all possible directions. In the present setup, the Hamilton and diffeomorphism constraint operators only involve three loops based in a given vertex. However, twelve loops are possible and it is these loops that we think needs to be included.

What we have in mind is a Hamilton constraint operator built from vertex operators of the form
$$
 \mathds{H}_{v_i}(N_{v_i}) =  \sum_{k=1}^{12}   \mathds{H}^{k}_{v_i}(N_{v_i})    
$$
where the sum runs over all 12 possible infinitesimal loops based in each vertex $v_i$, with
\begin{eqnarray}
 \mathds{H}^{k}_{v_i}(\bar{N}_{v_i}) 
 &=& 2^{2n-2}  N_{v_i}  \left\{ \cl^{(v_i,k)}_{{\bf e}^a_i} , \left\{ \cl^{(v_i,k)}_{{\bf e}^b_j}  , \mbox{Tr} \left(  \sigma^a \stackrel{\hspace{-3mm}\curvearrowleftright}{L^{v_i}_{k}  }   \sigma^b     \right) \right\} \right\}
\end{eqnarray}
where $\cl^{(v_i,k)}_{{\bf e}^c_j} $ is the vector-field associated to the edge $l_j$ in the $k$'th loop based in $v_i$ and where the edges $l_i$ and $l_j$ are the two edges in the loop, which are not connected to $v_i$.

It requires, however, a significant computational effort to check the complete operator constraint algebra with such an operator, and thus we leave it to be done elsewhere.\\

The commutator $[\mathds{D}(\bar{N}),\mathds{H}(N)]$ exhibits similar anomalies and we shall not write it down.

\end{appendix}

\end{document}

%% file: ogsaasvedig.pdf_t.tex
\begin{picture}(0,0)%
\includegraphics{ogsaasvedig.pdf}%
\end{picture}%
\setlength{\unitlength}{4144sp}%
\begingroup\makeatletter\ifx\SetFigFont\undefined%
\gdef\SetFigFont#1#2#3#4#5{%
  \reset@font\fontsize{#1}{#2pt}%
  \fontfamily{#3}\fontseries{#4}\fontshape{#5}%
  \selectfont}%
\fi\endgroup%
\begin{picture}(7029,5470)(1094,-5708)
\put(1666,-601){\makebox(0,0)[lb]{\smash{{\SetFigFont{25}{30.0}{\rmdefault}{\mddefault}{\updefault}{\color[rgb]{0,0,0}$\xi (m)$}%
}}}}
\put(6436,-871){\makebox(0,0)[lb]{\smash{{\SetFigFont{25}{30.0}{\rmdefault}{\mddefault}{\updefault}{\color[rgb]{0,0,0}$\hbox{Hol}(\gamma,\nabla) \xi(m)$}%
}}}}
\put(3151,-3886){\makebox(0,0)[lb]{\smash{{\SetFigFont{25}{30.0}{\rmdefault}{\mddefault}{\updefault}{\color[rgb]{0,0,0}$\gamma$}%
}}}}
\end{picture}%

%% file: svedigtegning1.pdf_t.tex
\begin{picture}(0,0)%
\includegraphics{svedigtegning1.pdf}%
\end{picture}%
\setlength{\unitlength}{4144sp}%
\begingroup\makeatletter\ifx\SetFigFont\undefined%
\gdef\SetFigFont#1#2#3#4#5{%
  \reset@font\fontsize{#1}{#2pt}%
  \fontfamily{#3}\fontseries{#4}\fontshape{#5}%
  \selectfont}%
\fi\endgroup%
\begin{picture}(9839,3644)(999,-3683)
\end{picture}%

%% file: AAAcubes.pdf_t
\begin{picture}(0,0)%
\includegraphics{AAAcubes.pdf}%
\end{picture}%
\setlength{\unitlength}{4144sp}%
\begingroup\makeatletter\ifx\SetFigFont\undefined%
\gdef\SetFigFont#1#2#3#4#5{%
  \reset@font\fontsize{#1}{#2pt}%
  \fontfamily{#3}\fontseries{#4}\fontshape{#5}%
  \selectfont}%
\fi\endgroup%
\begin{picture}(3894,1239)(3544,-5518)
\end{picture}%

%% file: virkning.pdf_t.tex
\begin{picture}(0,0)%
\includegraphics{virkning.pdf}%
\end{picture}%
\setlength{\unitlength}{4144sp}%
\begingroup\makeatletter\ifx\SetFigFont\undefined%
\gdef\SetFigFont#1#2#3#4#5{%
  \reset@font\fontsize{#1}{#2pt}%
  \fontfamily{#3}\fontseries{#4}\fontshape{#5}%
  \selectfont}%
\fi\endgroup%
\begin{picture}(7224,3624)(889,-3673)
\end{picture}%

%% file: ABOXIII.pdf_t
\begin{picture}(0,0)%
\includegraphics{ABOXIII.pdf}%
\end{picture}%
\setlength{\unitlength}{4144sp}%
\begingroup\makeatletter\ifx\SetFigFont\undefined%
\gdef\SetFigFont#1#2#3#4#5{%
  \reset@font\fontsize{#1}{#2pt}%
  \fontfamily{#3}\fontseries{#4}\fontshape{#5}%
  \selectfont}%
\fi\endgroup%
\begin{picture}(4544,3402)(-2271,-1160)
\put(-89,-1096){\makebox(0,0)[lb]{\smash{{\SetFigFont{12}{14.4}{\rmdefault}{\mddefault}{\updefault}{$\Delta x^1$}%
}}}}
\put(-269,929){\makebox(0,0)[lb]{\smash{{\SetFigFont{12}{14.4}{\rmdefault}{\mddefault}{\updefault}{$v_i$}%
}}}}
\put(271,659){\makebox(0,0)[lb]{\smash{{\SetFigFont{12}{14.4}{\rmdefault}{\mddefault}{\updefault}{$l_j$}%
}}}}
\put( 91,1379){\makebox(0,0)[lb]{\smash{{\SetFigFont{12}{14.4}{\rmdefault}{\mddefault}{\updefault}{$\Delta S^1_j$}%
}}}}
\end{picture}%

%% file: roevviktor.pdf_t
\begin{picture}(0,0)%
\includegraphics{roevviktor.pdf}%
\end{picture}%
\setlength{\unitlength}{4144sp}%
\begingroup\makeatletter\ifx\SetFigFont\undefined%
\gdef\SetFigFont#1#2#3#4#5{%
  \reset@font\fontsize{#1}{#2pt}%
  \fontfamily{#3}\fontseries{#4}\fontshape{#5}%
  \selectfont}%
\fi\endgroup%
\begin{picture}(4544,4094)(2004,-5258)
\put(4546,-4786){\makebox(0,0)[lb]{\smash{{\SetFigFont{14}{16.8}{\rmdefault}{\mddefault}{\updefault}{\color[rgb]{0,0,0}$v_2$}%
}}}}
\put(2386,-2761){\makebox(0,0)[lb]{\smash{{\SetFigFont{14}{16.8}{\rmdefault}{\mddefault}{\updefault}{\color[rgb]{0,0,0}$v_3$}%
}}}}
\put(3106,-4786){\makebox(0,0)[lb]{\smash{{\SetFigFont{14}{16.8}{\rmdefault}{\mddefault}{\updefault}{\color[rgb]{0,0,0}$x$}%
}}}}
\put(3286,-3616){\makebox(0,0)[lb]{\smash{{\SetFigFont{14}{16.8}{\rmdefault}{\mddefault}{\updefault}{\color[rgb]{0,0,0}$v_4$}%
}}}}
\put(2341,-4606){\makebox(0,0)[lb]{\smash{{\SetFigFont{14}{16.8}{\rmdefault}{\mddefault}{\updefault}{\color[rgb]{0,0,0}$v_1$}%
}}}}
\put(3196,-4246){\makebox(0,0)[lb]{\smash{{\SetFigFont{14}{16.8}{\rmdefault}{\mddefault}{\updefault}{\color[rgb]{0,0,0}$y$}%
}}}}
\put(2431,-4156){\makebox(0,0)[lb]{\smash{{\SetFigFont{14}{16.8}{\rmdefault}{\mddefault}{\updefault}{\color[rgb]{0,0,0}$z$}%
}}}}
\end{picture}%